%% file: HIN-14-005_temp.tex
\pdfoutput=1

\documentclass[11pt,twoside,a4paper,cmspaper,final,collab]{cms-tdr}
\def\svnVersion{403822}\def\cmsCernNoTag{CERN-EP/2016-243}\def\cmsCernDate{\today}\def\cmsMessage{Published in the European Physical Journal C as \href{http://dx.doi.org/10.1140/epjc/s10052-017-4781-1}{\doi{10.1140/epjc/s10052-017-4781-1.}}}

\begin{document}\cmsNoteHeader{HIN-14-005}

\hyphenation{had-ron-i-za-tion}
\hyphenation{cal-or-i-me-ter}
\hyphenation{de-vices}
\RCS$Revision: 398903 $
\RCS$HeadURL: svn+ssh://svn.cern.ch/reps/tdr2/papers/HIN-14-005/trunk/HIN-14-005.tex $
\RCS$Id: HIN-14-005.tex 398903 2017-04-12 13:31:51Z mironov $
\newlength\cmsFigWidth
\ifthenelse{\boolean{cms@external}}{\setlength\cmsFigWidth{0.40\textwidth}}{\setlength\cmsFigWidth{0.48\textwidth}}
\ifthenelse{\boolean{cms@external}}{\providecommand{\cmsLeft}{top\xspace}}{\providecommand{\cmsLeft}{left\xspace}}
\ifthenelse{\boolean{cms@external}}{\providecommand{\cmsRight}{bottom\xspace}}{\providecommand{\cmsRight}{right\xspace}}
\ifthenelse{\boolean{cms@external}}{\providecommand{\cmsLLeft}{top\xspace}}{\providecommand{\cmsLLeft}{upper left\xspace}}
\ifthenelse{\boolean{cms@external}}{\providecommand{\cmsRRight}{middle\xspace}}{\providecommand{\cmsRRight}{upper right\xspace}}
\ifthenelse{\boolean{cms@external}}{\providecommand{\cmsBottom}{bottom\xspace}}{\providecommand{\cmsbottom}{left\xspace}}
\providecommand{\mbinv} {\mbox{\ensuremath{\,\text{mb}^\text{$-$1}}}\xspace}

\providecommand{\PQQ}{\ensuremath{\cmsSymbolFace{Q}}\xspace}
\providecommand{\PAQQ}{\ensuremath{\overline{\cmsSymbolFace{Q}}}\xspace}
\providecommand{\QQbar}{\ensuremath{\PQQ\PAQQ}\xspace}
\providecommand{\Qqbar}{\ensuremath{\PQQ\PAQq}\xspace}
\newcommand{\mumu}{\ensuremath{\PGmp\PGmm}\xspace}
\newcommand{\dphi}{\ensuremath{\Delta\Phi}\xspace}
\newcommand {\npart}  {\ensuremath{N_{\text{part}}}\xspace}
\newcommand {\ncoll}  {\ensuremath{N_{\text{coll}}}\xspace}
\newcommand{\raa}{\ensuremath{R_{\mathrm{AA}}}\xspace}
\newcommand{\taa}{\ensuremath{T_{\mathrm{AA}}}\xspace}
\newcommand{\vtwo}{\ensuremath{v_{2}}\xspace}
\newcommand{\pp}{{\ensuremath{\Pp\Pp}}\xspace}
\newcommand{\PbPb}{\ensuremath{\mathrm{PbPb}}\xspace}
\newcommand{\sqrts}{\ensuremath{\sqrt{s}}\xspace}
\newcommand{\sqrtsnn}{\ensuremath{\sqrt{\smash[b]{s_{_{\text{NN}}}}}}\xspace}
\providecommand{\HYDJET} {{\textsc{hydjet}}\xspace}
\newcommand{\psiP}{\ensuremath{\psi\text{(2S)}}\xspace}
\newcommand{\nAA}{\ensuremath{N_{\text{PbPb}}^{\JPsi}}}
\newcommand{\sigPP}{\ensuremath{\sigma_{\text{pp}}^{\JPsi}}}

\cmsNoteHeader{HIN-14-005}
\title{Suppression and azimuthal anisotropy of prompt and nonprompt $\JPsi$ production in PbPb collisions at $\sqrtsnn=2.76$\TeV}
\titlerunning{Suppression and azimuthal anisotropy of $\JPsi$ production in PbPb collisions at 2.76\TeV}

\date{\today}

\abstract{The nuclear modification factor \raa and the azimuthal anisotropy coefficient \vtwo of prompt and nonprompt (\ie those from decays of b hadrons) \JPsi mesons, measured from PbPb and pp collisions at $\sqrtsnn=2.76$\TeV at the LHC, are reported. The results are presented in several event centrality intervals and several kinematic regions, for transverse momenta $\pt>6.5$\GeVc and rapidity $\abs{y}<2.4$, extending down to $\pt=3$\GeVc in the $1.6<\abs{y}<2.4$ range. The \vtwo of prompt \JPsi is found to be nonzero, but with no strong dependence on centrality, rapidity, or \pt over the full  kinematic range studied. The measured \vtwo of nonprompt \JPsi is consistent with zero. The \raa of prompt \JPsi exhibits a suppression that increases from peripheral to central collisions but does not vary strongly as a function of either $y$ or \pt in the fiducial range. The nonprompt \JPsi \raa shows a suppression which becomes stronger as rapidity or \pt increases. The \vtwo and \raa of open and hidden charm, and of open charm and beauty, are compared.}

\hypersetup{%
pdfauthor={CMS Collaboration},%
pdftitle={Suppression and azimuthal anisotropy of prompt and nonprompt J/psi production in PbPb collisions at sqrt(s[NN]) = 2.76 TeV},%
pdfsubject={CMS},%
pdfkeywords={physics, dimuons, heavy ions, charmonia, heavy flavour, suppression, quark gluon plasma }}

\maketitle
\section{Introduction}
\label{sec:introduction}

Recent data from RHIC and the CERN LHC for mesons containing charm and beauty quarks have allowed more detailed theoretical and experimental studies~\cite{Andronic:2015wma} of the phenomenology of these heavy quarks in a deconfined quark gluon plasma (QGP)~\cite{Shuryak:1977ut} at large energy densities and high temperatures~\cite{Karsch:2003jg}. Heavy quarks, whether as quarkonium states \QQbar (hidden heavy flavour)~\cite{Matsui:1986dk} or as mesons made of heavy-light quark-antiquark pairs \Qqbar (open heavy flavour)~\cite{Dokshitzer:2001zm}, are considered key probes of the QGP, since their short formation time allows them to probe all stages of the QGP evolution~\cite{Andronic:2015wma}.

At LHC energies, the inclusive \JPsi yield contains a significant
nonprompt contribution from \PQb hadron decays
\cite{Aaij:2011jh, Khachatryan:2010yr, Aad:2011sp},
offering the opportunity of studying both open beauty and hidden charm in the
same measurement. Because of the long lifetime ($\mathcal{O}(500)\mum/c$) of \PQb hadrons, compared to the QGP lifetime
($\mathcal{O}(10)\unit{fm}/c$), the nonprompt contribution should not suffer
from colour screening of the potential between the \PQQ and the \PAQQ by the surrounding
light quarks and gluons, which decreases the prompt quarkonium
yield~\cite{Mocsy:2007jz}. Instead, the nonprompt contribution should reflect the energy loss of \PQb quarks in the medium.
The importance of an unambiguous and detailed measurement of open beauty flavour is driven by the need to understand key features of the dynamics of parton interactions and hadron formation
in the QGP: the colour-charge and parton-mass
dependences for the in-medium interactions~\cite{Braaten:1991we,Dokshitzer:2001zm,Zhang:2003wk,Armesto:2005iq,vanHees:2005wb}, the relative contribution of radiative and collisional energy loss~\cite{Peigne:2008nd,Wicks:2007am,Gossiaux:2010yx}, and the effects of different hadron formation times~\cite{Adil:2006ra,Sharma:2009hn}. Another
aspect of the heavy-quark phenomenology in the QGP concerns
differences in the behaviour (energy loss mechanisms, amount and strength of interactions with the surrounding medium) of a \QQbar pair (the pre-quarkonium state) relative to that of a single heavy quark $Q$ (the pre-meson component)~\cite{Satz:2013ama,Riek:2010fk,Sharma:2012dy}.

Experimentally, modifications to the particle production are usually quantified by the ratio of
the yield measured in heavy ion collisions to that in proton-proton (\pp) collisions, scaled by the mean number of binary nucleon-nucleon (NN) collisions. This ratio is called the nuclear
modification factor \raa. In the absence of medium effects, one would
expect $\raa = 1$ for hard processes, which scale with the number of
NN collisions. The \raa for prompt and
nonprompt \JPsi have been previously measured in \PbPb at $\sqrtsnn=2.76$\TeV by CMS in bins of transverse momentum (\pt), rapidity ($y$) and collision centrality~\cite{Chatrchyan:2012np}. A strong centrality-dependent suppression has been observed for \JPsi with $\pt > 6.5\GeVc$. The
ALICE Collaboration has measured \JPsi down to $\pt=0$\GeVc in the electron channel at midrapidity ($\abs{y}<0.8$)~\cite{Adam:2015rba} and in the muon channel at forward rapidity ($2.5<y<4$)~\cite{Abelev:2012rv}. Except for the most peripheral event selection, a suppression of inclusive \JPsi meson production is observed for all collision centralities. However, the suppression is smaller than that at $\sqrtsnn=0.2$\TeV~\cite{Adare:2011yf}, smaller at midrapidity than at forward rapidity, and, in the forward region, smaller for $\pt<2$\GeVc than for $5<\pt<8$\GeVc~\cite{Adam:2015isa}. All these results were interpreted as evidence that the measured prompt \JPsi yield is the result of an interplay between a) primordial production (\JPsi produced in the initial hard-scattering of the collisions), b) colour screening and energy loss (\JPsi destroyed or modified by interactions with the surrounding medium), and c) recombination/regeneration mechanisms in a deconfined partonic medium, or at the time of hadronization (\JPsi created when a free charm and a free anti-charm quark come close enough to each other to form a bound state)~\cite{Zhao:2011cv, Andronic:2011yq, Ferreiro:2012rq}.

A complement to the \raa measurement is the elliptic anisotropy coefficient \vtwo. This is the
second Fourier coefficient in the expansion of the azimuthal angle ($\ensuremath{\Phi}$)
distribution of the \JPsi mesons,
$\rd N/\rd\Phi \propto 1+2\vtwo\cos[2(\Phi-\Psi_\mathrm{PP})]$
with respect to $\Psi_\mathrm{PP}$, the azimuthal angle
of the ``participant plane" calculated for each event. In a noncentral heavy ion collision,
the overlap region of the two colliding nuclei has a lenticular shape.
The participant plane is defined by the beam direction
and the direction of the shorter axis of the lenticular region.
Typical sources for a nonzero elliptic anisotropy are a path length difference
arising from energy loss of particles traversing the
reaction zone, or different pressure gradients  along the short and long axes. Both effects convert the initial spatial anisotropy into a momentum anisotropy $\vtwo$~\cite{Ollitrault:1992bk}.
The effect of energy loss is usually studied using high \pt and/or heavy particles (so-called ``hard probes'' of the medium), for which the parent parton is produced at an early stage of the collision. If the partons are emitted in the direction of the participant plane, they have on average a shorter in-medium path length than partons emitted orthogonally, leading to a smaller modification to their energy or, in the case of \QQbar and the corresponding onium state, a smaller probability of being destroyed. Pressure gradients drive in-medium interactions that can modify the direction of the partons. This effect is most important at low \pt.

The \vtwo of open charm (D mesons) and hidden charm (inclusive \JPsi mesons) was measured at the LHC by the ALICE Collaboration. The D mesons with $2<\pt<6$\GeVc~\cite{Abelev:2014ipa} were found to have a significant positive \vtwo, while for \JPsi mesons with $2<\pt<4$\GeVc there was an indication of nonzero \vtwo~\cite{ALICE:2013xna}. The precision of the results does not yet allow a determination of the origin of the observed anisotropy. One possible interpretation is that charm quarks at low \pt, despite  their much larger mass than those of the $u, s, d$ quarks, participate in the collective expansion of the medium. A second possibility is that there is no collective motion for the charm quarks, and the observed anisotropy is acquired via quark recombination~\cite{BraunMunzinger:2000px,Liu:2009nb,Zhao:2011cv}.

In this paper, the \raa and the \vtwo for prompt and nonprompt \JPsi mesons are
presented in several event centrality intervals and several kinematic regions. The results are based on event samples collected during the 2011 \PbPb and 2013 \pp LHC runs at a nucleon-nucleon centre-of-mass energy of 2.76\TeV, corresponding to integrated luminosities of 152\mubinv and 5.4\pbinv, respectively.

\section{Experimental setup and event selection}
\label{sec:cms}

A detailed description of the CMS detector, together with a definition of the
coordinate system and the relevant kinematic variables, can be found in
Ref.~\cite{bib_CMS}. The central feature of the CMS apparatus is a superconducting
solenoid, of 6\,m internal diameter and 15\unit{m} length.  Within the field volume are the
silicon tracker, the crystal electromagnetic
calorimeter, and the brass and scintillator hadron calorimeter. The CMS apparatus also has extensive forward calorimetry, including two steel and quartz-fiber Cherenkov hadron forward (HF) calorimeters, which cover the range $2.9<\abs{\eta_\text{det}}<5.2$, where $\eta_\text{det}$ is measured from the geometrical centre of the CMS detector. The calorimeter cells, in the $\eta$-$\phi$ plane, form towers projecting radially outwards from close to the nominal interaction point. These detectors are used in the present analysis for the event selection, collision impact parameter determination, and measurement of the azimuthal angle of the participant plane.

Muons are detected in the pseudorapidity window $\abs{\eta}< 2.4$,
by gas-ionization detectors made of three technologies: drift tubes, cathode
strip chambers, and resistive plate chambers, embedded in the steel flux-return yoke of the solenoid. The silicon tracker is composed of pixel detectors (three barrel
layers and two forward disks on either side of the detector, made of
66~million $100{\times}150\mum^2$ pixels) followed by microstrip detectors
(ten barrel layers plus three inner disks and nine forward disks on either side of the
detector, with strip pitch between 80 and 180\mum).

The measurements reported here are based on \PbPb and \pp events selected online (triggered) by a
hardware-based dimuon trigger without an explicit muon momentum threshold (i.e. the actual threshold is determined by the detector acceptance and efficiency of the muon trigger). The same trigger logic was used during the pp and PbPb data taking periods.

In order to select a sample of purely inelastic hadronic PbPb (pp) collisions, the contributions from ultraperipheral collisions and noncollision beam background are removed offline, as described in Ref.~\cite{Chatrchyan:2011sx}. Events are preselected if they contain a reconstructed primary vertex formed by at least two tracks and at least three (one in the case of pp events) HF towers on each side of the interaction point with an energy of at least 3\GeV deposited in each tower. To further suppress the beam-gas events, the distribution of hits in the pixel detector along the beam direction is required to be compatible with particles originating from the event vertex. These criteria select $(97\pm3)$\% ($>$99\%) of inelastic hadronic PbPb (pp) collisions with negligible contamination from non-hadronic interactions~\cite{Chatrchyan:2011sx}.
Using this efficiency it is calculated that the \PbPb sample corresponds to a number of minimum bias (MB) events $N_\mathrm{MB}=(1.16\pm0.04)\times10^9$. The pp data set corresponds to an integrated luminosity of 5.4\pbinv known to an accuracy of 3.7\% from the uncertainty in the calibration based on a van der Meer scan~\cite{CMS-PAS-LUM-13-002}. The two data sets correspond to approximately the same number of elementary NN collisions.

Muons are reconstructed offline using tracks in the muon detectors (``standalone
muons'') that are then matched to tracks in the silicon tracker, using an
algorithm optimized for the heavy ion environment~\cite{Roland:2006kz}.
In addition, an iterative track reconstruction algorithm~\cite{Chatrchyan:2014fea}
is applied to the \PbPb data, limited to regions defined by the standalone
muons. The \pp reconstruction algorithm includes an iterative tracking step
in the full silicon tracker. The final parameters of the muon trajectory are obtained from a global
fit of the standalone muon with a matching track in the silicon tracker.

The centrality of heavy ion collisions, i.e.~the geometrical overlap
of the incoming nuclei, is correlated to the energy released in the
collisions. In CMS, centrality is defined as percentiles of the
distribution of the energy deposited in the HFs. Using a Glauber model
calculation as described in Ref.~\cite{Chatrchyan:2011sx}, one can
estimate variables related to the centrality, such as the mean number
of nucleons participating in the collisions (\npart), the mean number
of binary NN collisions (\ncoll), and the average nuclear
overlap function (\taa)~\cite{Miller:2007ri}. The latter is equal to the number of NN binary collisions divided by the NN
cross section and can be interpreted as the NN-equivalent integrated
luminosity per heavy ion collision, at a given
centrality. In the following, \npart will be the variable used to show
the centrality dependence of the measurements, while \taa directly enters into the nuclear modification factor calculation.
It should be noted that the PbPb hadronic cross section ($7.65 \pm 0.42$\,b), computed with this Glauber simulation, results in an integrated luminosity of $152\pm9$\mubinv, compatible within 1.2~sigma with the integrated luminosity based on the van der Meer scan, which has been evaluated to be $166\pm8$\mubinv. All the \raa results presented in the paper have been obtained using the $N_{\mathrm{MB}}$ event counting that is equivalent to 152\mubinv expressed in terms of integrated luminosity.

Several Monte Carlo (MC) simulated event samples are used to model the signal shapes and evaluate reconstruction, trigger, and selection efficiencies. Samples of prompt and nonprompt \JPsi are
generated with \PYTHIA~6.424~\cite{Sjostrand:2006za} and decayed with
\EVTGEN 1.3.0~\cite{Lange:2001uf}, while the final-state bremsstrahlung is simulated
with \PHOTOS 2.0~\cite{Barberio:1993qi}. The prompt \JPsi is simulated unpolarized, a scenario in good agreement with \pp measurements~\cite{Abelev:2011md,Chatrchyan:2013cla,Aaij:2013nlm}.
For nonprompt \JPsi, the results are reported for the
polarization predicted by \EVTGEN, roughly $\lambda_{\theta} = -0.4$, however not a well-defined value, since in many $\PB \to \JPsi X$ modes the spin alignment is either forced by angular momentum conservation or given as input from measured values of helicity amplitudes in decays. If the acceptances were different in \pp and \PbPb, they would not perfectly cancel in the \raa. This would be the case if, for instance, some physics processes (such as polarization or energy loss) would affect the measurement in \PbPb collisions with a strong kinematic dependence within an analysis bin. As in previous analyses~\cite{Chatrchyan:2011pe,Chatrchyan:2012lxa,Chatrchyan:2013nza,Khachatryan:2014bva}, such possible physics effects are not considered as systematic uncertainties, but a quantitative estimate of this effect for two extreme polarization scenarios can be found in Ref.~\cite{Chatrchyan:2012np}. In the \PbPb case, the \PYTHIA signal events are further embedded in heavy ion events generated with \HYDJET~1.8~\cite{Lokhtin:2005px}, at the level of detector hits and with matching vertices. The detector response was simulated with \GEANTfour~\cite{Agostinelli:2002hh}, and the resulting information was processed through the full event reconstruction chain, including trigger emulation.

\section{Analysis}

Throughout this analysis the same methods for signal extraction and corrections are used for both the \pp and \PbPb data.

\subsection{Corrections}

For both \raa and \vtwo results, correction factors are applied event-by-event to each dimuon, to account for inefficiencies in the trigger, reconstruction, and selection of the \mumu pairs. They were evaluated, using MC samples, in
four dimensions (\pt, centrality, $y$, and $L_{xyz}$) for the \PbPb results, and in three-dimensions (\pt, $y$, and $L_{xyz}$) for the \pp results. After checking that the efficiencies on the prompt and nonprompt \JPsi MC samples
near $L_{xyz}=0$ are in agreement, two efficiency calculations are made. One
calculation is made on the prompt \JPsi MC sample, as a function of \pt, in 10 rapidity intervals between $y=-2.4$ and $y=2.4$, and 4 centrality bins (0--10\%,
10--20\%, 20--40\%, and 40--100\%). For each $y$ and centrality interval, the
\pt dependence of the efficiency is smoothed by fitting it with a Gaussian error
function. A second efficiency is calculated using the nonprompt \JPsi MC sample,
as a function of $L_{xyz}$, in the same $y$ binning, but for coarser \pt bins
and for centrality 0--100\%. This is done in two steps. The efficiency is first
calculated as a function of $L_{xyz}^{\text{true}}$, and then converted into an
efficiency versus measured $L_{xyz}$, using a 2D dispersion map of
$L_{xyz}^{\text{true}}$ \vs $L_{xyz}$. In the end, each dimuon candidate selected in data,
with transverse momentum \pt, rapidity $y$, centrality $c$, and $L_{xyz}=d$\,(mm),
is assigned an efficiency weight equal to
\ifthenelse{\boolean{cms@external}}{
\begin{multline}
w= \text{efficiency}^{\text{prompt \JPsi}}(\pt,y,c,L_{xyz}=0)\\ \times\frac{\text{efficiency}^{\text{nonprompt}\,\JPsi}(\pt,y,L_{xyz}=d)}{\text{efficiency}^{\text{nonprompt}\,\JPsi}(\pt,y,L_{xyz}=0)}.
\end{multline}
}{
\begin{equation}
w= \text{efficiency}^{\text{prompt \JPsi}}(\pt,y,c,L_{xyz}=0) \frac{\text{efficiency}^{\text{nonprompt}\,\JPsi}(\pt,y,L_{xyz}=d)}{\text{efficiency}^{\text{nonprompt}\,\JPsi}(\pt,y,L_{xyz}=0)}.
\end{equation}
}
The individual components of the MC efficiency
(tracking reconstruction, standalone muon reconstruction,
global muon fit, muon identification and selection, triggering) are
cross-checked using single muons from \JPsi decays in simulated and collision
data, with the \textit{tag-and-probe} technique (T\&P) ~\cite{Chatrchyan:2012xi}.
For all but the tracking reconstruction,
scaling factors (calculated as the ratios between the
data and MC T\&P obtained efficiencies), estimated as a function of the muon \pt in
several muon pseudorapidity regions, are used to scale the
dimuon MC-calculated efficiencies. They are applied
event-by-event, as a weight, to each muon that passes all analysis selections and
enter the mass and $\ell_{\JPsi}$ distributions. The weights are similar for the \pp and \PbPb samples, and range from
1.02 to 0.6 for single muons with $\pt>4-5$\GeVc and $\pt<3.5$\GeVc, respectively. For the tracking efficiency, which is above
99\% even in the case of PbPb events, the full difference between data and MC
T\&P results (integrated over all the kinematic region probed) is propagated as
a global (common to all points) systematic uncertainty.

\subsection{Signal extraction}
\label{sec:sig_extraction}
The single-muon acceptance and identification criteria are the same as in
Ref.~\cite{Chatrchyan:2012np}. Opposite-charge muon pairs, with invariant mass between 2.6 and 3.5\GeVcc, are fitted with a common
vertex constraint and are kept if the fit $\chi^2$ probability is larger
than 1\%. Results are presented in up to six bins of absolute \JPsi meson rapidity (equally spaced between 0 and 2.4) integrated over $6.5<\pt<30$\GeVc, up to six bins in \pt
([6.5,8.5], [8.5,9.5], [9.5,11], [11,13], [13,16], [16,30] \GeVc) integrated
over rapidity ($\abs{y}<2.4$), and up to three additional low-\pt bins ([3,4.5], [4.5,5.5], [5.5,6.5]\GeVc) at forward rapidity ($1.6<\abs{y}<2.4$). The lower \pt limit for which the results are reported is imposed by the detector acceptance, the muon reconstruction algorithm, and the selection criteria used in the analysis.
The \PbPb sample is split in bins of collision centrality,
defined using fractions of the inelastic hadronic cross
section where 0\% denotes the most central collisions. This
fraction is determined from the HF energy distribution~\cite{Chatrchyan:2011pb}.
The most central (highest HF energy deposit)
and most peripheral (lowest HF energy deposit) centrality bins used in
the analysis are 0--5\% and 60--100\%, and 0--10\% and 50--100\%, for
prompt and nonprompt \JPsi results, respectively. The rest of the centrality
bins are in increments of 5\% up to 50\% for the high \pt prompt \JPsi results
integrated over $y$, and in increments of 10\% for all other cases.
The \npart values, computed for events with a flat centrality distribution,
range from 381$\pm$2 in the 0--5\% bin to 14$\pm$2 in the 60--100\% bin.
If the events would be distributed according to the number of NN collisions, \ncoll, which is expected for initially produced
hard probes, the average \npart would become 25 instead of 14 for the most peripheral bin, and 41 instead of 22 in the case of the 50--100\% bin. For the other finer bins, the difference is negligible (less than 3\%).

\begin{figure}[bth]\centering
    \includegraphics[width=\cmsFigWidth]{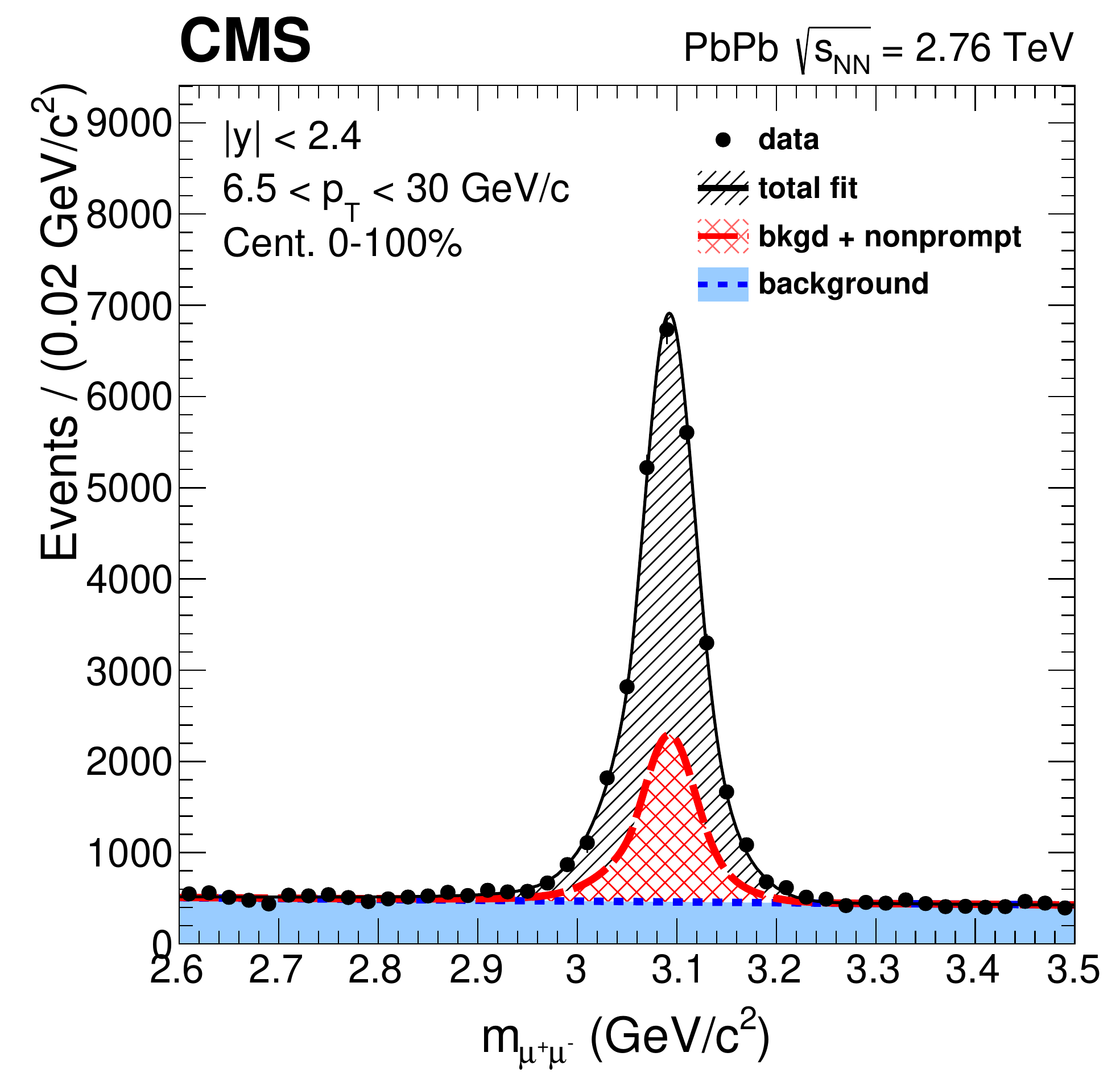}\hspace{1em}
    \includegraphics[width=\cmsFigWidth]{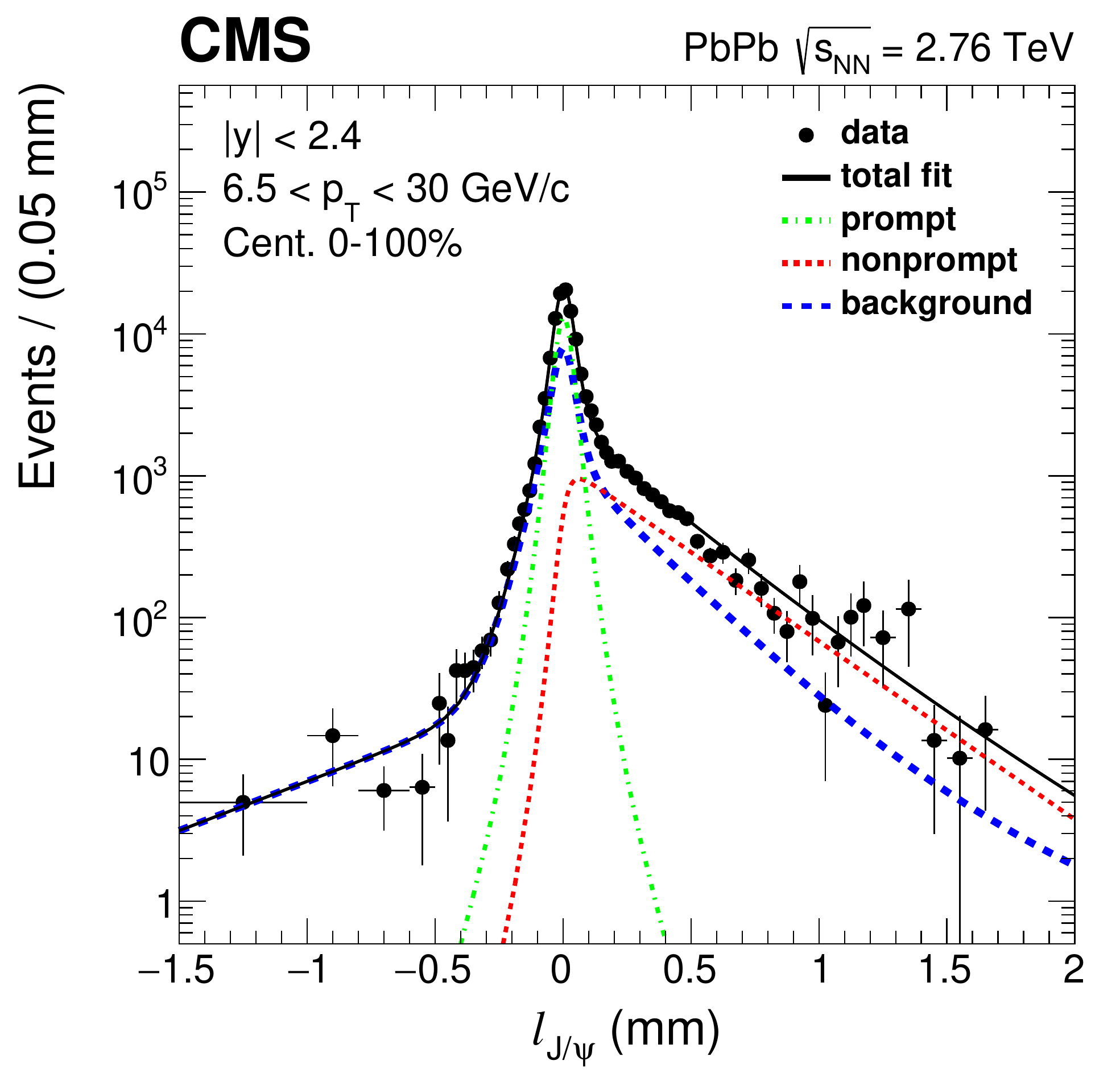}\\
    \caption{Invariant mass spectra (\cmsLeft) and pseudo-proper decay
      length distribution (\cmsRight) of \mumu pairs in centrality 0--100\%
      and integrated over the rapidity range
      $\abs{y}<2.4$ and the \pt range $6.5<\pt<30\GeVc$. The error bars on each point represent statistical uncertainties. The projections
      of the two-dimensional fit onto the respective axes are overlaid
      as solid black lines. The dashed green and red lines show the fitted
      contribution of prompt and nonprompt \JPsi. The fitted background
      contributions are shown as dotted blue lines.}
    \label{fig:jpsi_2dmassfits_ethfp}
\end{figure}

The same method for signal extraction is used in both the \vtwo and
the \raa analyses, for both the \PbPb and \pp samples. The separation of prompt \JPsi mesons from those coming
from \PQb hadron decays relies on the measurement of a secondary \mumu vertex
displaced from the primary collision vertex. The displacement $\vec{r}$ between the \mumu
vertex and the primary vertex is measured first. Then, the most
probable decay length of \PQb hadron in the laboratory
frame~\cite{Buskulic:1993vi} is calculated as
\begin{equation}\label{eq:lxy}
  L_{xyz} = \frac{\hat{u}^TS^{-1}\vec{r}}{\hat{u}^TS^{-1}\hat{u}},
\end{equation}
where $\hat{u}$ is the unit vector in the direction of the \JPsi meson
momentum ($\vec{p}$) and $S$ is the sum of the primary and secondary vertex covariance
matrices. From this quantity,
the pseudo-proper decay length $\ell_{\JPsi} = L_{xyz}\, m_{\JPsi}/p$ (which is the decay length of the \JPsi meson) is computed as an estimate of the \PQb hadron decay length.

To measure the fraction of the \JPsi mesons coming from b hadron decays (the so-called \textit{b fraction}),
the invariant-mass spectrum of \mumu pairs and their $\ell_{\JPsi}$ distribution
are fitted sequentially in an extended unbinned maximum
likelihood fit. The fits are performed for each \pt, $\abs{y}$, and centrality bin
of the analysis, and in addition in the case of the \PbPb \vtwo analysis, in four bins in $\abs{\dphi} = \abs{\phi-\Psi_2}$, equally spaced between 0 and $\pi/2$. The second-order ``event plane'' angle $\Psi_2$, measured as explained below, corresponds to the event-by-event azimuthal angle of maximum particle density. It is an approximation of the participant plane angle $\Psi_\mathrm{PP}$, which is not directly observable.

The fitting procedure is similar to the one used in earlier analyses of pp collisions at \sqrts =
7\TeV~\cite{Chatrchyan:2011kc}, and PbPb collisions at \sqrtsnn =
2.76\TeV~\cite{Chatrchyan:2012np}. The \JPsi meson mass distribution is modelled by the sum of a
Gaussian function and a Crystal Ball (CB) function~\cite{Oreglia:1980cs}, with a
common mean $m_0$ and independent widths. The CB radiative tail parameters are fixed to the values obtained in fits to simulated
distributions for different kinematic regions~\cite{Khachatryan:2014bva}. The invariant mass background probability density
function (PDF) is an exponential function whose parameters are allowed to float in each fit.
Since the mass resolution depends on $y$ and \pt, all resolution-related parameters are left free when binning as a function of $\abs{y}$ or \pt.
In the case of centrality
binning, the width of the CB function is left free, while the rest of the
parameters are fixed to the centrality-integrated results, 0--100\%, for a given
\pt and $\abs{y}$ bin. When binning in $\abs{\dphi}$, all signal parameters are fixed to their values in the $\abs{\dphi}$-integrated fit.

The $\ell_{\JPsi}$ distribution is modeled by a prompt signal component
represented by a resolution function, a nonprompt component given by an
exponential function convoluted with the resolution function, and the continuum
background component represented by the sum of the resolution function
plus three exponential decay functions to take into account long-lived background
components~\cite{Chatrchyan:2011kc}. The resolution function is comprised of
the sum of two Gaussian functions, which depend upon the per-event uncertainty of the measured $\ell_{\JPsi}$, determined from the covariance matrices of the primary
and secondary vertex fits. The fit parameters of the $\ell_{\JPsi}$ distribution were determined through a series of fits. Pseudo-proper decay length background function parameters are fixed using dimuon events
in data located on each side of the \JPsi resonance peak. In all  cases, the b fraction is a free fit parameter. An example of 2D fits is given in Fig.~\ref{fig:jpsi_2dmassfits_ethfp}.

\begin{figure}[bthp]\centering
    \includegraphics[width=\cmsFigWidth]{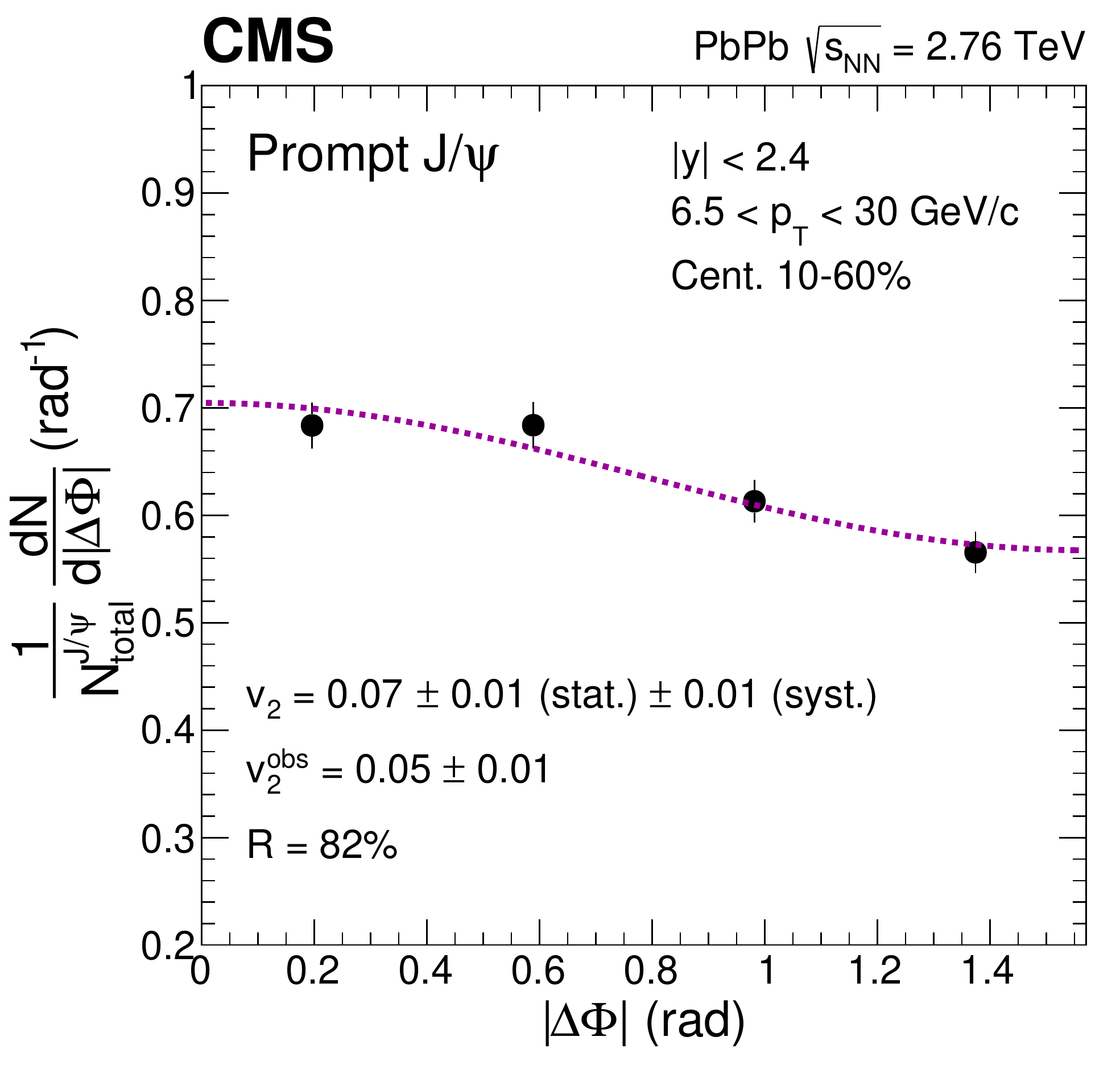}\hspace{1em}
    \caption{The $\abs{\Delta\Phi}$ distribution of high \pt prompt \JPsi mesons, $6.5<\pt<30$\GeVc, measured in the rapidity range $\abs{y}<2.4$ and event centrality 10--60\%, normalized by the bin width and the sum of the prompt yields in all four $\Delta\Phi$ bins. The dashed line represents the function $1+2\vtwo^{\text{obs}} \cos(\abs{2\dphi})$ used to extract the $\vtwo^{\text{obs}}$. The event-averaged resolution correction factor, corresponding to this event centrality, is also listed, together with the calculated final \vtwo for this kinematic bin. The systematic uncertainty listed in the legend includes the 2.7\% global uncertainty from the event plane measurement.}
    \label{fig:dphiFit_prompt}
\end{figure}

The \vtwo analysis follows closely the event plane method described in Ref.~\cite{Chatrchyan:2012ta}.
The \JPsi mesons reconstructed with $y>0$ ($y<0$) are correlated with the event plane
$\Psi_2$ found using energy deposited in a region of
the HF spanning $-5 < \eta < -3$ ($3 < \eta < 5$). This is chosen to introduce a rapidity gap between the particles used in the event plane
determination and the \JPsi meson, in order to reduce the effect of other correlations that
might exist, such as those from dijet production. To account for nonuniformities in the
detector acceptance that can lead to artificial asymmetries in the event plane
angle distribution and thereby affect the deduced \vtwo values,
a Fourier analysis ``flattening'' procedure~\cite{Poskanzer:1998yz} is
used, where each calculated event plane angle is shifted
slightly to recover a uniform azimuthal distribution, as described in Ref.~\cite{Chatrchyan:2012ta}. The event plane has a resolution that depends on centrality, and is caused by the finite number of particles used in its determination.

The corrections applied event-by-event ensure that the prompt and nonprompt yields extracted from fitting the invariant mass and $\ell_{\JPsi}$ distributions account for reconstruction and selection inefficiencies. As such, after extracting the yields in each $\abs{y}$, \pt, centrality
(and $\abs{\dphi}$) bin, the \vtwo and \raa can be calculated directly. The \raa is defined by
\begin{equation}
\raa = \frac{\nAA}{(\taa \, \sigPP)},
\label{eq:raa}
\end{equation}
where $N_{\text{PbPb}}^{\JPsi}$ is the number of prompt or nonprompt \JPsi mesons produced per \PbPb collision, $\sigma_{\text{pp}}^{\JPsi}$ is the corresponding pp cross section, and \taa is the nuclear overlap function.

{\sloppy
The \vtwo is calculated by fitting the
$[1/N_{\text{total}}^{\JPsi}][\rd N^{\JPsi}/\rd\abs{\Delta\Phi}]$ distributions
with the function $1+2\vtwo^{\text{obs}} \cos(\abs{2\dphi})$, where the
$N_{\text{total}}^{\JPsi}$ is the prompt or nonprompt \JPsi yield integrated over azimuth for each kinematic
bin. An example of such a fit is shown in Fig.~\ref{fig:dphiFit_prompt}. The final
\vtwo coefficient in the event plane method is evaluated by dividing
the observed value $v_2^{\text{obs}}$ by an event-averaged resolution-correction $R$, \ie $v_2=v_2^{\text{obs}}/R$, as described in Ref.~\cite{Chatrchyan:2012xq}. The factor $R$, calculated experimentally as described in Ref.~\cite{Chatrchyan:2012ta}, can
range from 0 to 1, with a better resolution corresponding to a larger value of
$R$. No difference is observed when determining $R$ using the dimuon-triggered
events analysed here, compared to the values used
in Ref.~\cite{Chatrchyan:2012ta} for the analysis of charged hadrons. For this paper, the \vtwo analysis is restricted to the centrality interval 10-60\% to ensure a nonsymmetric overlap region in the colliding nuclei, while maintaining a good event plane resolution ($R \gtrsim 0.8$ in the event centrality ranges in which results are reported: 10--20\%, 20--30\%, and 30--60\%).

\par}

\subsection{Estimation of uncertainties}

Several sources of systematic uncertainties are considered for both \raa and \vtwo analyses. They are mostly common, thus calculated and propagated in a similar way.

The systematic uncertainties in the signal extraction method (fitting) are
evaluated by varying the analytical form of each component of the PDF
hypotheses. For the invariant mass PDF, as an alternative signal shape, a sum of
two Gaussian functions is used, with shared mean and both widths as free parameters in the fit.
For the same PDF, the uncertainty in the background shape is evaluated using
a first order Chebychev polynomial. For the differential centrality bins, with
the invariant mass signal PDF parameters fixed to the 0--100\% bin, an
uncertainty is calculated by performing fits in which the
constrained parameters are allowed to vary with a Gaussian PDF. The mean of the
constraining Gaussian function and the initial value of the constrained
parameters come from the fitting in the 0--100\% bin with no fixed parameters.
The uncertainties of the parameters
in the 0--100\% bin is used as a width of the constraining Gaussian. For the
lifetime PDF components, the settings that could potentially affect the
b fraction are changed. The $\ell_{\JPsi}$ shape of the nonprompt \JPsi is taken
directly from the reconstructed one in simulation and converted to a PDF. Tails
of this PDF, where the MC statistics are insufficient, are mirrored from
neighboring points, weighted with the corresponding efficiency. The sum in
quadrature of all yield variations with respect to the nominal fit is propagated
in the calculation of the systematic uncertainty in the final results.
The variations across all \raa (\vtwo) analysis bins are between 0.7 and 16\%
(2.6 and 38\%) for prompt \JPsi, and 1.4 and 19\% (20 and 81\%) for nonprompt \JPsi.
They increase from mid to forward rapidity, from high- to low-\pt, and for \PbPb results also from
central to peripheral bins.

Three independent uncertainties are assigned for the dimuon efficiency corrections.
One addresses the uncertainty on the parametrization of the efficiency \vs \pt, $y$, and
centrality. For the \raa results, it is estimated, in each signal $y$ and
centrality bin, by randomly moving 100 times, each individual efficiency versus \pt point within its statistical uncertainty, re-fitting with the Gaussian error function, and recalculating each time a corrected MC signal yield. For the \vtwo results, this procedure is not practical: it requires re-weighting and re-fitting many times the full data sample. So in this case, the uncertainty is estimated by changing two settings in the nominal efficiency, and re-fitting data only once, with the modified efficiency: (a)~using binned efficiency instead of fits, and
(b)~using only the nonprompt \JPsi MC sample, integrated over all event centralities.
The relative uncertainties for this source, propagated into the final results,
are calculated for \raa as the root-mean-square of the 100 yield variations with respect to the yield obtained with the nominal efficiency parametrization, and for the \vtwo analysis as the full difference between the nominal
and the modified-efficiency results.
Across all \raa (\vtwo) analysis bins, the values are
between 0.6 and 20\% (1.5 and 54\%) for prompt \JPsi, and 0.7 and 24\%
(6.1 and 50\%) for nonprompt \JPsi results. These uncertainties increase from high to low
\pt, and from mid to forward rapidity but do not have a strong centrality dependence.

A second uncertainty addresses the
accuracy of the efficiency \vs $L_{xyz}$ calculation, and is estimated by
changing the $L_{xyz}$ resolution. It is done in several steps: (a)~the binning
in the $L_{xyz}^{\text{true}}$ \vs $L_{xyz}$ maps is changed; (b)~the dimuon
efficiency weights are recalculated; c) the data is reweighed and refitted to
extract the signal yields. The variations across all \raa (\vtwo) analysis bins
are between 0.025 and 3.7\% (0.1 and 16\%) for prompt \JPsi, and 0.1 and 13\%
(29 and 32\%) for nonprompt \JPsi results. In the case of the prompt \JPsi, the variations are small and rather constant across all bins, around 2-3\%, with the 16\% variation being reached only in the lowest-\pt bin in the \vtwo analysis. For nonprompt \JPsi the variations increase from mid to forward rapidity, and for \PbPb also from peripheral to central bins.

Finally, a third class of uncertainty arises from the scaling factors. For the
\vtwo analysis, the full difference between results with and without T\&P
corrections is propagated to the final systematic uncertainty.
It varies between 0.4 and 7.4\% for prompt \JPsi, and
5.4 and 8.8\% for nonprompt \JPsi results. For the \raa analysis, this uncertainty comprises two contributions. A parametrization
uncertainty was estimated by randomly moving each of the data T\&P efficiency points within their statistical uncertainty, recalculating each time the scaling factors and the dimuon efficiencies in all the analysis bins, and propagating the root-mean-square of all variations to the total T\&P uncertainty. In addition, a systematic uncertainty was estimated by changing different settings of the T\&P method. The contributions are similar for the prompt and nonprompt \JPsi results,
and vary between 1.4 and 13\% across all bins, for the combined trigger,
identification, and reconstruction efficiencies, with the largest
uncertainties in the forward and low \pt regions. On top of these bin-by-bin
T\&P uncertainties, an uncertainty in the tracking reconstruction efficiency,
0.3 and 0.6\% for each muon track, for \pp and \PbPb, respectively,
is doubled for dimuon candidates, and considered as a global uncertainty in the final results.

There is one additional source of uncertainty that is particular to each analysis. For the
\raa results, it is the \taa uncertainty, which varies between 16 and 4.1\%
from most peripheral (70--100\%) to most central (0--5\%) events,
and it has a value of 5.6\% for the 0--100\% case,
estimated as described in~Ref.~\cite{Chatrchyan:2011sx}.
For the \vtwo analysis, uncertainties are assigned for the event plane measurement.
A systematic uncertainty is associated with the event plane flattening procedure
and the resolution correction determination ($\pm$1\%~\cite{Chatrchyan:2012xq}),
and another with the sensitivity of the measured \vtwo
values to the size of the minimum $\eta$ gap
(2.5\%, following Ref.~\cite{Chatrchyan:2012xq}).
The two uncertainties are added quadratically to a total
of 2.7\% global uncertainty in the \vtwo measurement.

The total systematic uncertainty in the \raa is estimated by summing in
quadrature the uncertainties from the signal extraction and efficiency
weighting. The range of the final uncertainties on prompt and nonprompt \JPsi
\raa is between 2.1 and 22\%, and 2.8 and 28\%, respectively, across bins of
the analysis. The uncertainty in the
integrated luminosity of the \pp data (3.7\%), $N_{\mathrm{MB}}$ events in \PbPb
data (3\%), and tracking efficiency (0.6\% for pp and 1.2\% for \PbPb data) are
considered as global uncertainties.

The total systematic uncertainty for \vtwo is estimated by summing in
quadrature the contributions from the yield extraction and efficiency corrections.
The range of the final uncertainties on prompt and
nonprompt \JPsi \vtwo results is between 10 and 57\%, and 37 and 100\%, respectively.

\subsection{Displaying uncertainties}

In all the results shown, statistical uncertainties are represented by
error bars, and systematic uncertainties by boxes centered on the points. For the \vtwo results, the global uncertainty from the event plane measurement is not included in the point-by-point uncertainties. 
Boxes plotted at $\raa=1$ represent the scale of the global uncertainties.
For \raa results plotted as a function of \pt or $\abs{y}$,
the statistical and systematic uncertainties include
the statistical and systematic components from both \PbPb and \pp samples, added in quadrature.
For these types of results, the systematic uncertainty on \taa, the \pp sample integrated luminosity
uncertainty, the uncertainty in the $N_{\mathrm{MB}}$ of \PbPb events, and the tracking efficiency are added in quadrature and shown as a global uncertainty.

For \raa results shown as a function of \npart, the uncertainties on \taa are included in the systematic uncertainty, point-by-point. The global uncertainty plotted at $\raa=1$ as a grey box includes in this case the statistical and systematic uncertainties from the \pp measurement, the integrated luminosity uncertainty for the \pp data, the uncertainty in the $N_{\mathrm{MB}}$ of \PbPb events, and the tracking efficiency uncertainty, added in quadrature. When showing \raa \vs \npart separately for different \pt or $\abs{y}$ intervals, the statistical and systematic uncertainties from the \pp measurement are added together in quadrature and plotted as a coloured box at $\raa=1$. In addition, a second global uncertainty, that is common for all the \pt and $\abs{y}$ bins, is calculated as the quadratic sum of the integrated luminosity uncertainty for pp data, the uncertainty in $N_{\mathrm{MB}}$ of \PbPb events, and the tracking efficiency uncertainty, and is plotted as an empty box at $\raa=1$.

\section{Results}

For all results plotted versus \pt or $\abs{y}$, the abscissae of the points correspond to the centre of the respective bin, and the horizontal error bars reflect the width of the bin. When plotted as a function of centrality, the abscissae are average \npart values corresponding to events flatly distributed across centrality. For the \raa results, the numerical values of the numerator and denominator of Eq.~\ref{eq:raa} are available in tabulated form in Appendix~\ref{app_supplementalMaterial}.

\subsection{Prompt \texorpdfstring{\JPsi}{Jpsi}}

\begin{figure}[tbhp]\centering
    {\includegraphics[width=\cmsFigWidth]{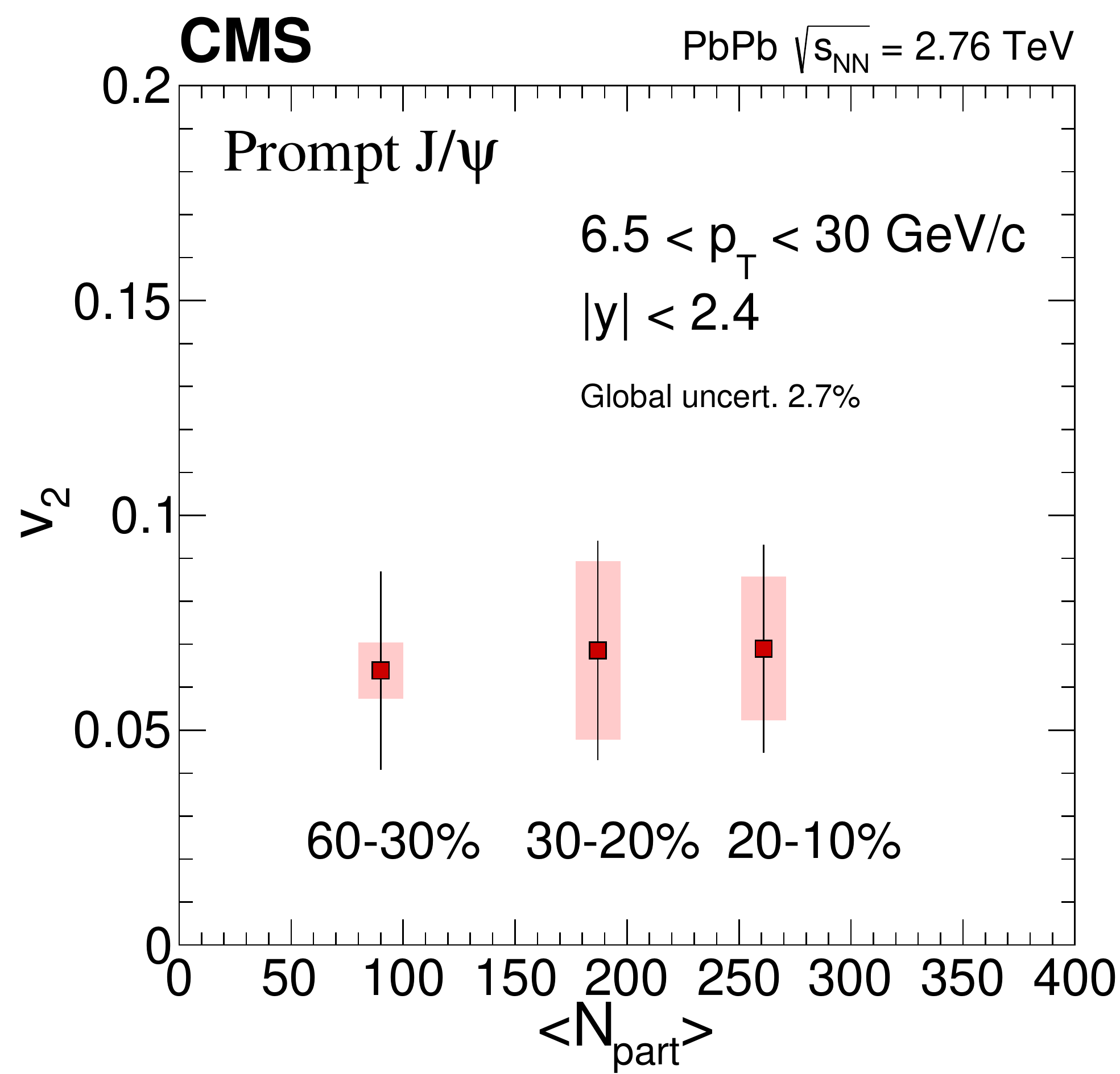}}
    {\includegraphics[width=\cmsFigWidth]{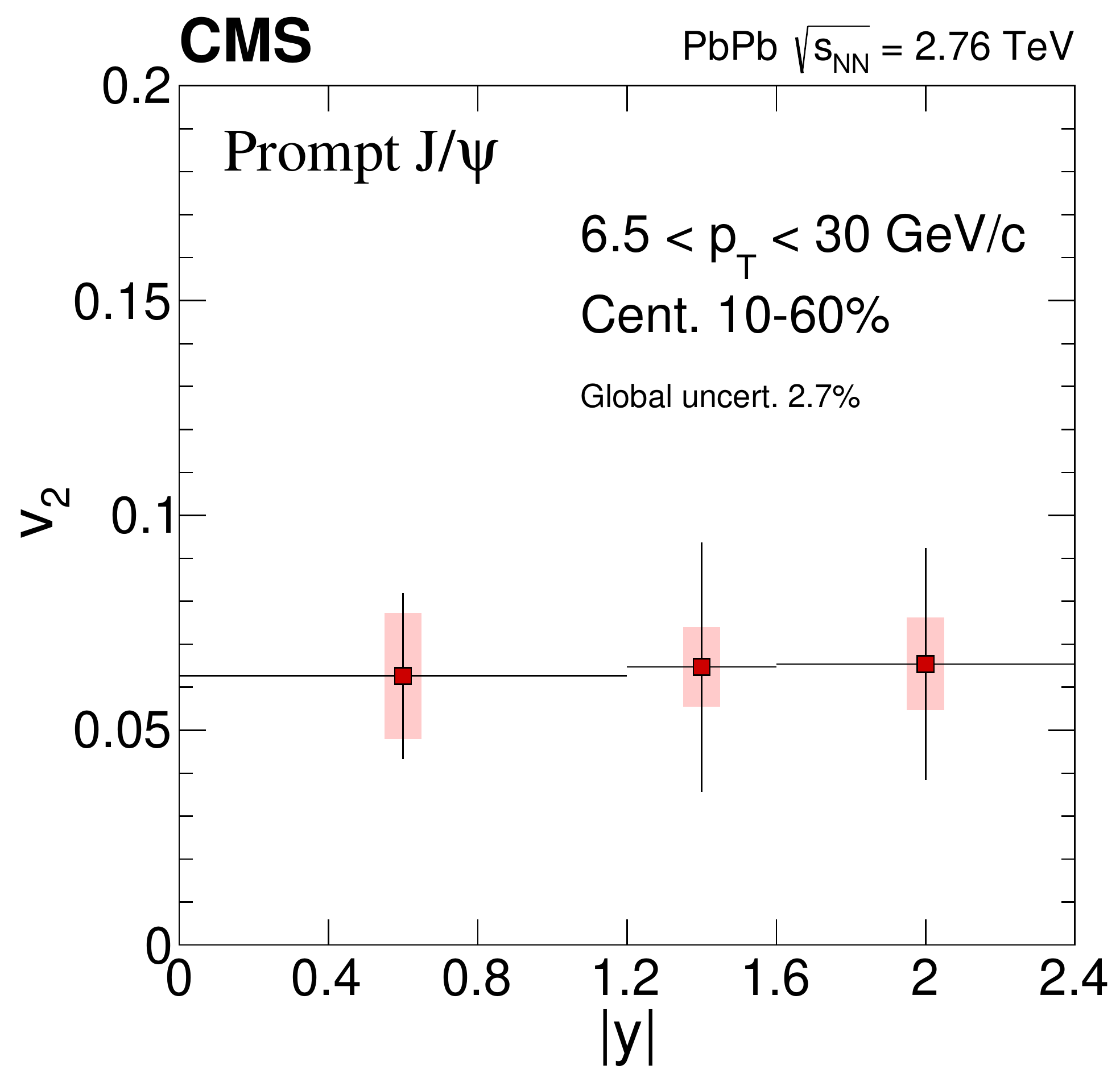}}\\		
    {\includegraphics[width=\cmsFigWidth]{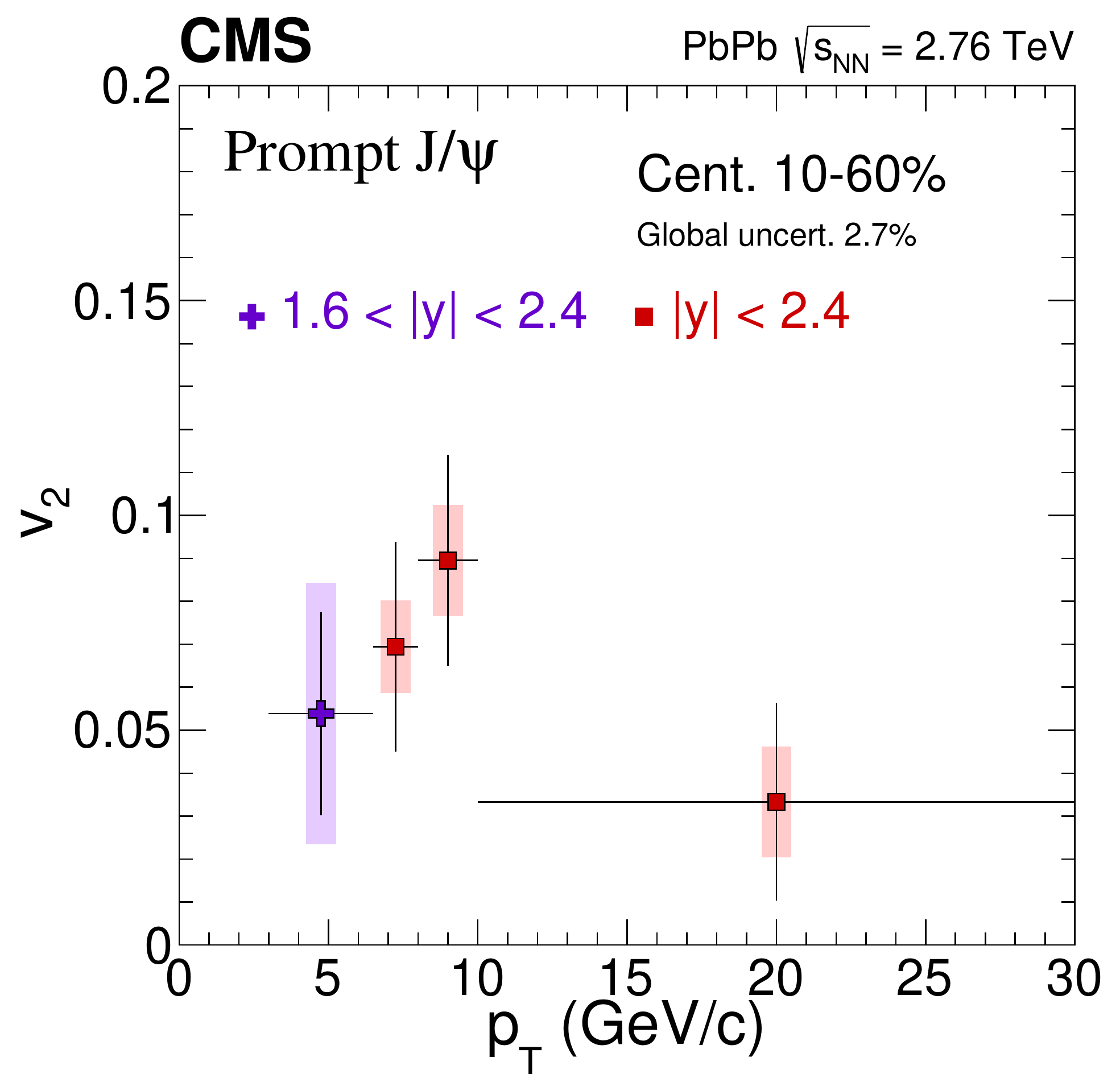}}		
    \caption{Prompt \JPsi \vtwo as a function of centrality (\cmsLLeft), rapidity (\cmsRRight), and \pt (bottom).
	   The bars (boxes) represent statistical (systematic) point-by-point
           uncertainties. The global uncertainty, listed in the legend, is not
           included in the point-by-point uncertainties.
    Horizontal bars indicate the bin width. The average \npart values correspond to events flatly distributed across centrality.}
    \label{fig:v2_pr}
\end{figure}

The measured prompt \JPsi \vtwo, for 10--60\% event centrality and integrated over
$6.5<\pt<30$\GeVc and $\abs{y}<2.4$, is
$0.066 \pm 0.014\stat \pm 0.014\syst \pm 0.002\,(\text{global})$. The significance corresponding to a deviation from a $\vtwo=0$ value is 3.3~sigma. Figure~\ref{fig:v2_pr} shows the dependence of \vtwo on centrality, $\abs{y}$, and \pt.
For each of these results, the dependence on one variable is studied by
integrating over the other two. A nonzero \vtwo value is measured in all the kinematic
bins studied. The observed anisotropy shows no strong centrality, rapidity, or \pt dependence.

\begin{figure}[tbhp]\centering
    \includegraphics[width=\cmsFigWidth]{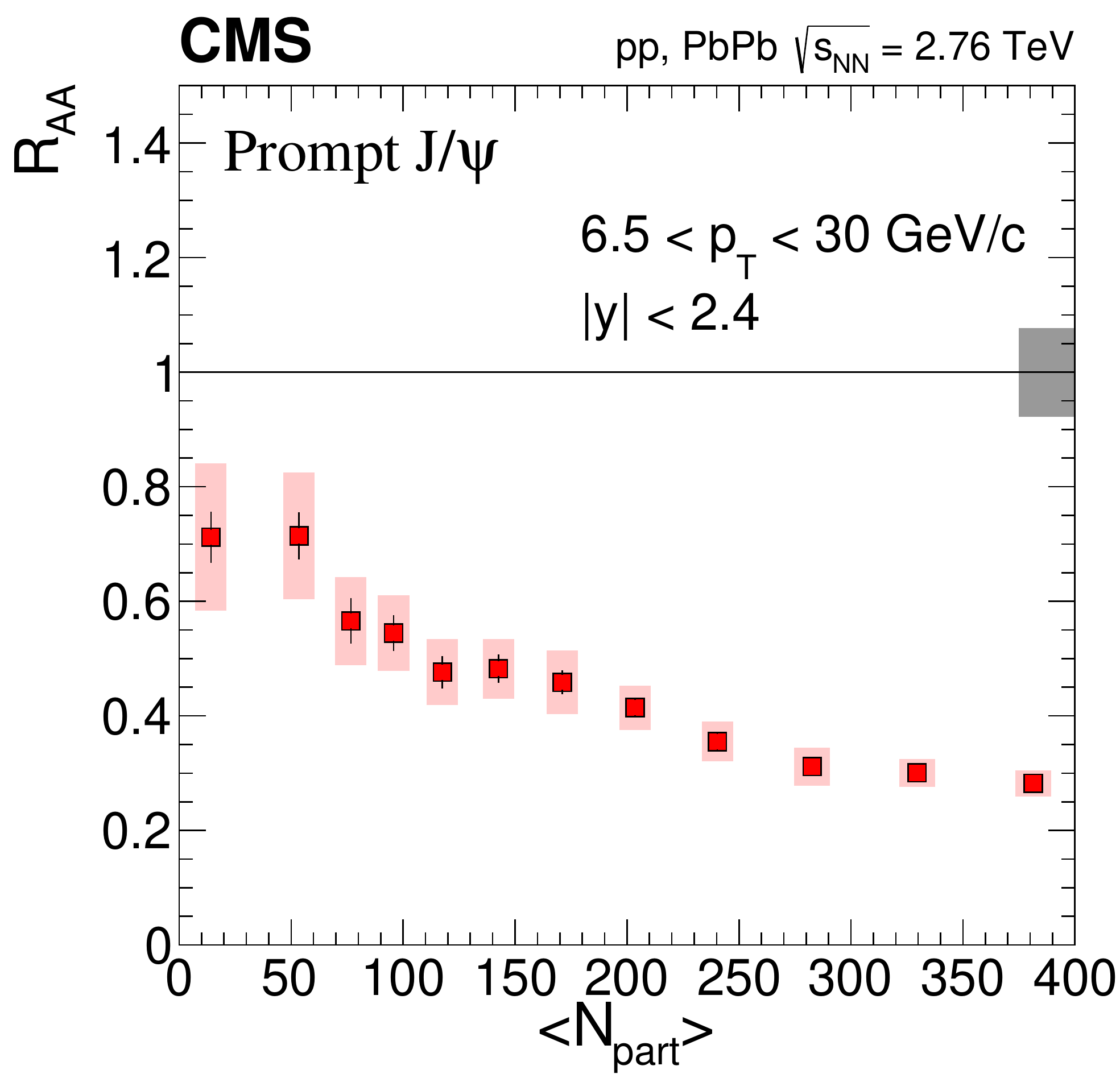}
    \includegraphics[width=\cmsFigWidth]{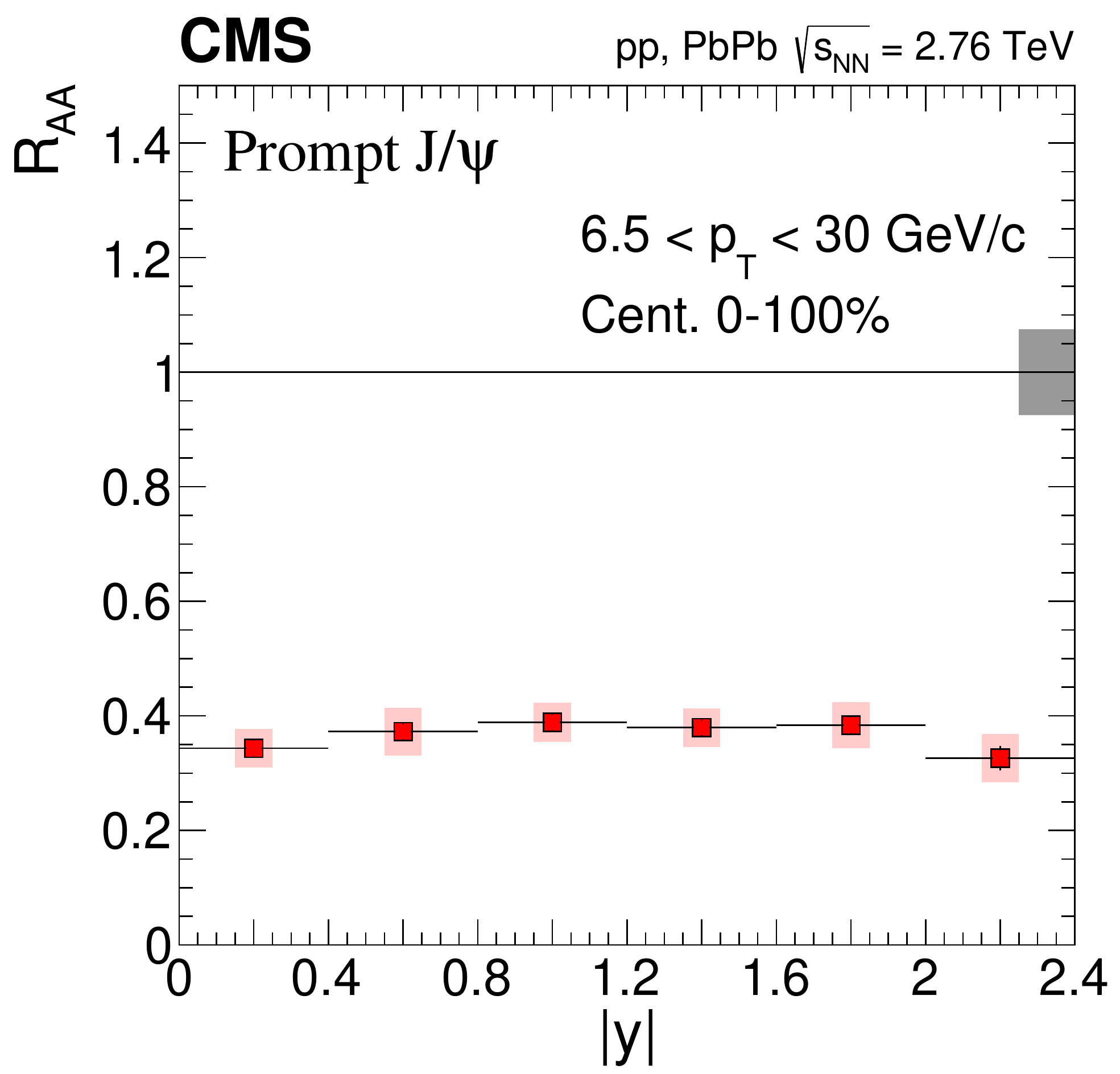}\\
    \includegraphics[width=\cmsFigWidth]{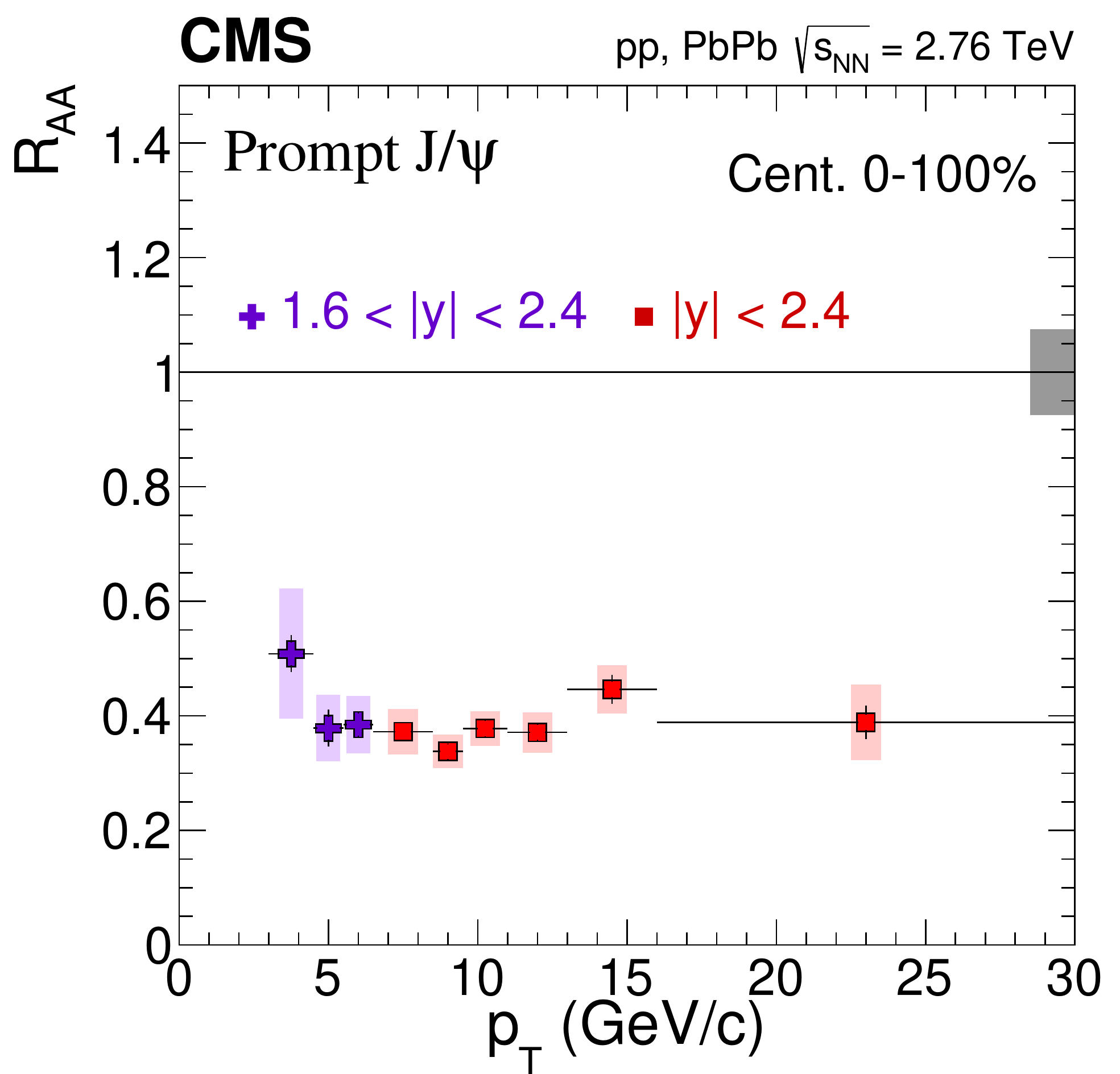}	
    \caption{Prompt \JPsi \raa as a function of centrality
    (\cmsLLeft), rapidity (\cmsRRight), and \pt (bottom). The bars (boxes) represent statistical (systematic) point-by-point uncertainties. The gray boxes plotted on the right side at $\raa=1$ represent the scale of the global uncertainties. The average \npart values correspond to events flatly distributed across centrality.}
    \label{fig:promptJpsi_RAA}
\end{figure}

In Fig.~\ref{fig:promptJpsi_RAA}, the \raa of prompt \JPsi
as a function of centrality, $\abs{y}$, and \pt are shown, integrating in each
case over the other two variables. The \raa is suppressed even for the most peripheral bin (60--100\%), with the suppression slowly increasing with \npart. The \raa for the most central events (0--5\%) is
measured for $6.5<\pt<30$\GeVc and $\abs{y}<2.4$ to be
$0.282 \pm 0.010\stat\pm 0.023\syst$. No strong rapidity or \pt dependence of the suppression is observed.

\begin{figure}[htbp]\centering
    \includegraphics[width=\cmsFigWidth]{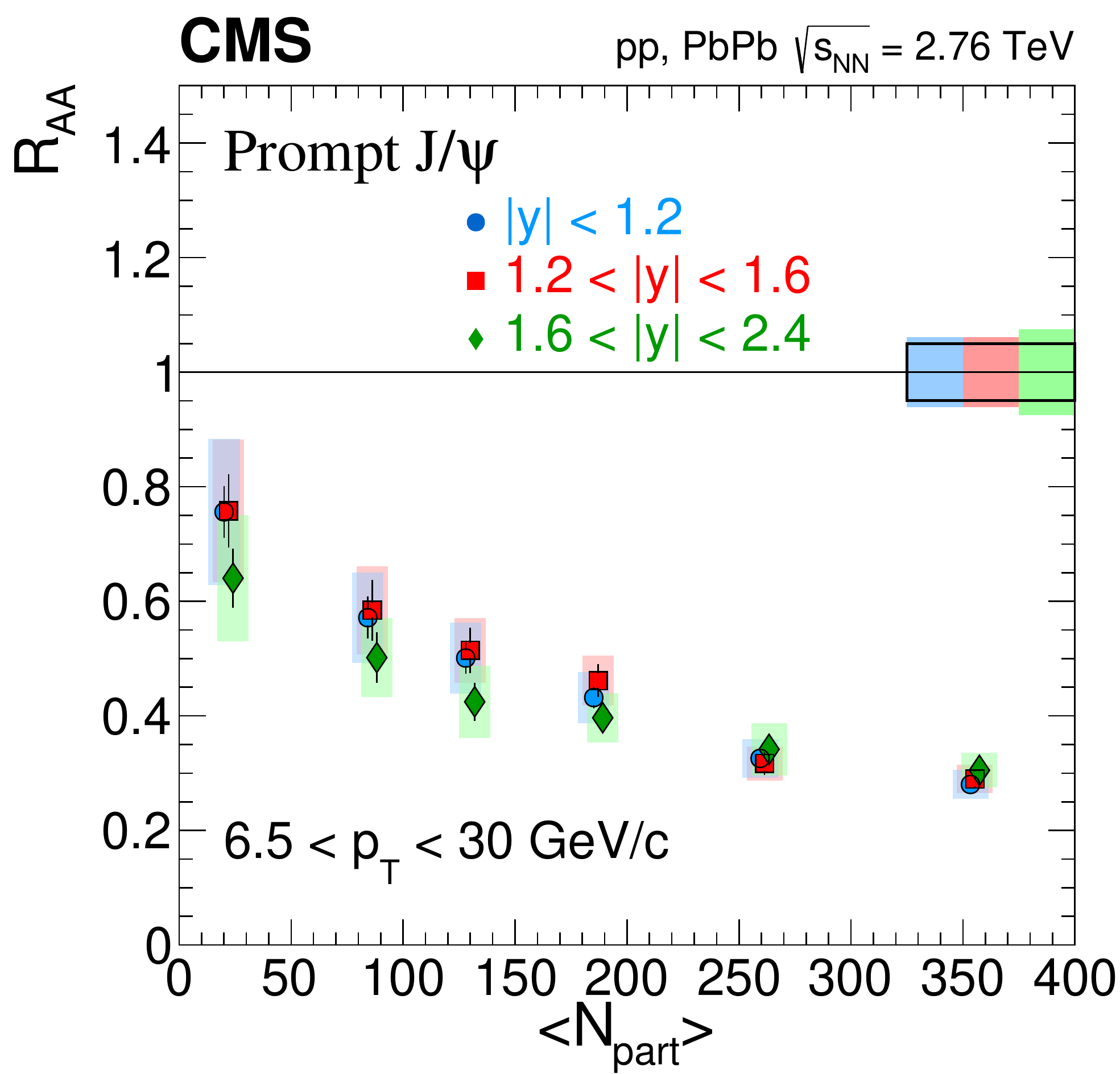}
    \includegraphics[width=\cmsFigWidth]{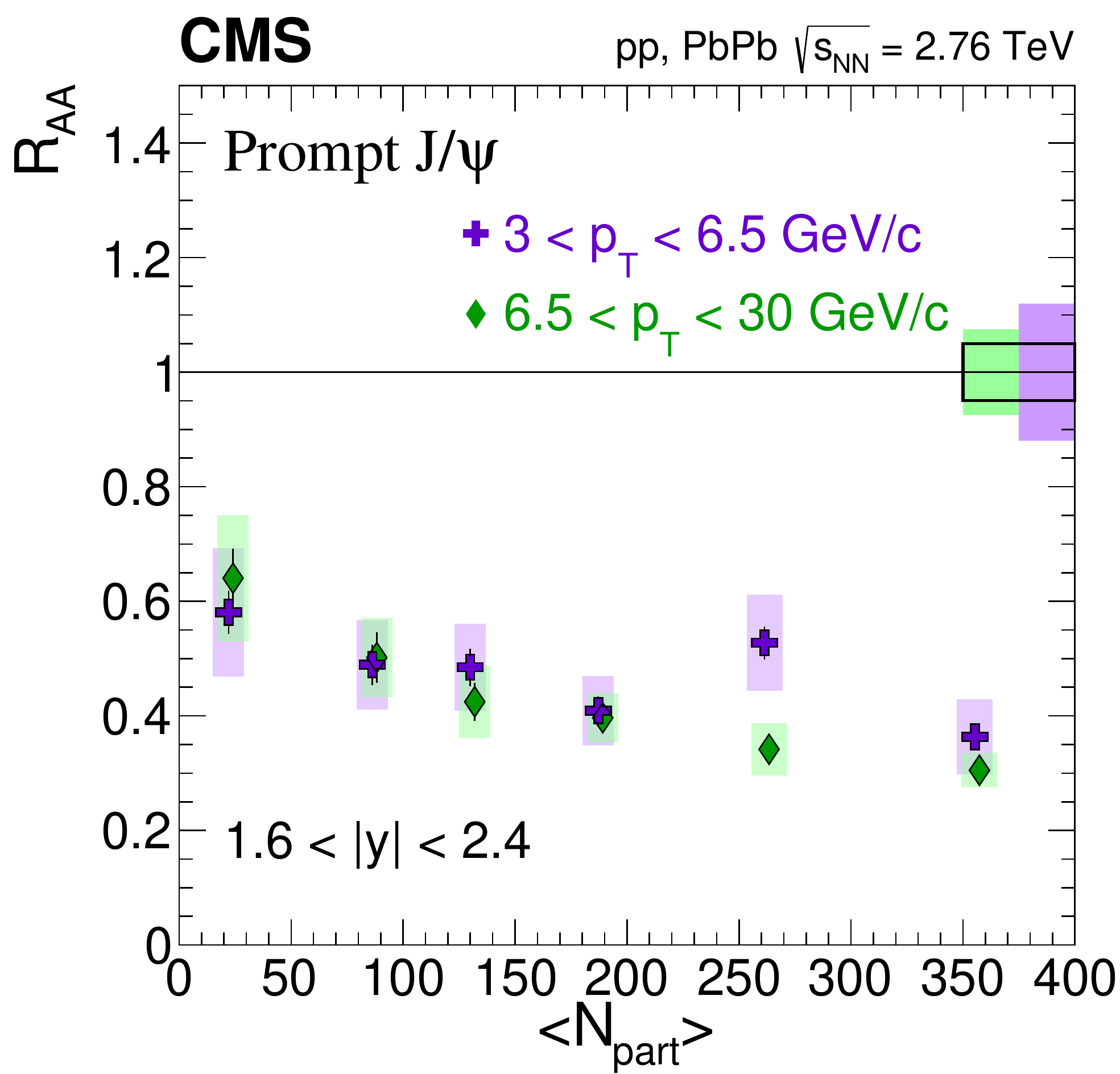}	
    \caption{(\cmsLeft) Prompt \JPsi \raa as a function of centrality at high \pt, $6.5<\pt<30$\GeVc, for three different $\abs{y}$ regions. The high-\pt mid- and forward-rapidity points are shifted horizontally by $\Delta\npart=2$ for better visibility.
    (\cmsRight) Prompt \JPsi \raa as a function of centrality, at forward rapidity, $1.6<\abs{y}<2.4$, for two different \pt regions.
    	   The bars (boxes) represent statistical (systematic) point-by-point uncertainties. The boxes plotted on the right side at $\raa=1$ represent the scale of the global uncertainties: the coloured boxes show the statistical and systematic uncertainties from \pp measurement, and the open box shows the global uncertainties common to all data points.
The average \npart values correspond to events flatly distributed across centrality.
}
    \label{fig:promptJpsi_RAA_pteta}
\end{figure}

Two double-differential studies are also made, in which a simultaneous binning in
centrality and $\abs{y}$, or in centrality and \pt is done.
Figure~\ref{fig:promptJpsi_RAA_pteta} (\cmsLeft) shows the centrality dependence of
high \pt ($6.5<\pt<30$\GeVc) prompt \JPsi  \raa measured in three $\abs{y}$ intervals. A similar suppression pattern is observed for all rapidities. Figure~\ref{fig:promptJpsi_RAA_pteta} (\cmsRight) shows, for $1.6<\abs{y}<2.4$, the \pt dependence of \raa \vs \npart. The suppression at low \pt ($3<\pt<6.5$\GeVc) is consistent with that at high \pt ($6.5<\pt<30$\GeVc).

\subsection{Nonprompt \texorpdfstring{\JPsi}{Jpsi}}
\begin{figure}[htbp]\centering
    \includegraphics[width=\cmsFigWidth]{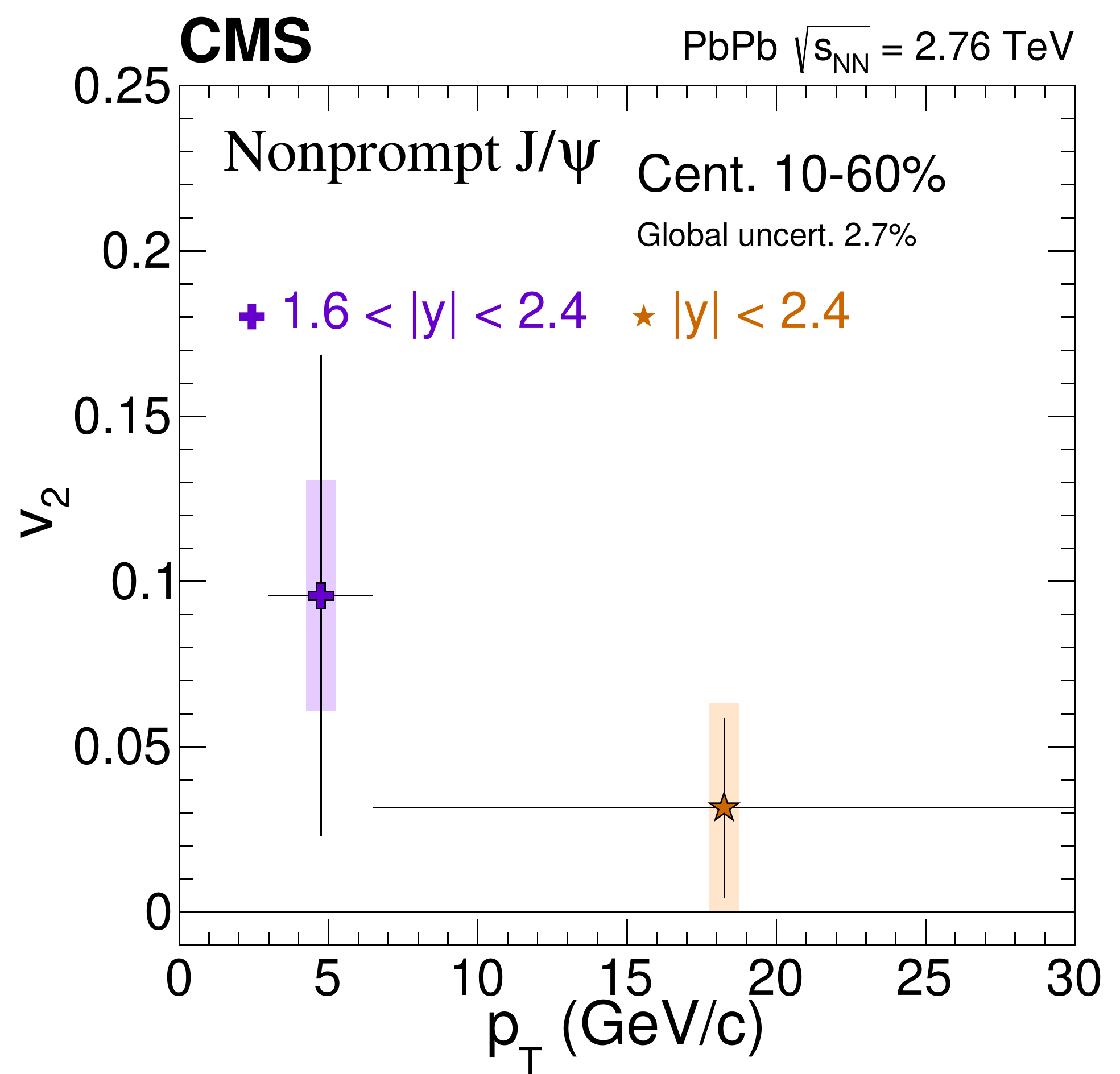}
    \caption{Nonprompt \JPsi \vtwo as a function of \pt. The bars (boxes)
      represent statistical (systematic) point-by-point uncertainties. The
      global uncertainty, listed in the legend, is not included in the point-by-point uncertainties.
    Horizontal bars indicate the bin width.}
    \label{fig:v2_npr}
\end{figure}

Figure~\ref{fig:v2_npr} shows the nonprompt \JPsi \vtwo \vs \pt for 10--60\% event centrality, in two kinematic regions:  $6.5<\pt<30$\GeVc and $\abs{y}<2.4$, and $3<\pt<6.5$\GeVc and $1.6<\abs{y}<2.4$.  The measured \vtwo for the high-(low-) \pt is $0.032 \pm 0.027 \stat \pm 0.032 \syst \pm 0.001 \,(\text{global})$ ($0.096 \pm 0.073 \stat \pm 0.035 \syst \pm 0.003 \,(\text{global})$). This is obtained from the fit to the $|\Delta\Phi|$ distribution (as described in Section~\ref{sec:sig_extraction}) with a $\chi^2$ probability of 22(20)\%. Fitting the same distribution with a constant (corresponding to the $\vtwo=0$ case) the $\chi^2$ probability is 11(8)\%. Both measurements are consistent with each other and with a \vtwo value of zero, though both nominal values are positive.

\begin{figure}[htbp]\centering
    \includegraphics[width=\cmsFigWidth]{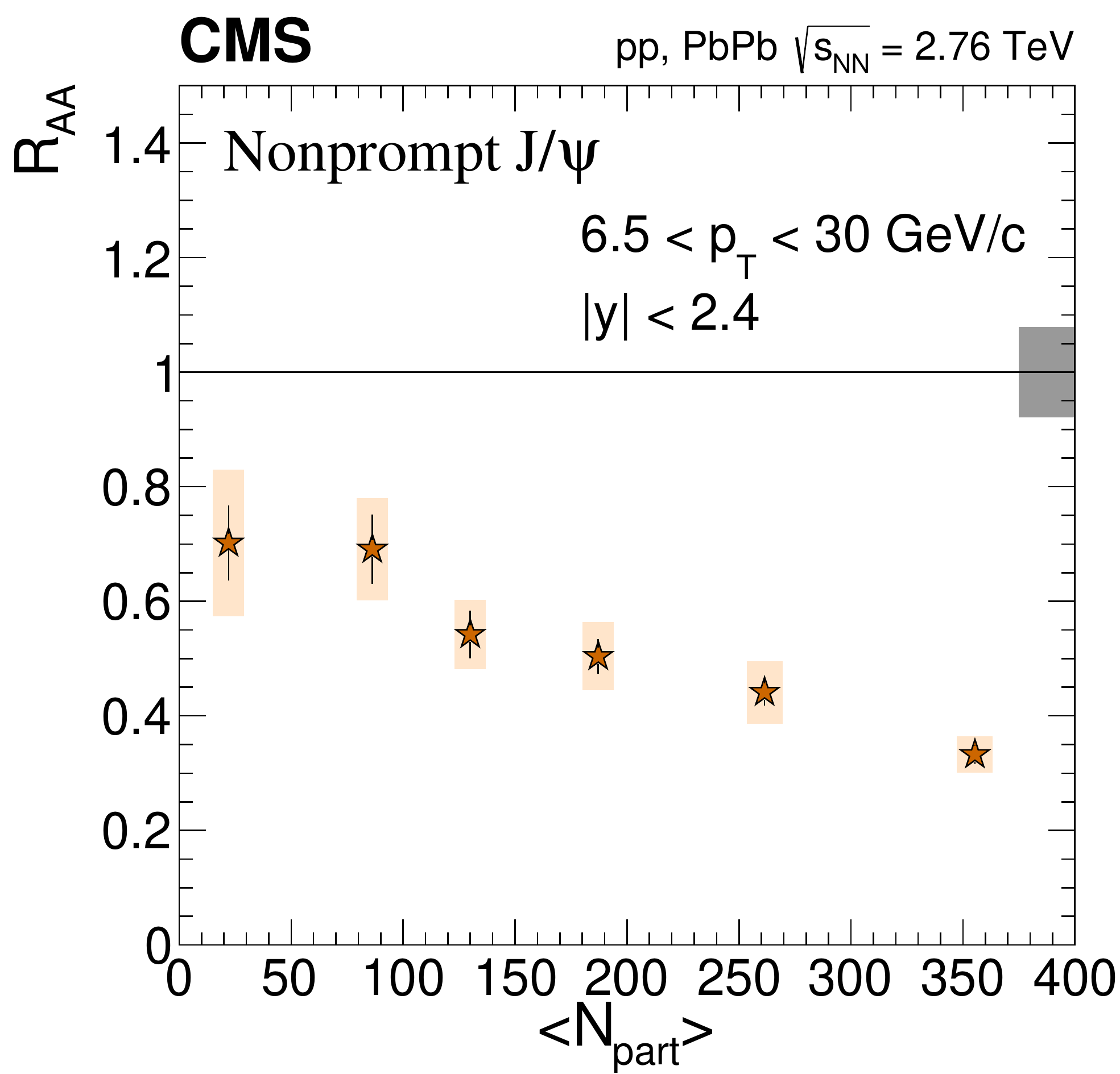}
    \includegraphics[width=\cmsFigWidth]{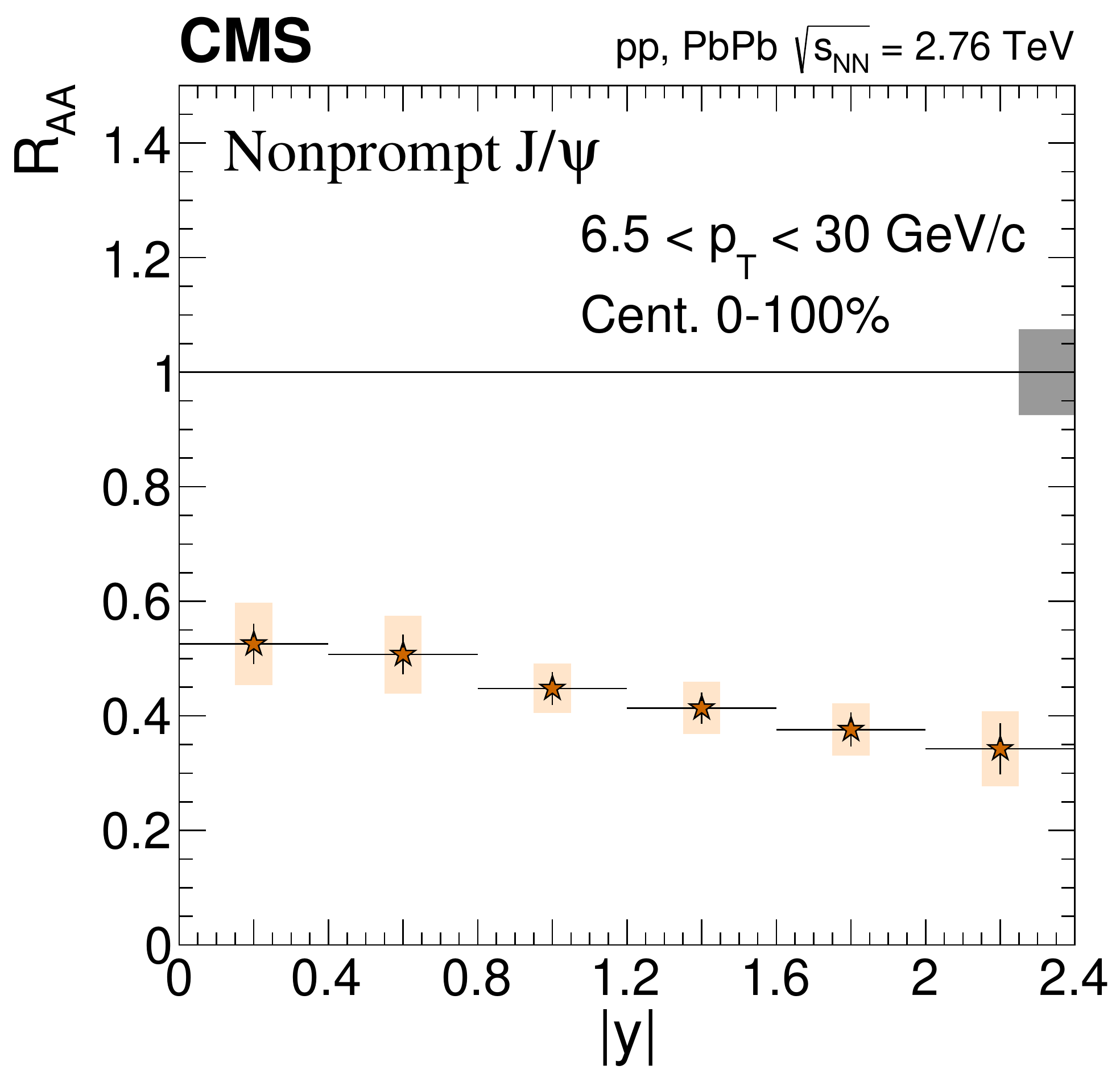}\\
    \includegraphics[width=\cmsFigWidth]{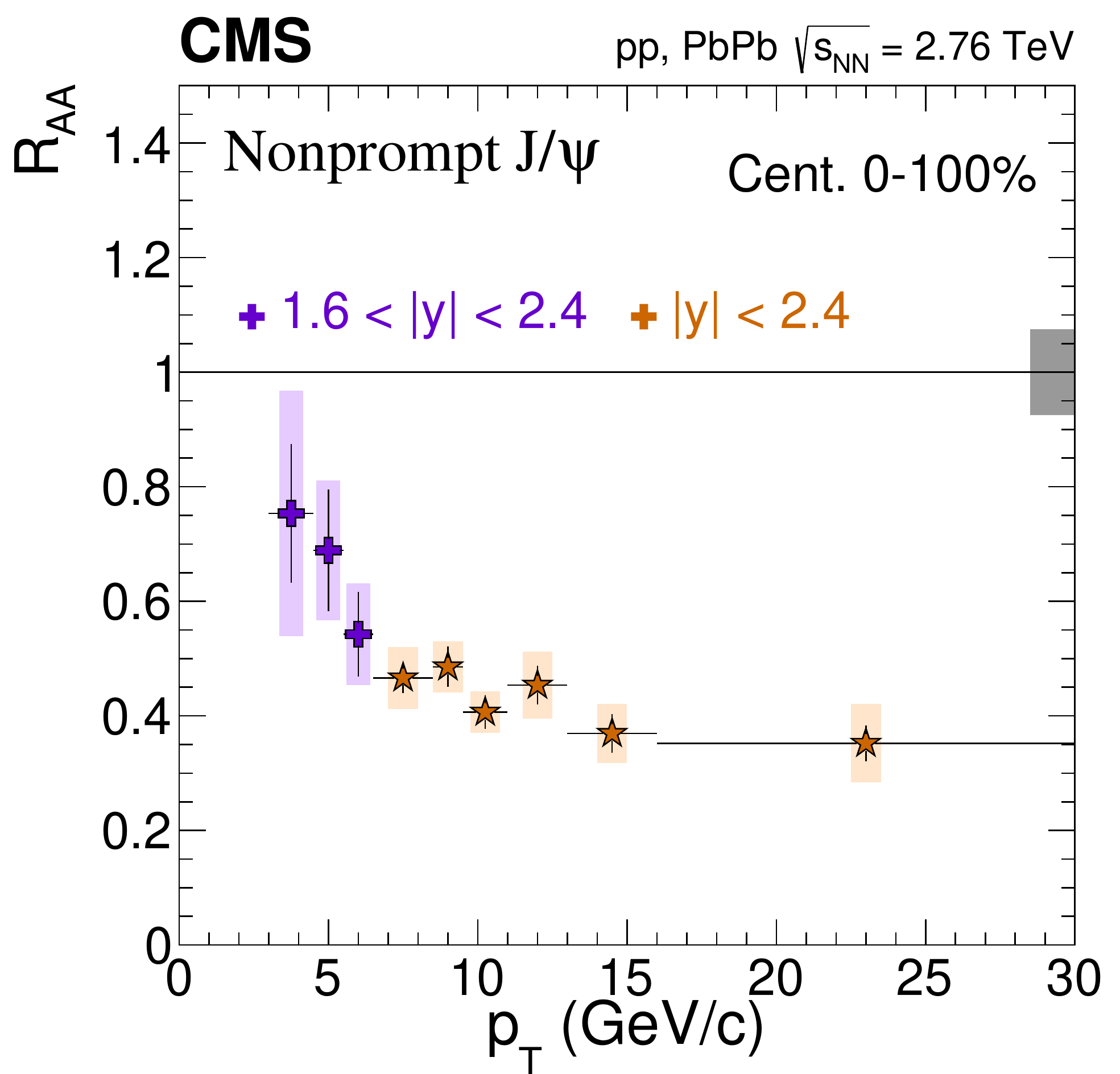}
    \caption{Nonprompt \JPsi \raa as a function of centrality (\cmsLLeft),
    rapidity (\cmsRRight), and \pt (bottom). The bars (boxes) represent statistical (systematic) point-by-point uncertainties.
The gray boxes plotted on the right side at $\raa=1$ represent the scale of the global
    uncertainties. For \raa \vs \npart, the average \npart values correspond to events flatly distributed across centrality.
}
    \label{fig:nonPromptJpsi_RAA}
\end{figure}

In Fig.~\ref{fig:nonPromptJpsi_RAA}, the \raa of nonprompt \JPsi
as a function of centrality, $\abs{y}$, and \pt are shown, integrating in
each case over the other two variables. A steady
increase of the suppression is observed with increasing centrality of
the collision. The \raa for the most central events (0--10\%) measured for
$6.5<\pt<30$\GeVc and $\abs{y}<2.4$ is
$0.332\pm 0.017\stat\pm0.028\syst$. Stronger suppression is observed with both increasing rapidity and \pt.

\begin{figure}[htbp]\centering
    \includegraphics[width=\cmsFigWidth]{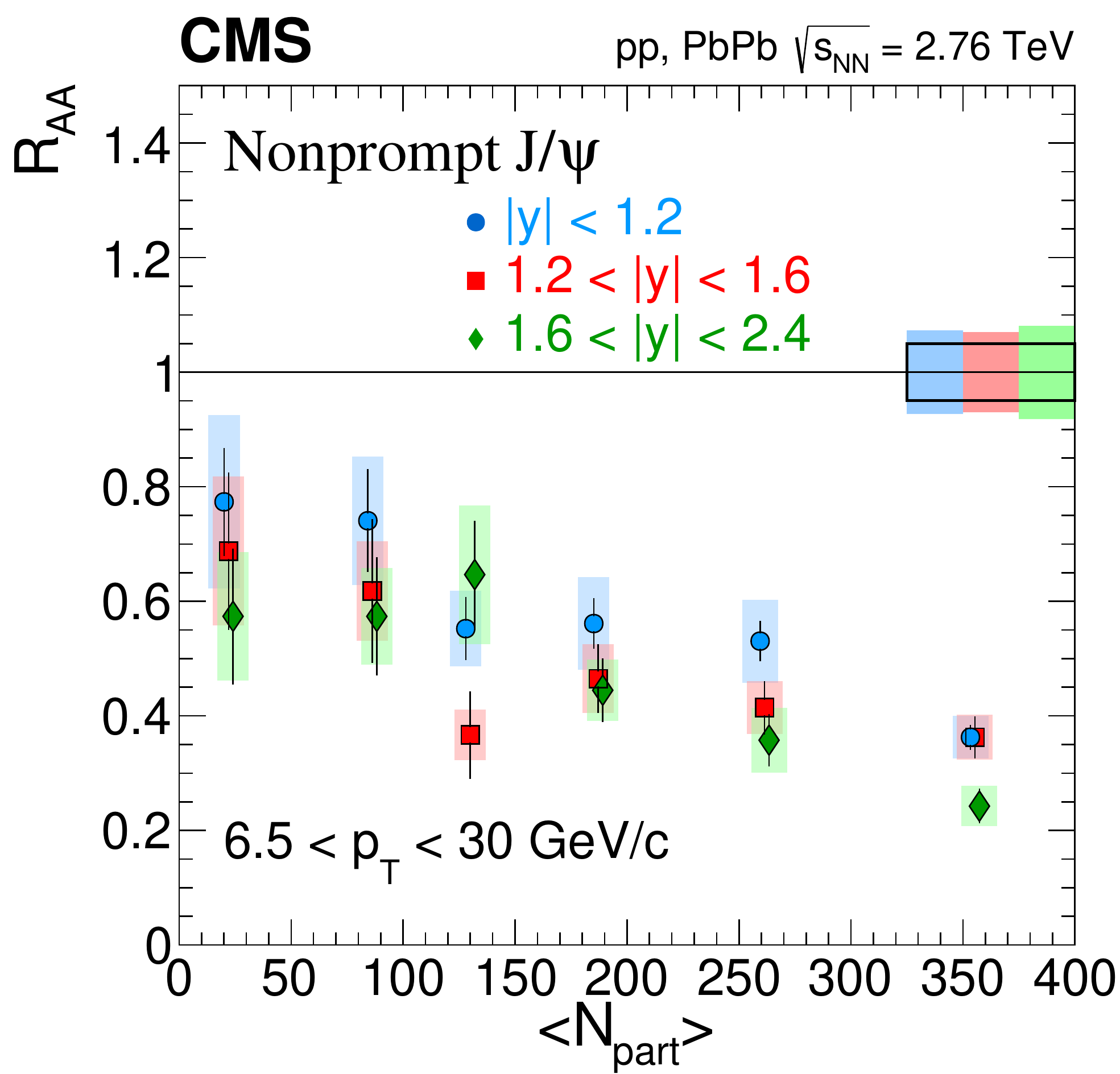}
    \includegraphics[width=\cmsFigWidth]{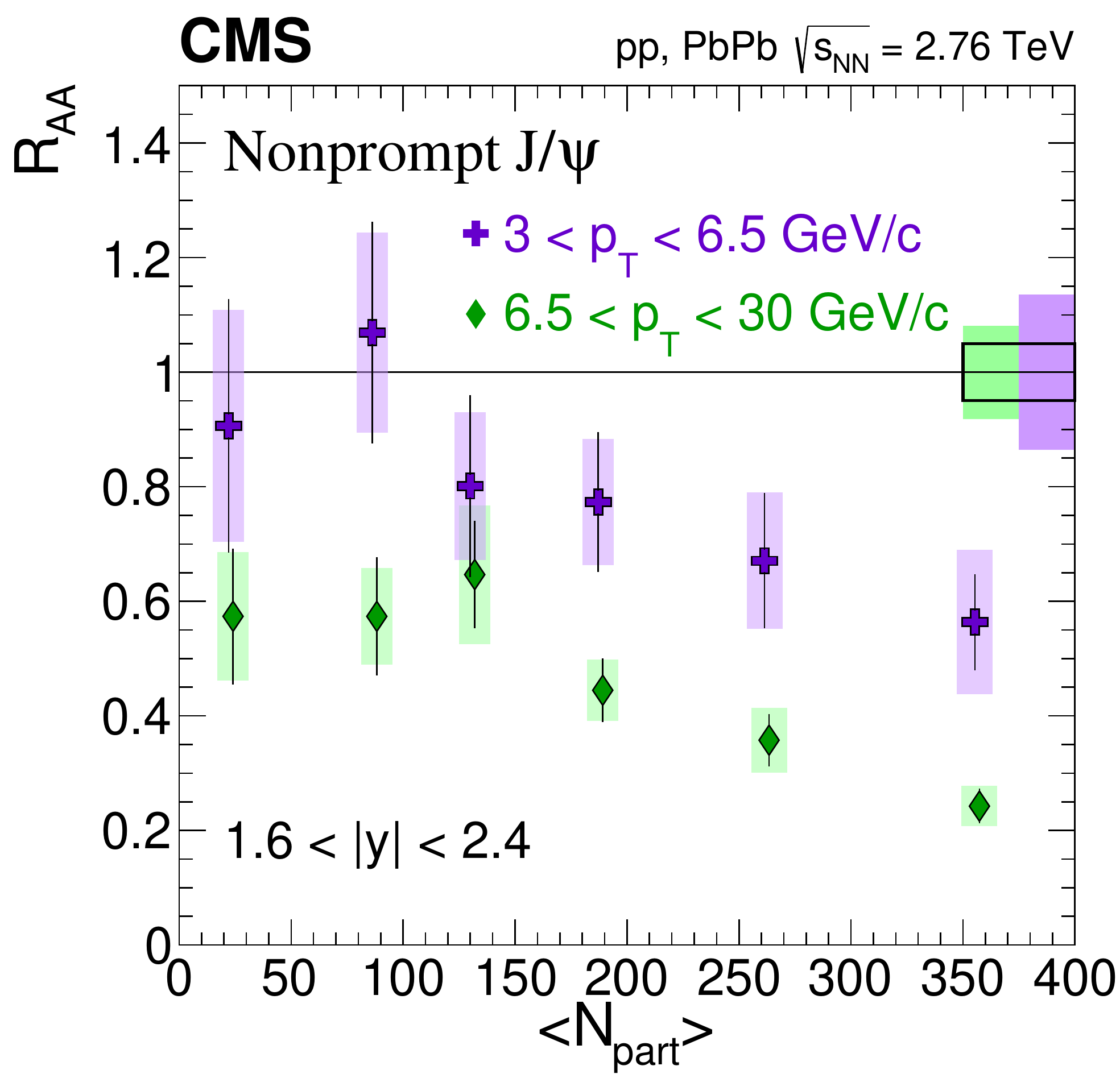}
    \caption{(\cmsLeft) Nonprompt \JPsi \raa as a function of centrality at high \pt, $6.5<\pt<30$\GeVc, for three different $\abs{y}$ regions. The high-\pt mid- and forward-rapidity points are shifted horizontally by $\Delta\npart=2$ for better visibility. (\cmsRight) Nonprompt \JPsi \raa as a function of centrality, at forward rapidity, $1.6<\abs{y}<2.4$, for two different \pt regions. The bars (boxes) represent statistical (systematic) point-by-point uncertainties. The boxes plotted on the right side at $\raa=1$ represent the scale of the global uncertainties: the coloured boxes show the statistical and systematic uncertainties from \pp measurement, and the open box shows the global uncertainties common to all data points. The average \npart values correspond to events flatly distributed across centrality. }
    \label{fig:nonPromptJpsi_RAA_pteta}
\end{figure}

{\tolerance=1200
As for the prompt production case, two double-differential studies were done, simultaneously
binning in centrality and $\abs{y}$ or \pt. Figure~\ref{fig:nonPromptJpsi_RAA_pteta} (\cmsLeft) shows the rapidity dependence of \raa \vs \npart for high \pt nonprompt \JPsi. Figure~\ref{fig:nonPromptJpsi_RAA_pteta} (\cmsRight) shows, for $1.6<\abs{y}<2.4$, the \pt dependence of \raa \vs \npart. The centrality dependences of the three $\abs{y}$ intervals are quite similar, and the same is true for the two \pt ranges. As was also seen in Fig.~\ref{fig:nonPromptJpsi_RAA}, smaller suppression is observed at lower $\abs{y}$ and lower \pt.
\par}

\section{Discussion}

In this section, the \raa and \vtwo results are compared first for open and hidden charm, and then for open charm and beauty, using data from the ALICE experiment~\cite{ALICE:2012ab, Adam:2015nna, Abelev:2014ipa}. For open charm, the measurements of \raa \vs \npart of prompt D$^{0}$ mesons, and of averaged prompt $\PD$ mesons (\PDz, $\PDp$ and $\PD^{*+}$ combined), measured in $\abs{y}<0.5$ at low \pt ($2<\pt<5$\GeVc), and high \pt ($6<\pt<12$\GeVc)~\cite{ALICE:2012ab} are used. These are compared to hidden charm data from the prompt \JPsi results described in this paper, in two \pt regions that are similar to the D measurement, \ie ($3<\pt<6.5$\GeVc, $1.6<\abs{y}<2.4$) and ($6.5<\pt<30$\GeVc, $\abs{y}<1.2$). For the \raa comparison of open charm \vs beauty, the averaged prompt D mesons measured in $\abs{y}<0.5$~\cite{Adam:2015nna} are compared to the nonprompt \JPsi results reported in this paper for $\abs{y}<1.2$. The \pt interval ($8<\pt<16$\GeVc) for the D is chosen to correspond to that of the parent B mesons of the CMS nonprompt \JPsi result~\cite{Adam:2015nna}.

For the \vtwo results, the \pt dependence reported in this paper for both prompt and nonprompt \JPsi in the centrality 10--60\% bin are compared with the \vtwo of the averaged D mesons~\cite{Abelev:2014ipa} measured in the 30--50\% centrality bin. In addition, the CMS charged-hadron \vtwo results, measured for $\abs{\eta}<0.5$, derived for 10--60\% centrality bin from Refs.~\cite{Chatrchyan:2012xq} and~\cite{Chatrchyan:2012ta}, are added to the comparison.

\subsection{Open versus hidden charm}

\begin{figure}[htbp]\centering
    {\includegraphics[width=\cmsFigWidth]{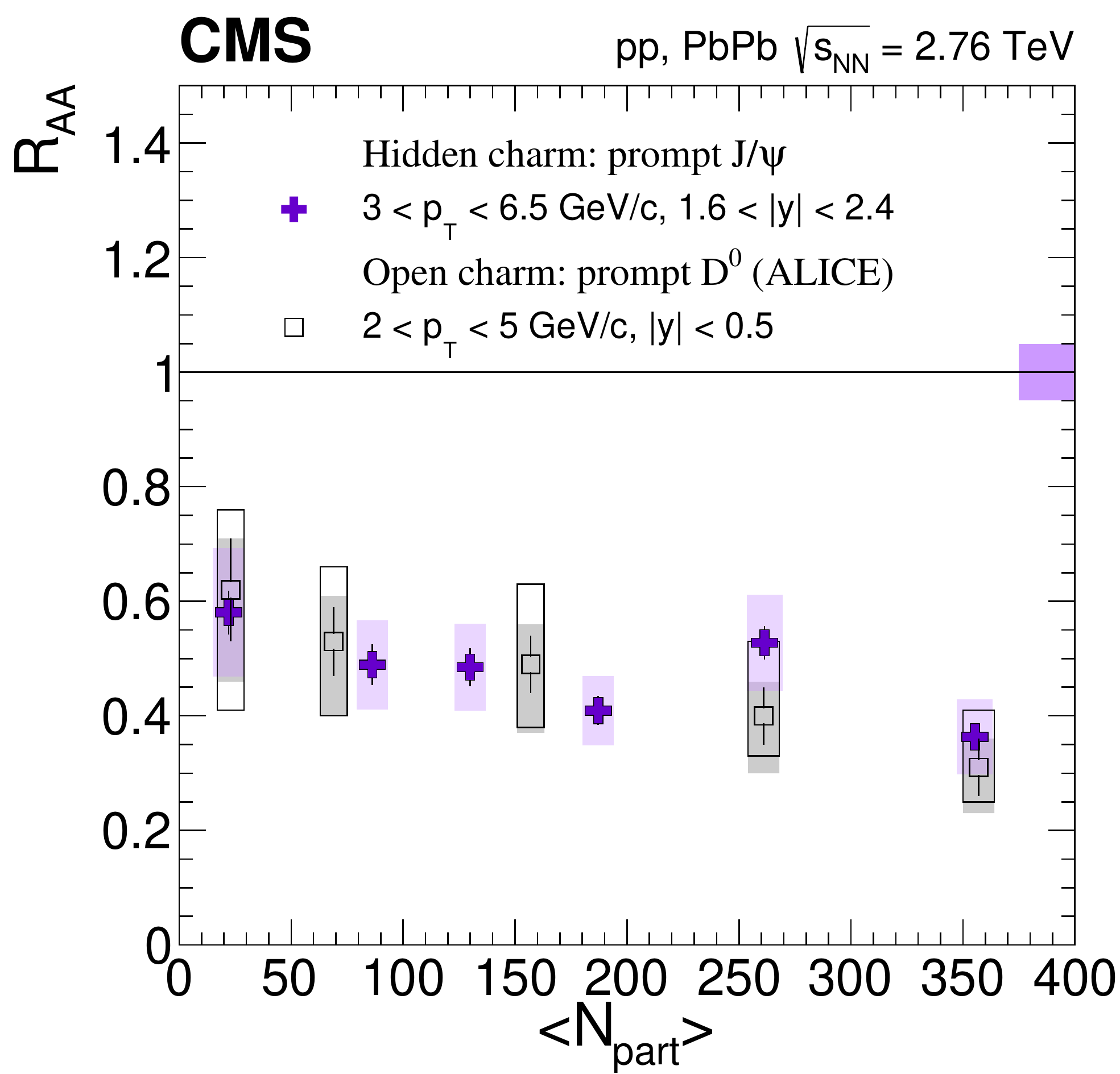}}
    {\includegraphics[width=\cmsFigWidth]{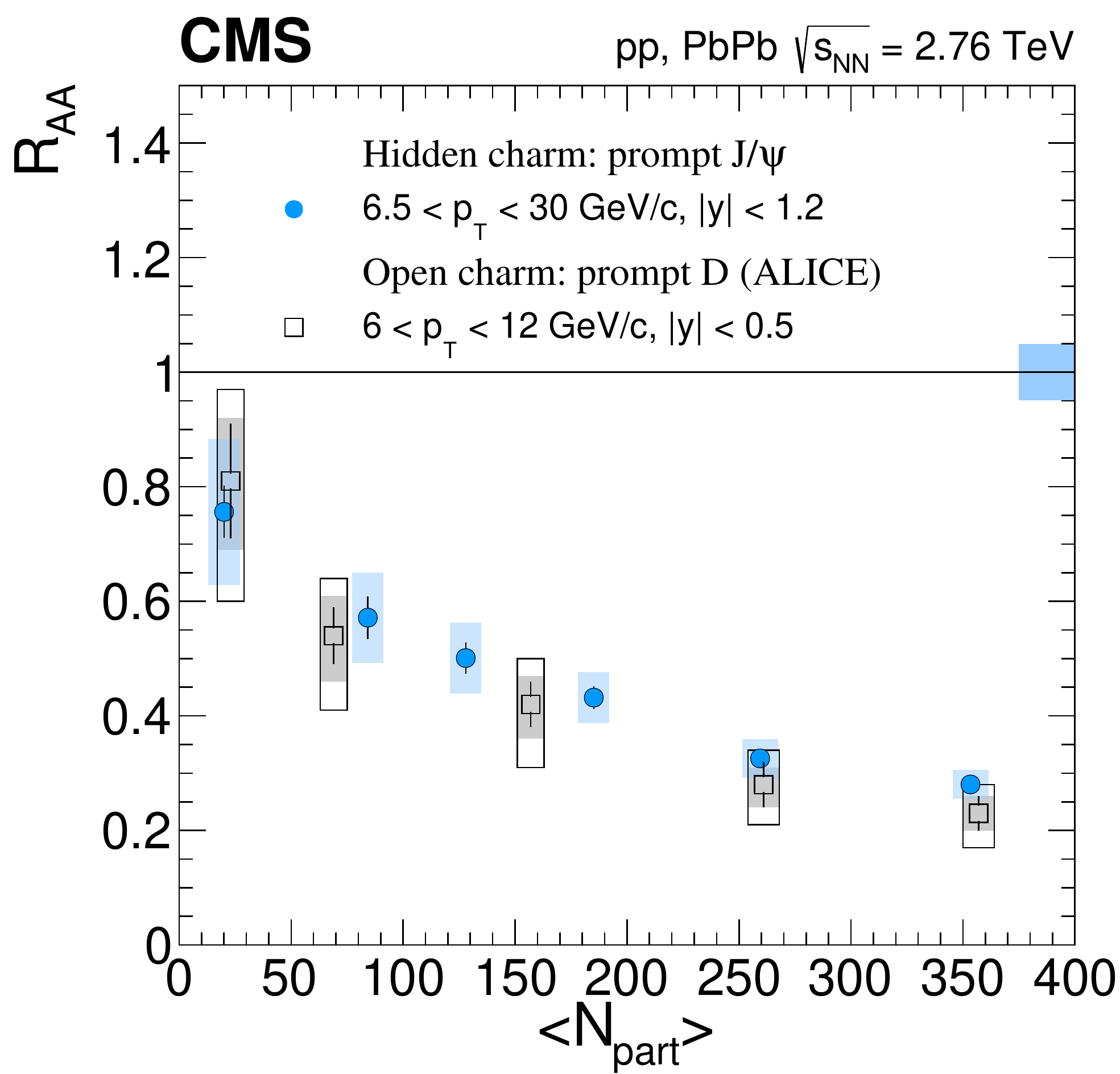}}\\
    {\includegraphics[width=\cmsFigWidth]{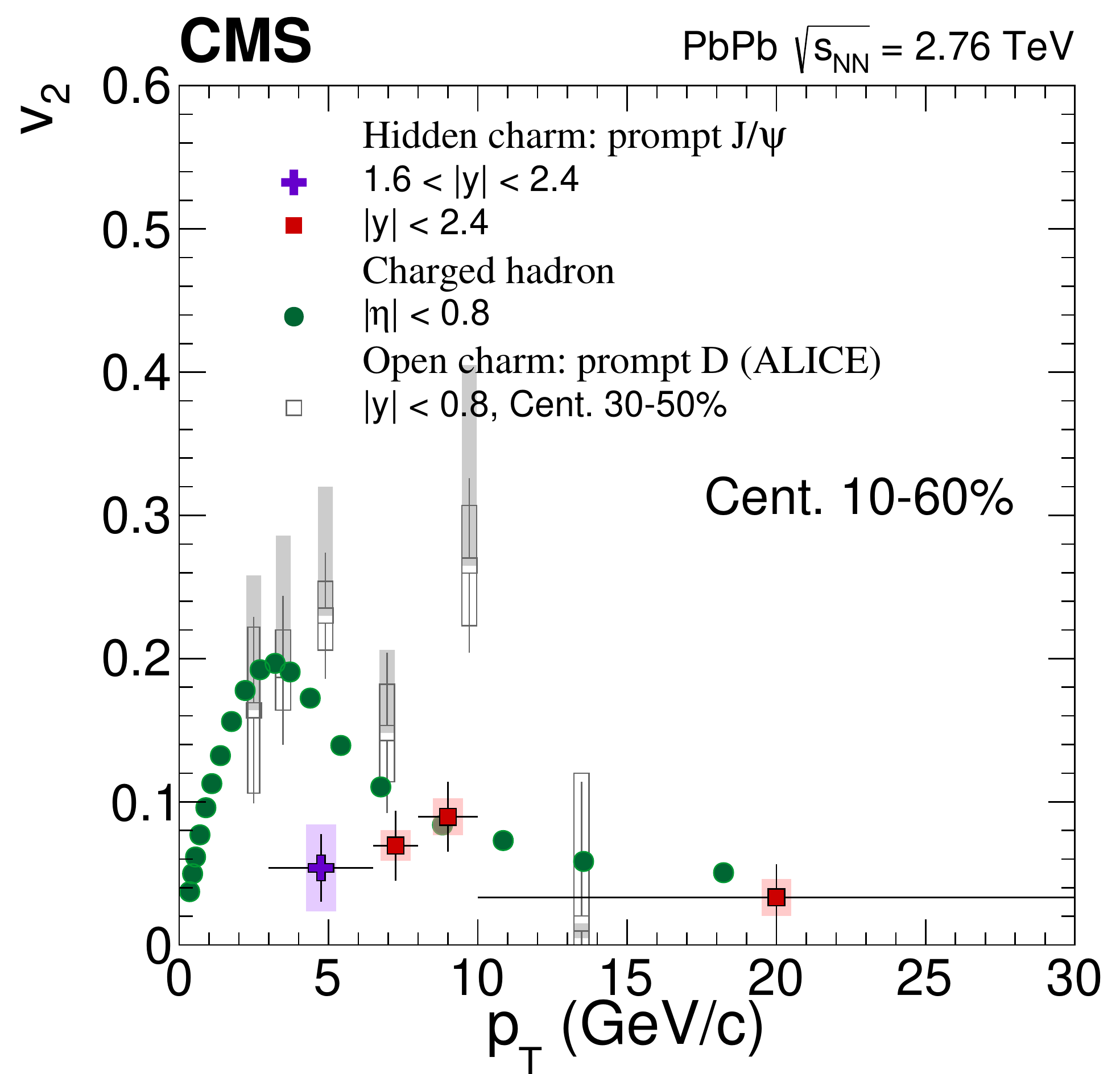}}
    \caption{Prompt \JPsi and D meson (\cite{ALICE:2012ab}) \raa \vs centrality for low \pt (\cmsLLeft) and high \pt (\cmsRRight). The average \npart values correspond to events flatly distributed across centrality.
(bottom) Prompt \JPsi and D meson (\cite{ Abelev:2014ipa}), and charged hadron (\cite{Chatrchyan:2012xq,Chatrchyan:2012ta}) \vtwo \vs \pt.}
    \label{fig:end_charm}
\end{figure}

The top two panels of Fig.~\ref{fig:end_charm} show the \raa dependence on the centrality of the prompt \JPsi (bound \QQbar state) and of prompt D (charm-light states \Qqbar) mesons, for low- (\cmsLLeft) and high- (\cmsRRight) \pt selections. In both cases, the mesons suffer a similar suppression, over the whole \npart range, even though the charmonium yield should be affected by colour screening~\cite{Matsui:1986dk, Chatrchyan:2012lxa}, potentially by final-state nuclear interactions unrelated to the QGP \cite{Arleo:2013zua, Ferreiro:2014bia, Fujii:2013gxa, Adam:2015jsa, Adam:2015iga}, and by rather large feed-down contributions from excited states~\cite{Faccioli:2008ir,LHCb:2012af}. Moreover, common processes (\ie recombination or energy loss effects) are expected to affect differently the open and hidden charm~\cite{Zhao:2011cv, Adam:2015isa, He:2014cla, Adam:2015sza}. While the present results cannot resolve all these effects, the comparison of open and hidden charm could help to determine their admixture.

A comparison of the \pt dependence of the azimuthal anisotropy \vtwo between the prompt \JPsi and D mesons is made in the bottom panel of Fig.~\ref{fig:end_charm}. While the \raa is similar both at low and high \pt, the \vtwo of prompt \JPsi at low \pt is lower than that of both D mesons and charged hadrons. At high \pt, all three results, within the uncertainties, are similar: the prompt \JPsi results seem to point to a similar anisotropy as the light-quarks hadrons, hinting at a flavour independence of the energy-loss path-length dependence. The prompt \JPsi results could help advance the theoretical knowledge on the relative contribution of the regenerated charmonium yield, as this is the only type of \JPsi expected to be affected by the collective expansion of the medium. Such prompt \JPsi should have higher \vtwo values, closer to those of light-quark hadrons~\cite{Zhao:2011cv}.

\subsection{Open charm versus beauty}

\begin{figure}[tbhp]\centering
    \includegraphics[width=\cmsFigWidth]{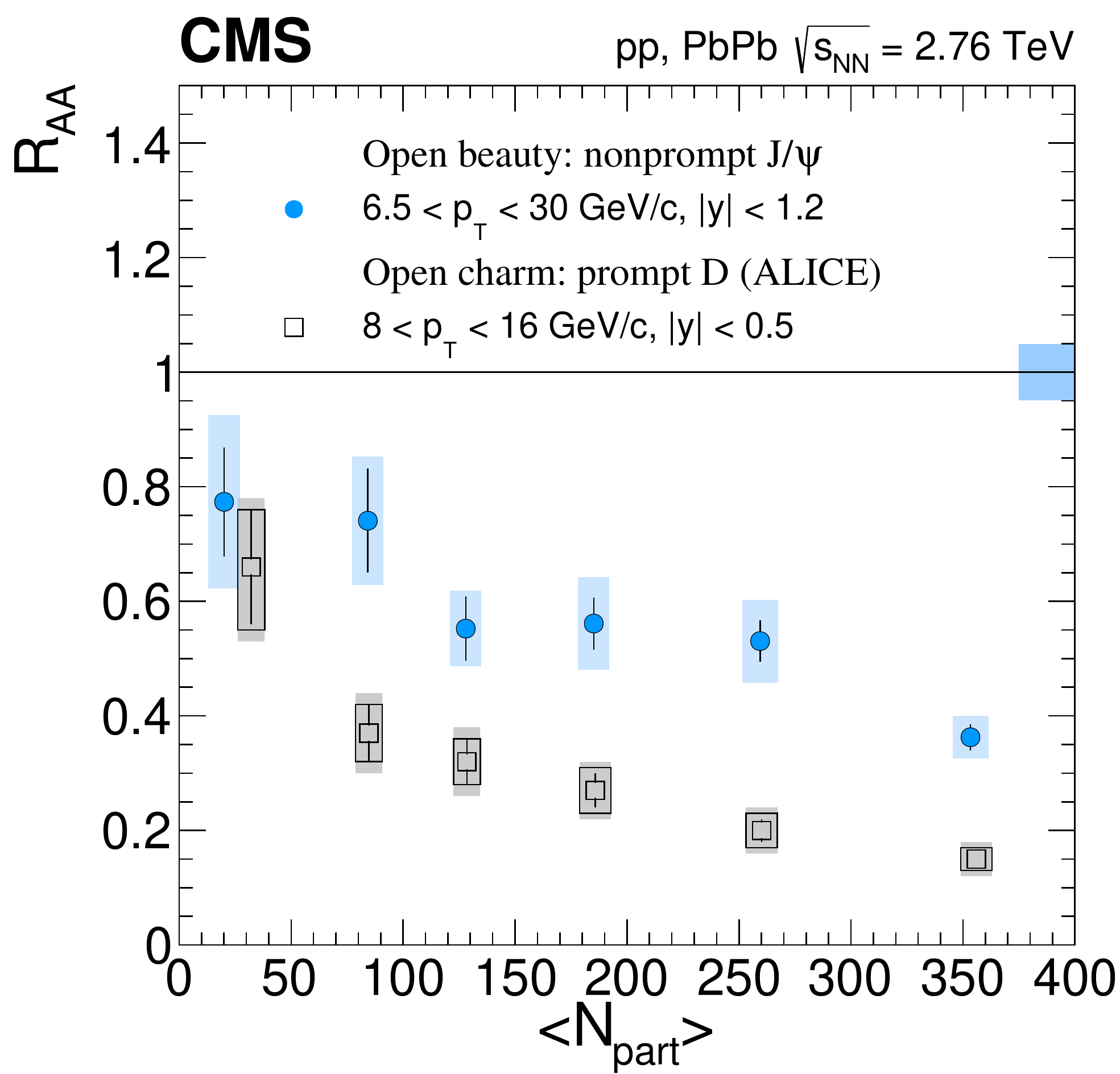}
    \includegraphics[width=\cmsFigWidth]{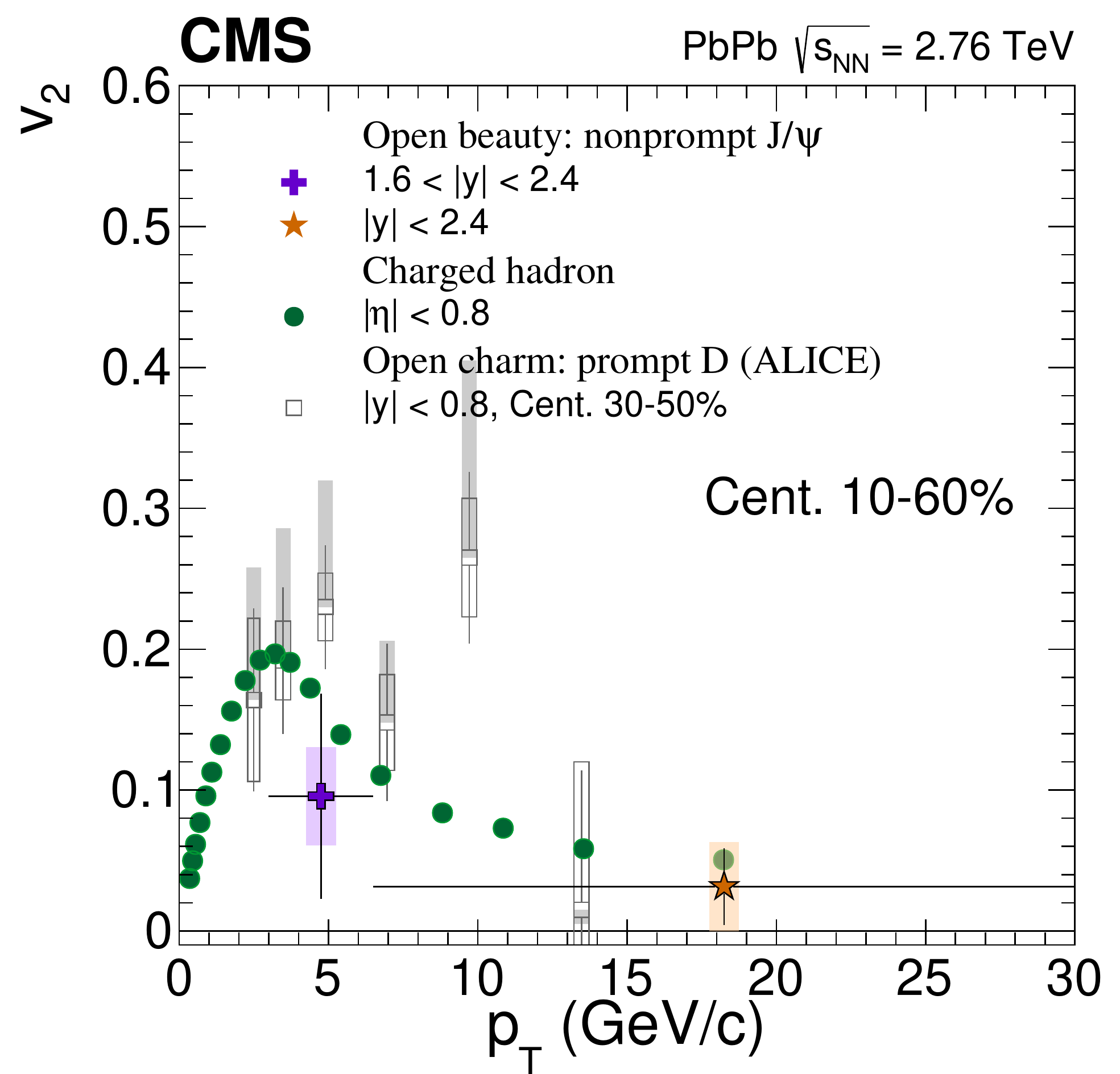}
    \caption{Nonprompt \JPsi and prompt D meson (\cite{Adam:2015nna, Abelev:2014ipa}), and charged hadron (\cite{Chatrchyan:2012xq, Chatrchyan:2012ta}) \raa \vs centrality (\cmsLeft), and \vtwo \vs \pt (\cmsRight). For the \cmsLeft plot, the average \npart values correspond to events flatly distributed across centrality.}
    \label{fig:end_open}
\end{figure}

The \cmsLeft panel of Fig.~\ref{fig:end_open} shows the \raa dependence on centrality of the nonprompt \JPsi (decay product of B~mesons originating from b quarks) and for D~mesons (originating from c quarks). The D~mesons are more suppressed than the nonprompt \JPsi. This is expected in models that assume less radiative energy loss for the b quark compared to that of a c quark because of the `dead-cone effect' (the suppression of gluon bremsstrahlung of a quark with mass $m$ and energy $E$, for angles $\theta < m/E$~\cite{Dokshitzer:1991fd,Armesto:2003jh}), and smaller collisional energy loss for the much heavier b quark than for the c quark~\cite{Djordjevic:2003zk,Wicks:2007am}. The results bring extra information in a kinematic phase space not accessible with fully reconstructed b jet measurements, which show that for $\pt>80$\GeVc the \raa of b jets is compatible to that of light-quark or gluon jets~\cite{Chatrchyan:2013exa}. However, assessing and quantifying the parton mass dependence of the in-medium phenomena is not trivial: one has to account among other things for different starting kinematics (different unmodified vacuum spectra of the beauty and charm quarks in the medium), and the effect of different fragmentation functions (and extra decay kinematics)~\cite{Djordjevic:2013pba}. Also, when considering the parton mass dependence, it should be noted that at high-\pt, the \raa of D~mesons was found to be similar to that of charged pions over a wide range of event centrality~\cite{Abelev:2014ipa}.

The \cmsRight panel of Fig.~\ref{fig:end_open} shows the \pt dependence of the measured \vtwo for nonprompt \JPsi, prompt D mesons, and charged hadrons. The precision and statistical reach of the present LHC open beauty and charm \vtwo results can not answer: (a)~at low \pt, whether the b quarks, with their mass much larger than that of the charm quarks, participate or not in the collective expansion of the medium as the charm quarks seem to do; (b)~at high \pt, whether there is a difference in path-length dependence of energy loss between b and c quarks.

\section{Summary}

The production of prompt and nonprompt (coming from b hadron decay) \JPsi has been studied in pp and PbPb collisions at $\sqrtsnn= 2.76$\TeV.
The \raa of the prompt \JPsi mesons, integrated over the rapidity range $\abs{y}<2.4$ and
high \pt, $6.5<\pt<30$\GeVc, is measured in 12 centrality bins. The \raa is less than unity even in the most peripheral bin, and the suppression becomes steadily stronger as centrality increases. Integrated over rapidity (\pt) and centrality, no strong evidence for a \pt (rapidity) dependence of the
suppression is found. The azimuthal anisotropy of prompt \JPsi mesons shows a nonzero \vtwo value in all studied bins, while no strong dependence on centrality, rapidity, or \pt is observed.

The \raa of nonprompt \JPsi mesons shows a slow decrease with increasing centrality and rapidity. The results show less suppression at low \pt. The first measurement of the nonprompt \JPsi \vtwo is also reported in two \pt bins for 10--60\% event centrality, and the values are consistent with zero elliptical azimuthal anisotropy, though both nominal values are positive.

\begin{acknowledgments}
\hyphenation{Bundes-ministerium Forschungs-gemeinschaft Forschungs-zentren Rachada-pisek} We congratulate our colleagues in the CERN accelerator departments for the excellent performance of the LHC and thank the technical and administrative staffs at CERN and at other CMS institutes for their contributions to the success of the CMS effort. In addition, we gratefully acknowledge the computing centres and personnel of the Worldwide LHC Computing Grid for delivering so effectively the computing infrastructure essential to our analyses. Finally, we acknowledge the enduring support for the construction and operation of the LHC and the CMS detector provided by the following funding agencies: the Austrian Federal Ministry of Science, Research and Economy and the Austrian Science Fund; the Belgian Fonds de la Recherche Scientifique, and Fonds voor Wetenschappelijk Onderzoek; the Brazilian Funding Agencies (CNPq, CAPES, FAPERJ, and FAPESP); the Bulgarian Ministry of Education and Science; CERN; the Chinese Academy of Sciences, Ministry of Science and Technology, and National Natural Science Foundation of China; the Colombian Funding Agency (COLCIENCIAS); the Croatian Ministry of Science, Education and Sport, and the Croatian Science Foundation; the Research Promotion Foundation, Cyprus; the Secretariat for Higher Education, Science, Technology and Innovation, Ecuador; the Ministry of Education and Research, Estonian Research Council via IUT23-4 and IUT23-6 and European Regional Development Fund, Estonia; the Academy of Finland, Finnish Ministry of Education and Culture, and Helsinki Institute of Physics; the Institut National de Physique Nucl\'eaire et de Physique des Particules~/~CNRS, and Commissariat \`a l'\'Energie Atomique et aux \'Energies Alternatives~/~CEA, France; the Bundesministerium f\"ur Bildung und Forschung, Deutsche Forschungsgemeinschaft, and Helmholtz-Gemeinschaft Deutscher Forschungszentren, Germany; the General Secretariat for Research and Technology, Greece; the National Scientific Research Foundation, and National Innovation Office, Hungary; the Department of Atomic Energy and the Department of Science and Technology, India; the Institute for Studies in Theoretical Physics and Mathematics, Iran; the Science Foundation, Ireland; the Istituto Nazionale di Fisica Nucleare, Italy; the Ministry of Science, ICT and Future Planning, and National Research Foundation (NRF), Republic of Korea; the Lithuanian Academy of Sciences; the Ministry of Education, and University of Malaya (Malaysia); the Mexican Funding Agencies (BUAP, CINVESTAV, CONACYT, LNS, SEP, and UASLP-FAI); the Ministry of Business, Innovation and Employment, New Zealand; the Pakistan Atomic Energy Commission; the Ministry of Science and Higher Education and the National Science Centre, Poland; the Funda\c{c}\~ao para a Ci\^encia e a Tecnologia, Portugal; JINR, Dubna; the Ministry of Education and Science of the Russian Federation, the Federal Agency of Atomic Energy of the Russian Federation, Russian Academy of Sciences, and the Russian Foundation for Basic Research; the Ministry of Education, Science and Technological Development of Serbia; the Secretar\'{\i}a de Estado de Investigaci\'on, Desarrollo e Innovaci\'on and Programa Consolider-Ingenio 2010, Spain; the Swiss Funding Agencies (ETH Board, ETH Zurich, PSI, SNF, UniZH, Canton Zurich, and SER); the Ministry of Science and Technology, Taipei; the Thailand Center of Excellence in Physics, the Institute for the Promotion of Teaching Science and Technology of Thailand, Special Task Force for Activating Research and the National Science and Technology Development Agency of Thailand; the Scientific and Technical Research Council of Turkey, and Turkish Atomic Energy Authority; the National Academy of Sciences of Ukraine, and State Fund for Fundamental Researches, Ukraine; the Science and Technology Facilities Council, UK; the US Department of Energy, and the US National Science Foundation.

Individuals have received support from the Marie-Curie programme and the European Research Council and EPLANET (European Union); the Leventis Foundation; the A. P. Sloan Foundation; the Alexander von Humboldt Foundation; the Belgian Federal Science Policy Office; the Fonds pour la Formation \`a la Recherche dans l'Industrie et dans l'Agriculture (FRIA-Belgium); the Agentschap voor Innovatie door Wetenschap en Technologie (IWT-Belgium); the Ministry of Education, Youth and Sports (MEYS) of the Czech Republic; the Council of Science and Industrial Research, India; the HOMING PLUS programme of the Foundation for Polish Science, cofinanced from European Union, Regional Development Fund, the Mobility Plus programme of the Ministry of Science and Higher Education, the National Science Center (Poland), contracts Harmonia 2014/14/M/ST2/00428, Opus 2013/11/B/ST2/04202, 2014/13/B/ST2/02543 and 2014/15/B/ST2/03998, Sonata-bis 2012/07/E/ST2/01406; the Thalis and Aristeia programmes cofinanced by EU-ESF and the Greek NSRF; the National Priorities Research Program by Qatar National Research Fund; the Programa Clar\'in-COFUND del Principado de Asturias; the Rachadapisek Sompot Fund for Postdoctoral Fellowship, Chulalongkorn University and the Chulalongkorn Academic into Its 2nd Century Project Advancement Project (Thailand); and the Welch Foundation, contract C-1845.
\end{acknowledgments}

\bibliography{auto_generated}
\appendix
\section{Supplemental Material}
\label{app_supplementalMaterial}

The nominator and denominator of the \raa, defined in Eq.~\ref{eq:raa}, and presented in this paper in Figs.~\ref{fig:promptJpsi_RAA} and ~\ref{fig:promptJpsi_RAA_pteta} for prompt \JPsi, and Figs.~\ref{fig:nonPromptJpsi_RAA} and ~\ref{fig:nonPromptJpsi_RAA_pteta} for nonprompt \JPsi, are tabulated. They represent the efficiency-corrected signal yield within the single muon kinematic region used in this paper. This kinematic region is defined in Eq.~\ref{eq:singleMuonAcc}. These \sqrtsnn=2.76\TeV pp and PbPb fiducial cross sections do not depend on the acceptance, or the associated uncertainties. The corresponding \taa values used in each case are also tabulated.
\begin{linenomath}
  \begin{align}\label{eq:singleMuonAcc}
    \pt^{\mu} &> 3.4\GeVc &\text{ for } |\eta^{\mu}| < 1.0, \notag\\
    \pt^{\mu} &> (5.8 - 2.4\,|\eta^{\mu}|)\GeVc &\text{ for } 1.0 < |\eta^{\mu}| < 1.5,\\
    \pt^{\mu} &> (3.4 - 0.78\,|\eta^{\mu}|)\GeVc &\text{ for } 1.5 < |\eta^{\mu}| < 2.4. \notag
  \end{align}
\end{linenomath}

\subsection{Prompt \texorpdfstring{\JPsi}{Jpsi}}

\begin{table*}[h]
  \begin{center}
    \caption{The prompt \JPsi fiducial cross section in bins of centrality, measured in
    \PbPb and \pp collisions at 2.76\TeV within the muon acceptance defined by Eq.~\ref{eq:singleMuonAcc}, and the nuclear overlap function (\taa, with its systematic uncertainty). Listed uncertainties are statistical
     first and systematic second. A global systematic uncertainty of 3.2\% (3.7\%) affects all \PbPb (\pp) fiducial cross sections. The table corresponds to the top panel of Fig.~\ref{fig:promptJpsi_RAA}.}
    \vspace{1em}
    \label{tab:yield_prompt_cent}
    \begin{tabular}{r@{$-$}l r@{$\,\pm\,$}l r@{$\,\pm\,$}c@{$\,\pm\,$}l c}
      \hline
      \multicolumn{4}{c}{\,} & \multicolumn{3}{c}{PbPb} & pp\\
      \hline
        \multicolumn{2}{c}{centrality} & \multicolumn{2}{c}{\taa} &
         \multicolumn{3}{c}{$\frac{1}{\taa} \, \frac{\rd^3\nAA}{\rd y \rd \pt \rd \text{Cent.}}$} &
          $\frac{\rd^2\sigPP}{\rd y \rd \pt}$  \\
              \multicolumn{2}{c}{[\%]}      & \multicolumn{2}{c}{[\mbinv]}  &  \multicolumn{3}{c}{[pb/\GeVc]} & [pb/\GeVc] \\
      \hline
              \multicolumn{8}{c}{$\abs{y}<2.4$, $6.5<\pt<30$\GeVc} \\
      \hline
        60 & 100  &  0.246 & 0.041   &   50 & 3 & 9   &  \multirow{12}{*}{69.6 $\pm$ 0.6 $\pm$ 4.1}\\
        50 & 60    &  1.36  & 0.19      &   50 & 3 & 8   &  \\
        45 & 50    &  2.29  & 0.26      &   39 & 3 & 5   &  \\
        40 & 45    &  3.20  & 0.34       &   38 & 2 & 5   &  \\
        35 & 40    &  4.4     & 0.4        &   33 & 2 & 4   &  \\
        30 & 35   &  5.8      & 0.5       &   34 & 2 & 4   &  \\
        25 & 30   &  7.7      & 0.5       &   32 & 1 & 4   &  \\
        20 & 25   &  9.9      & 0.6       &   29 & 1 & 3   &  \\
        15 & 20   &  12.7 & 0.7          &  25  & 1 & 2   & \\
        10 & 15   &  16.2 & 0.8          &  21.7 & 0.9 & 2.3  & \\
        5 &10     &  20.5 & 0.9           &   20.9 & 0.8 & 1.7  & \\
        0 & 5      &  25.9 & 1.1           &  19.6  & 0.7 & 1.6  &  \\
      \hline
    \end{tabular}
  \end{center}
\end{table*}

\begin{table*}[h]
  \begin{center}
    \caption{The prompt \JPsi fiducial cross section in bins of absolute rapidity, measured in
    \PbPb and \pp collisions at 2.76\TeV within the muon acceptance defined by Eq.~\ref{eq:singleMuonAcc}, and the nuclear overlap function (\taa, with its systematic uncertainty). Listed uncertainties are statistical
     first and systematic second. A global systematic uncertainty of 6.5\% (3.7\%) affects all \PbPb (\pp) fiducial cross sections. The table corresponds to the middle panel of Fig.~\ref{fig:promptJpsi_RAA}.}
    \vspace{1em}
    \label{tab:yield_prompt_y}
    \begin{tabular}{r@{$-$}l c r@{$\,\pm\,$}c@{$\,\pm\,$}l r@{$\,\pm\,$}c@{$\,\pm\,$}l}
      \hline
     \multicolumn{3}{c}{\,} & \multicolumn{3}{c}{PbPb} & \multicolumn{3}{c}{pp}\\
      \hline
        \multicolumn{2}{c}{$\abs{y}$}  & \taa &
        \multicolumn{3}{c}{$\frac{1}{\taa} \, \frac{\rd^2\nAA}{\rd y \rd \pt}$}&
        \multicolumn{3}{c}{$\frac{\rd^2\sigPP}{\rd y \rd \pt}$} \\
         \multicolumn{2}{c}{}   & [\mbinv]  & \multicolumn{3}{c}{[pb/\GeVc]} & \multicolumn{3}{c}{[pb/\GeVc]} \\
      \hline
            \multicolumn{9}{c}{Cent. 0--100\%, $6.5<\pt<30$\GeVc}\\
      \hline
        0.0 & 0.4     &  \multirow{6}{*}{5.67$\pm$0.32} & 18.1 & 0.6 & 1.4    &  53 & 1 & 3  \\
        0.4 & 0.8     &                                                    & 21.1 & 0.7 & 1.8    &  57 & 1 & 4\\
        0.8 & 1.2     &                                                    & 28.7 & 0.9 & 2.0    &  74 & 1 & 4\\
        1.2 & 1.6     &   						      & 36    & 1     & 2      &  94 & 2 & 6\\
        1.6 & 2 .0    &    					      & 38    & 1     & 3      &  98 & 2 & 7\\
        2.0 & 2.4     &    				              & 14.4 & 0.8 & 1.4    &  44 & 1 & 4\\
      \hline
    \end{tabular}
  \end{center}
\end{table*}

\begin{table*}[h]
  \begin{center}
    \caption{The prompt \JPsi fiducial cross section in bins of \pt, measured in
    \PbPb and \pp collisions at 2.76\TeV within the muon acceptance defined by Eq.~\ref{eq:singleMuonAcc}, and the nuclear overlap function (\taa, with its systematic uncertainty). Listed uncertainties are statistical
     first and systematic second. A global systematic uncertainty of 6.5\% (3.7\%) affects all \PbPb (\pp) fiducial cross sections. The table corresponds to the bottom panel of Fig.~\ref{fig:promptJpsi_RAA}.}
    \vspace{1em}
    \label{tab:yield_prompt_pt}
    \begin{tabular}{r@{$-$}l c r@{$\,\pm\,$}c@{$\,\pm\,$}l r@{$\,\pm\,$}c@{$\,\pm\,$}l}
      \hline
     \multicolumn{2}{c}{}&\multicolumn{3}{c}{PbPb} & \multicolumn{3}{c}{pp}\\
     \hline
       \multicolumn{2}{c}{\pt} & \taa &
       \multicolumn{3}{c}{ $\frac{1}{\taa} \, \frac{\rd^2\nAA}{\rd y \rd \pt}$} &
        \multicolumn{3}{c}{$\frac{\rd^2\sigPP}{\rd y \rd \pt}$} \\
                \multicolumn{2}{c}{[\GeVc]}      & [\mbinv]  & \multicolumn{3}{c}{[pb/\GeVc]} & \multicolumn{3}{c}{[pb/\GeVc]} \\
      \hline
               \multicolumn{9}{c}{Cent. 0--100\%, $1.6<\abs{y}<2.4$}\\
      \hline
      3    & 4.5    &  \multirow{3}{*}{5.67$\pm$0.32}  & 272  & 16 & 40    &  534  & 10    & 90  \\
      4.5 & 5.5    &  					             &  181 & 15 & 23    &  478  & 10    & 41  \\
      5.5 & 6.5    &   						     &  137 & 7   &14     &  355  & 8      & 28  \\
       \hline
        \multicolumn{9}{c}{Cent. 0--100\%, $\abs{y}<2.4$}\\
      \hline
    6.5 & 8.5    &  \multirow{6}{*}{5.67$\pm$0.32}     &  169 & 4   &14     &  455  & 5      & 33  \\
    8.5  & 9.5   &   						     &  85   & 3   &   5    &  252  & 5      & 15  \\
    9.5  & 11    &   						     &  55   & 2   &  3     &  147  & 3      & 8  \\
    11   & 13    &   						     &  26   & 1   &  2     &  70    & 2      & 4  \\
    13   & 16    &   						     &  11.5   &  0.5  &  0.9     &  25.8 & 0.8   & 1.2  \\
    16   & 30    &   						     &  1.25   & 0.08 &0.20     &  3.23 & 0.14 & 0.14  \\
      \hline
    \end{tabular}
  \end{center}
\end{table*}

\begin{table*}[h]
  \begin{center}
    \caption{The prompt \JPsi fiducial cross section in bins of centrality, for three $\abs{y}$ and two \pt intervals, measured in
    \PbPb and \pp collisions at 2.76\TeV within the muon acceptance defined by Eq.~\ref{eq:singleMuonAcc}, and the nuclear overlap function (\taa, with its systematic uncertainty). Listed uncertainties are statistical
     first and systematic second. A global systematic uncertainty of 3.2\% (3.7\%) affects all \PbPb (\pp) fiducial cross sections. The table corresponds to Fig.~\ref{fig:promptJpsi_RAA_pteta}.}
    \vspace{1em}
    \label{tab:yield_prompt_pteta}
    \begin{tabular}{r@{$-$}l r@{$\,\pm\,$}l r@{$\,\pm\,$}c@{$\,\pm\,$}l c}
      \hline
      \multicolumn{4}{c}{\,} & \multicolumn{3}{c}{PbPb} & pp\\
      \hline
        \multicolumn{2}{c}{centrality} & \multicolumn{2}{c}{\taa} &
         \multicolumn{3}{c}{$\frac{1}{\taa} \, \frac{\rd^3\nAA}{\rd y \rd \pt \rd \text{Cent.}}$} &
          $\frac{\rd^2\sigPP}{\rd y \rd \pt}$  \\
              \multicolumn{2}{c}{[\%]}      & \multicolumn{2}{c}{[\mbinv]}  &  \multicolumn{3}{c}{[pb/\GeVc]} & [pb/\GeVc] \\
      \hline
      \multicolumn{8}{c}{$0<\abs{y}<1.2$, $6.5<\pt<30$\GeVc} \\
      \hline
        50&100 &   0.468 & 0.070 &  47 & 3 & 8  & \multirow{6}{*}{61.4$\pm$0.7$\pm$3.7}  \\
        40&50   &   2.75 & 0.30     &  35 & 2 & 5  &    \\
        30&40   &   5.1 & 0.4         &  31 & 2 & 4  &   \\
        20&30   &   8.8 & 0.6         &  27 & 1 & 3  &   \\
        10&20   &   14.5 & 0.8       &  20.0 & 0.8 & 2.1  &   \\
        0&10     &   23  & 1            &  17.2 & 0.7 & 1.6  &    \\
      \hline
      \multicolumn{8}{c}{$1.2<\abs{y}<1.6$, $6.5<\pt<30$\GeVc} \\
      \hline
        50&100 &  0.468 & 0.070 &  71 & 6 & 12  &  \multirow{6}{*}{94$\pm$2$\pm$6}  \\
        40&50   &  2.75 & 0.30     &  55 & 5 & 7    &    \\
        30&40   &  5.1 & 0.4         &  48 & 4 & 5    &    \\
        20&30   &  8.8 & 0.6         &  43 & 3 & 4    &    \\
        10&20   &  14.5 & 0.8       &  30 & 2 & 3    &    \\
        0&10     &  23  & 1            &  27 & 1 & 2    &    \\
      \hline
      \multicolumn{8}{c}{$1.6<\abs{y}<2.4$, $6.5<\pt<30$\GeVc} \\
      \hline
        50&100 &  0.468 & 0.070  &  46 & 4 & 8   &  \multirow{6}{*}{71$\pm$1$\pm$5}  \\
        40&50   &  2.75 & 0.30      &  36 & 3 & 5   &    \\
        30&40   &  5.1 & 0.4          &  30 & 2 & 5   &    \\
        20&30   &  8.8 & 0.6          &  28 & 2 & 3   &   \\
        10&20   &  14.5 & 0.8        &  24 & 1 & 3   &    \\
        0&10     &  23  & 1             &  22 & 1 & 2   &    \\
      \hline
      \multicolumn{8}{c}{$1.6<\abs{y}<2.4$, $3<\pt<6.5$\GeVc} \\
      \hline
        50&100 &  0.468 & 0.070  &  815 & 53 & 158   &  \multirow{6}{*}{1397$\pm$16$\pm$166}  \\
        40&50   &  2.75 & 0.30      &  685 & 50 & 109   &  \\
        30&40   &  5.1 & 0.4          &  677 & 46 & 107   &  \\
        20&30   &  8.8 & 0.6          &  572 & 35 & 85     &  \\
        10&20   &  14.5 & 0.8        &  737 & 40 & 117   &  \\
        0&10     &  23  & 1             &  508 & 29 & 92     &  \\
      \hline
    \end{tabular}
  \end{center}
\end{table*}

\clearpage
\subsection{Nonprompt \texorpdfstring{\JPsi}{Jpsi}}

\begin{table*}[h]
  \begin{center}
    \caption{The nonprompt \JPsi fiducial cross section in bins of centrality, measured in
    \PbPb and \pp collisions at 2.76\TeV within the muon acceptance defined by Eq.~\ref{eq:singleMuonAcc}, and the nuclear overlap function (\taa, with its systematic uncertainty). Listed uncertainties are statistical
     first and systematic second. A global systematic uncertainty of 3.2\% (3.7\%) affects all \PbPb (\pp) fiducial cross sections. The table corresponds to the top panel of Fig.~\ref{fig:nonPromptJpsi_RAA}.}
    \vspace{1em}
    \label{tab:yield_nonprompt_cent}
    \begin{tabular}{r@{$-$}l r@{$\,\pm\,$}l r@{$\,\pm\,$}c@{$\,\pm\,$}l c}
      \hline
      \multicolumn{4}{c}{\,} & \multicolumn{3}{c}{PbPb} & pp\\
      \hline
        \multicolumn{2}{c}{centrality} & \multicolumn{2}{c}{\taa} &
         \multicolumn{3}{c}{$\frac{1}{\taa} \, \frac{\rd^3\nAA}{\rd y \rd \pt \rd \text{Cent.}}$} &
          $\frac{\rd^2\sigPP}{\rd y \rd \pt}$  \\
              \multicolumn{2}{c}{[\%]}      & \multicolumn{2}{c}{[\mbinv]}  &  \multicolumn{3}{c}{[pb/\GeVc]} & [pb/\GeVc] \\
      \hline
              \multicolumn{8}{c}{$\abs{y}<2.4$, $6.5<\pt<30$\GeVc} \\
      \hline
        50 & 100  &  0.468 & 0.070  &  17    & 2     & 3   &  \multirow{6}{*}{23.57$\pm$0.33$\pm$1.41}  \\
        40 & 50    &  2.75 & 0.30      &  16    & 1     & 2   &     \\
        30 & 40    &  5.1 & 0.4          &  13    & 1     & 1   &     \\
        20 & 30    &  8.8 & 0.6          &  11.9 & 0.7  & 1.4   &   \\
        10 & 20    &  14.5 & 0.8        &  10.4 & 0.5  & 1.3  &    \\
        0 & 10      &  23  & 1             &  7.8   & 0.4  & 0.7   &   \\
      \hline
    \end{tabular}
  \end{center}
\end{table*}

\begin{table*}[h]
  \begin{center}
    \caption{The nonprompt \JPsi fiducial cross section in bins of absolute rapidity, measured in
    \PbPb and \pp collisions at 2.76\TeV within the muon acceptance defined by Eq.~\ref{eq:singleMuonAcc}, and the nuclear overlap function (\taa, with its systematic uncertainty). Listed uncertainties are statistical
     first and systematic second. A global systematic uncertainty of 6.5\% (3.7\%) affects all \PbPb (\pp) fiducial cross sections. The table corresponds to the middle panel of Fig.~\ref{fig:nonPromptJpsi_RAA}.}
    \vspace{1em}
    \label{tab:yield_nonprompt_y}
    \begin{tabular}{r@{$-$}l c r@{$\,\pm\,$}c@{$\,\pm\,$}l r@{$\,\pm\,$}c@{$\,\pm\,$}l}
      \hline
     \multicolumn{3}{c}{\,} & \multicolumn{3}{c}{PbPb} & \multicolumn{3}{c}{pp}\\
      \hline
        \multicolumn{2}{c}{$\abs{y}$}  & \taa &
        \multicolumn{3}{c}{$\frac{1}{\taa} \, \frac{\rd^2\nAA}{\rd y \rd \pt}$}&
        \multicolumn{3}{c}{$\frac{\rd^2\sigPP}{\rd y \rd \pt}$} \\
         \multicolumn{2}{c}{}   & [\mbinv]  & \multicolumn{3}{c}{[pb/\GeVc]} & \multicolumn{3}{c}{[pb/\GeVc]} \\
      \hline
            \multicolumn{9}{c}{Cent. 0--100\%, $6.5<\pt<30$\GeVc}\\
      \hline
        0.0 & 0.4  &\multirow{6}{*}{5.67$\pm$0.32}  &  10.5 & 0.6 & 1.3    &  20.0 & 0.7 & 1.3  \\
        0.4 & 0.8 &   						  &  12.1 & 0.7 & 1.3   &  23.8 & 0.8 & 1.9  \\
        0.8 & 1.2 &   						  &  11.3 & 0.6 & 0.9   &  25.2 & 0.8 & 1.4  \\
        1.2 & 1.6 &   						  &  13.1 & 0.8 & 1.2   &  32 & 1 & 2  \\
        1.6 & 2.0 &   						  &  10.7  & 0.8 & 1.0  &  29 & 1 & 2  \\
        2.0 & 2.4 &   						  &  4.2    & 0.5 & 0.7  &  12.2 & 0.7 & 1.2  \\
      \hline
    \end{tabular}
  \end{center}
\end{table*}

\begin{table*}[h]
  \begin{center}
    \caption{The nonprompt \JPsi fiducial cross section in bins of \pt, measured in
    \PbPb and \pp collisions at 2.76\TeV within the muon acceptance defined by Eq.~\ref{eq:singleMuonAcc}, and the nuclear overlap function (\taa, with its systematic uncertainty). Listed uncertainties are statistical
     first and systematic second. A global systematic uncertainty of 6.5\% (3.7\%) affects all \PbPb (\pp) fiducial cross sections. The table corresponds to the bottom panel of Fig.~\ref{fig:nonPromptJpsi_RAA}.}
    \vspace{1em}
    \label{tab:yield_nonprompt_pt}
    \begin{tabular}{r@{$-$}l c r@{$\,\pm\,$}c@{$\,\pm\,$}l r@{$\,\pm\,$}c@{$\,\pm\,$}l}
      \hline
     \multicolumn{2}{c}{}&\multicolumn{3}{c}{PbPb} & \multicolumn{3}{c}{pp}\\
     \hline
      \multicolumn{2}{c}{\pt} & \taa &
       \multicolumn{3}{c}{ $\frac{1}{\taa} \, \frac{\rd^2\nAA}{\rd y \rd \pt}$} &
        \multicolumn{3}{c}{$\frac{\rd^2\sigPP}{\rd y \rd \pt}$} \\
                \multicolumn{2}{c}{[\GeVc]}      & [\mbinv]  & \multicolumn{3}{c}{[pb/\GeVcns{}]} & \multicolumn{3}{c}{[pb/\GeVc]} \\
      \hline
             \multicolumn{9}{c}{Cent. 0--100\%, $1.6<\abs{y}<2.4$}\\
      \hline
       3 & 4.5    &\multirow{3}{*}{5.67$\pm$0.32}  &   46 & 7 & 8               &  61 & 4 & 14  \\
       4.5 &5.5  &   						  &  43 & 6 & 6                &  63 & 4 & 6  \\
       5.5 & 6.5  &   						  &  31 & 4 & 4                &  57 & 3 & 5  \\
        \hline
        \multicolumn{9}{c}{Cent. 0--100\%, $\abs{y}<2.4$}\\
      \hline
       6.5 & 8.5  & \multirow{6}{*}{5.67$\pm$0.32}  &   52 & 3 & 4              &  111 & 3 & 9  \\
       8.5 & 9.5  &   					           &   39 & 2 & 3               &  80 & 3 & 5  \\
       9.5 & 11   &   						  &   22 & 1 & 1                 &  55 & 2 & 3  \\
      11 & 13    &   						  &   16 & 1 & 2                 &  35 & 1 & 2  \\
      13 & 16    &   					           & 6.0 & 0.5 & 0.8           &16.3 & 0.7 & 0.8  \\
      16 & 30    &   						  & 1.071 & 0.082 & 0.203   & 3.04 & 0.13 & 0.14  \\
      \hline
    \end{tabular}
  \end{center}
\end{table*}

\begin{table*}[h]
  \begin{center}
    \caption{The nonprompt \JPsi fiducial cross section in bins of centrality, for three $\abs{y}$ and two \pt intervals, measured in
    \PbPb and \pp collisions at 2.76\TeV within the muon acceptance defined by Eq.~\ref{eq:singleMuonAcc}, and the nuclear overlap function (\taa, with its systematic uncertainty). Listed uncertainties are statistical
     first and systematic second. A global systematic uncertainty of 3.2\% (3.7\%) affects all \PbPb (\pp) fiducial cross sections. The table corresponds to Fig.~\ref{fig:nonPromptJpsi_RAA_pteta}.}
    \vspace{1em}
    \label{tab:yield_nonprompt_pteta}
    \begin{tabular}{r@{$-$}l r@{$\,\pm\,$}l r@{$\,\pm\,$}c@{$\,\pm\,$}l c}
      \hline
      \multicolumn{4}{c}{\,} & \multicolumn{3}{c}{PbPb} & pp\\
      \hline
        \multicolumn{2}{c}{centrality} & \multicolumn{2}{c}{\taa} &
         \multicolumn{3}{c}{$\frac{1}{\taa} \, \frac{\rd^3\nAA}{\rd y \rd \pt \rd \text{Cent.}}$} &
          $\frac{\rd^2\sigPP}{\rd y \rd \pt}$  \\
              \multicolumn{2}{c}{[\%]}      & \multicolumn{2}{c}{[\mbinv]}  &  \multicolumn{3}{c}{[pb/\GeVc{}]} & [pb/\GeVc{}] \\
      \hline
      \multicolumn{8}{c}{$0<\abs{y}<1.2$, $6.5<\pt<30$\GeVc} \\
      \hline
        50&100  &  0.468 & 0.070&  18    & 2 & 4   &  \multirow{6}{*}{23.3$\pm$0.4$\pm$1.6}  \\
        40&50    &  2.75 & 0.30    &  17    & 2 & 3   &     \\
        30&40    &  5.1 & 0.4        &  13    & 1 & 2   &     \\
        20&30    &  8.8 & 0.6        &  13    & 1 & 2   &     \\
        10&20    &  14.5 & 0.8      &  12.4 & 0.8 & 1.7   &     \\
        0&10      &  23  & 1           &  8.5   & 0.5 & 0.9   &     \\
      \hline
      \multicolumn{8}{c}{$1.2<\abs{y}<1.6$, $6.5<\pt<30$\GeVc} \\
      \hline
        50&100  &  0.468 & 0.070    &  22 & 4 & 4   &  \multirow{6}{*}{32$\pm$1$\pm$2}  \\
        40&50    &  2.75 & 0.30        &  20 & 4 & 3   &     \\
        30&40    &  5.1 & 0.4            &  12 & 2 & 1   &     \\
        20&30    &  8.8 & 0.6            &  15 & 2 & 2   &     \\
        10&20    &  14.5 & 0.8          &  13 & 1 & 1   &     \\
        0&10      &  23  & 1               &  11 & 1 & 1   &     \\
      \hline
      \multicolumn{8}{c}{$1.6<\abs{y}<2.4$, $6.5<\pt<30$\GeVc} \\
      \hline
        50&100  &  0.468 & 0.070  &  12 & 2 & 2   &  \multirow{6}{*}{20.3$\pm$0.6$\pm$1.5}  \\
        40&50    &  2.75 & 0.30      &  12 & 2 & 2   &     \\
        30&40    &  5.1 & 0.4          &  13 & 2 & 2   &     \\
        20&30    &  8.8 & 0.6          &    9 & 1 & 1   &     \\
        10&20    &  14.5 & 0.8        & 7.3 & 0.9 & 1.1   &     \\
        0&10      &  23  & 1             & 4.9 & 0.6 & 0.7   &     \\
      \hline
      \multicolumn{8}{c}{$1.6<\abs{y}<2.4$, $3<\pt<6.5$\GeVc} \\
      \hline
        50&100  &  0.468 & 0.070   &  163 & 40 & 37   &  \multirow{6}{*}{179$\pm$7$\pm$23}  \\
        40&50    &  2.75 & 0.30       &  192 & 35 & 31   &     \\
        30&40    &  5.1 & 0.4           &  144 & 29 & 23   &     \\
        20&30    &  8.8 & 0.6           &  139 & 22 & 20   &     \\
        10&20    &  14.5 & 0.8         &  120 & 21 & 21   &     \\
        0&10     &  23  & 1               &  101 & 15 & 23  &     \\
      \hline
    \end{tabular}
  \end{center}
\end{table*}

\cleardoublepage \section{The CMS Collaboration \label{app:collab}}\begin{sloppypar}\hyphenpenalty=5000\widowpenalty=500\clubpenalty=5000\input{HIN-14-005-authorlist.tex}\end{sloppypar}
\end{document}

%% file: HIN-14-005-authorlist.tex
\textbf{Yerevan Physics Institute,  Yerevan,  Armenia}\\*[0pt]
V.~Khachatryan, A.M.~Sirunyan, A.~Tumasyan
\vskip\cmsinstskip
\textbf{Institut f\"{u}r Hochenergiephysik,  Wien,  Austria}\\*[0pt]
W.~Adam, E.~Asilar, T.~Bergauer, J.~Brandstetter, E.~Brondolin, M.~Dragicevic, J.~Er\"{o}, M.~Flechl, M.~Friedl, R.~Fr\"{u}hwirth\cmsAuthorMark{1}, V.M.~Ghete, C.~Hartl, N.~H\"{o}rmann, J.~Hrubec, M.~Jeitler\cmsAuthorMark{1}, A.~K\"{o}nig, I.~Kr\"{a}tschmer, D.~Liko, T.~Matsushita, I.~Mikulec, D.~Rabady, N.~Rad, B.~Rahbaran, H.~Rohringer, J.~Schieck\cmsAuthorMark{1}, J.~Strauss, W.~Waltenberger, C.-E.~Wulz\cmsAuthorMark{1}
% ...
\vskip\cmsinstskip
% ...
1:~~Also at Vienna University of Technology, Vienna, Austria\\
% ...


%% file: HIN-14-005_temp.bbl
\providecommand{\href}[2]{#2}\begingroup\raggedright\begin{thebibliography}{10}%
\makeatletter
\providecommand{\hrefCMSnoop }[0]{\@secondoftwo}%
\makeatother
\providecommand{\doi}{\texttt{doi:}\begingroup \urlstyle{tt}\Url}

\bibitem{Andronic:2015wma}
\hrefCMSnoop {}{A.~Andronic {et~al.}, ``{Heavy-flavour and quarkonium
  production in the LHC era: from proton-proton to heavy-ion collisions}'',}
  \textit{ Eur. Phys. J. C} \textbf{ 76} (2016) 107,
  \href{http://dx.doi.org/10.1140/epjc/s10052-015-3819-5}{\doi{10.1140/epjc/s10052-015-3819-5}},
\href{http://www.arXiv.org/abs/1506.03981}{\texttt{arXiv:1506.03981}}.
%%CITATION = ARXIV:1506.03981;%%.

\bibitem{Shuryak:1977ut}
\href {http://jetp.ac.ru/cgi-bin/dn/e_047_02_0212.pdf}{E.~V. Shuryak, ``Theory
  of Hadronic Plasma'',} \textit{ Sov. Phys. JETP} \textbf{ 47} (1978)
212.
%%CITATION = SPHJA,47,212;%%.

\bibitem{Karsch:2003jg}
\hrefCMSnoop {}{F.~Karsch and E.~Laermann, ``Thermodynamics and in-medium
  hadron properties from lattice {QCD}'',} in \textit{ Quark-Gluon Plasma III},
  R.~C. Hwa and X.-N. Wang, eds.
\newblock World Scientific Publishing Co. Pte. Ltd., 2004.
\newblock
\href{http://www.arXiv.org/abs/hep-lat/0305025}{\texttt{arXiv:hep-lat/0305025}}.
\newblock
%%CITATION = HEP-LAT/0305025;%%.

\bibitem{Matsui:1986dk}
\hrefCMSnoop {}{T.~Matsui and H.~Satz, ``\JPsi suppression by quark-gluon
  plasma formation'',} \textit{ Phys. Lett. B} \textbf{ 178} (1986) 416,
\href{http://dx.doi.org/10.1016/0370-2693(86)91404-8}{\doi{10.1016/0370-2693(86)91404-8}}.
%%CITATION = PHLTA,B178,416;%%.

\bibitem{Dokshitzer:2001zm}
\hrefCMSnoop {}{Y.~L. Dokshitzer and D.~E. Kharzeev, ``Heavy quark colorimetry
  of {QCD} matter'',} \textit{ Phys. Lett. B} \textbf{ 519} (2001) 199,
  \href{http://dx.doi.org/10.1016/S0370-2693(01)01130-3}{\doi{10.1016/S0370-2693(01)01130-3}},
  \href{http://www.arXiv.org/abs/hep-ph/0106202}{\texttt{arXiv:hep-ph/0106202}}.

\bibitem{Aaij:2011jh}
\hrefCMSnoop {}{{LHCb} Collaboration, ``Measurement of \JPsi production in \pp
  collisions at $\sqrts = 7$\TeV'',} \textit{ Eur. Phys. J. C} \textbf{ 71}
  (2011) 1645,
  \href{http://dx.doi.org/10.1140/epjc/s10052-011-1645-y}{\doi{10.1140/epjc/s10052-011-1645-y}},
  \href{http://www.arXiv.org/abs/1103.0423}{\texttt{arXiv:1103.0423}}.

\bibitem{Khachatryan:2010yr}
\hrefCMSnoop {}{{CMS} Collaboration, ``Prompt and non-prompt \JPsi production
  in \pp collisions at $\sqrts = 7$\TeV'',} \textit{ Eur. Phys. J. C} \textbf{
  71} (2011) 1575,
  \href{http://dx.doi.org/10.1140/epjc/s10052-011-1575-8}{\doi{10.1140/epjc/s10052-011-1575-8}},
\href{http://www.arXiv.org/abs/1011.4193}{\texttt{arXiv:1011.4193}}.
%%CITATION = 1011.4193;%%.

\bibitem{Aad:2011sp}
\hrefCMSnoop {}{{ATLAS} Collaboration, ``Measurement of the differential
  cross-sections of inclusive, prompt and non-prompt \JPsi production in \pp
  collisions at $\sqrts = 7$\TeV'',} \textit{ Nucl. Phys. B} \textbf{ 850}
  (2011) 387,
  \href{http://dx.doi.org/10.1016/j.nuclphysb.2011.05.015}{\doi{10.1016/j.nuclphysb.2011.05.015}},
  \href{http://www.arXiv.org/abs/1104.3038}{\texttt{arXiv:1104.3038}}.

\bibitem{Mocsy:2007jz}
\hrefCMSnoop {}{{\'A}.~M{\'o}csy and P.~Petreczky, ``{Color screening melts
  quarkonium}'',} \textit{ Phys. Rev. Lett.} \textbf{ 99} (2007) 211602,
  \href{http://dx.doi.org/10.1103/PhysRevLett.99.211602}{\doi{10.1103/PhysRevLett.99.211602}},
  \href{http://www.arXiv.org/abs/0706.2183}{\texttt{arXiv:0706.2183}}.

\bibitem{Braaten:1991we}
\hrefCMSnoop {}{E.~Braaten and M.~H. Thoma, ``{Energy loss of a heavy quark in
  the quark-gluon plasma}'',} \textit{ Phys. Rev. D} \textbf{ 44} (1991) R2625,
\href{http://dx.doi.org/10.1103/PhysRevD.44.R2625}{\doi{10.1103/PhysRevD.44.R2625}}.
%%CITATION = PHRVA,D44,R2625;%%.

\bibitem{Zhang:2003wk}
\hrefCMSnoop {}{B.-W. Zhang, E.~Wang, and X.-N. Wang, ``{Heavy quark energy
  loss in nuclear medium}'',} \textit{ Phys. Rev. Lett.} \textbf{ 93} (2004)
  072301,
  \href{http://dx.doi.org/10.1103/PhysRevLett.93.072301}{\doi{10.1103/PhysRevLett.93.072301}},
\href{http://www.arXiv.org/abs/nucl-th/0309040}{\texttt{arXiv:nucl-th/0309040}}.
%%CITATION = NUCL-TH/0309040;%%.

\bibitem{Armesto:2005iq}
\hrefCMSnoop {}{N.~Armesto, A.~Dainese, C.~A. Salgado, and U.~A. Wiedemann,
  ``{Testing the color charge and mass dependence of parton energy loss with
  heavy-to-light ratios at BNL RHIC and CERN LHC}'',} \textit{ Phys. Rev. D}
  \textbf{ 71} (2005) 054027,
  \href{http://dx.doi.org/10.1103/PhysRevD.71.054027}{\doi{10.1103/PhysRevD.71.054027}},
  \href{http://www.arXiv.org/abs/hep-ph/0501225}{\texttt{arXiv:hep-ph/0501225}}.

\bibitem{vanHees:2005wb}
\hrefCMSnoop {}{H.~van Hees, V.~Greco, and R.~Rapp, ``{Heavy-quark probes of
  the quark-gluon plasma at RHIC}'',} \textit{ Phys. Rev. C} \textbf{ 73}
  (2006) 034913,
  \href{http://dx.doi.org/10.1103/PhysRevC.73.034913}{\doi{10.1103/PhysRevC.73.034913}},
\href{http://www.arXiv.org/abs/nucl-th/0508055}{\texttt{arXiv:nucl-th/0508055}}.
%%CITATION = NUCL-TH/0508055;%%.

\bibitem{Peigne:2008nd}
\hrefCMSnoop {}{S.~Peigne and A.~Peshier, ``{Collisional energy loss of a fast
  heavy quark in a quark-gluon plasma}'',} \textit{ Phys. Rev. D} \textbf{ 77}
  (2008) 114017,
  \href{http://dx.doi.org/10.1103/PhysRevD.77.114017}{\doi{10.1103/PhysRevD.77.114017}},
  \href{http://www.arXiv.org/abs/0802.4364}{\texttt{arXiv:0802.4364}}.

\bibitem{Wicks:2007am}
\hrefCMSnoop {}{S.~Wicks, W.~Horowitz, M.~Djordjevic, and M.~Gyulassy, ``{Heavy
  quark jet quenching with collisional plus radiative energy loss and path
  length fluctuations}'',} \textit{ Nucl. Phys. A} \textbf{ 783} (2007) 493,
  \href{http://dx.doi.org/10.1016/j.nuclphysa.2006.11.102}{\doi{10.1016/j.nuclphysa.2006.11.102}},
\href{http://www.arXiv.org/abs/nucl-th/0701063}{\texttt{arXiv:nucl-th/0701063}}.
%%CITATION = NUCL-TH/0701063;%%.

\bibitem{Gossiaux:2010yx}
\hrefCMSnoop {}{P.~B. Gossiaux, J.~Aichelin, T.~Gousset, and V.~Guiho,
  ``{Competition of heavy-quark radiative and collisional energy loss in
  deconfined matter}'',} \textit{ J. Phys. G} \textbf{ 37} (2010) 094019,
  \href{http://dx.doi.org/10.1088/0954-3899/37/9/094019}{\doi{10.1088/0954-3899/37/9/094019}},
\href{http://www.arXiv.org/abs/1001.4166}{\texttt{arXiv:1001.4166}}.
%%CITATION = ARXIV:1001.4166;%%.

\bibitem{Adil:2006ra}
\hrefCMSnoop {}{A.~Adil and I.~Vitev, ``{Collisional dissociation of heavy
  mesons in dense QCD matter}'',} \textit{ Phys. Lett. B} \textbf{ 649} (2007)
  139,
  \href{http://dx.doi.org/10.1016/j.physletb.2007.03.050}{\doi{10.1016/j.physletb.2007.03.050}},
\href{http://www.arXiv.org/abs/hep-ph/0611109}{\texttt{arXiv:hep-ph/0611109}}.
%%CITATION = HEP-PH/0611109;%%.

\bibitem{Sharma:2009hn}
\hrefCMSnoop {}{R.~Sharma, I.~Vitev, and B.-W. Zhang, ``{Light-cone wave
  function approach to open heavy flavor dynamics in QCD matter}'',} \textit{
  Phys. Rev. C} \textbf{ 80} (2009) 054902,
  \href{http://dx.doi.org/10.1103/PhysRevC.80.054902}{\doi{10.1103/PhysRevC.80.054902}},
\href{http://www.arXiv.org/abs/0904.0032}{\texttt{arXiv:0904.0032}}.
%%CITATION = ARXIV:0904.0032;%%.

\bibitem{Satz:2013ama}
\hrefCMSnoop {}{H.~Satz, ``{Calibrating the in-medium behavior of
  quarkonia}'',} \textit{ Adv. High Energy Phys.} \textbf{ 2013} (2013) 242918,
  \href{http://dx.doi.org/10.1155/2013/242918}{\doi{10.1155/2013/242918}},
\href{http://www.arXiv.org/abs/1303.3493}{\texttt{arXiv:1303.3493}}.
%%CITATION = ARXIV:1303.3493;%%.

\bibitem{Riek:2010fk}
\hrefCMSnoop {}{F.~Riek and R.~Rapp, ``{Quarkonia and heavy-quark relaxation
  times in the quark-gluon plasma}'',} \textit{ Phys. Rev. C} \textbf{ 82}
  (2010) 035201,
  \href{http://dx.doi.org/10.1103/PhysRevC.82.035201}{\doi{10.1103/PhysRevC.82.035201}},
\href{http://www.arXiv.org/abs/1005.0769}{\texttt{arXiv:1005.0769}}.
%%CITATION = ARXIV:1005.0769;%%.

\bibitem{Sharma:2012dy}
\hrefCMSnoop {}{R.~Sharma and I.~Vitev, ``{High transverse momentum quarkonium
  production and dissociation in heavy ion collisions}'',} \textit{ Phys. Rev.
  C} \textbf{ 87} (2013) 044905,
  \href{http://dx.doi.org/10.1103/PhysRevC.87.044905}{\doi{10.1103/PhysRevC.87.044905}},
\href{http://www.arXiv.org/abs/1203.0329}{\texttt{arXiv:1203.0329}}.
%%CITATION = ARXIV:1203.0329;%%.

\bibitem{Chatrchyan:2012np}
\hrefCMSnoop {}{{CMS} Collaboration, ``{Suppression of non-prompt \JPsi, prompt
  \JPsi, and \PgUa\ in \PbPb collisions at $\sqrtsnn=2.76$\TeV}'',} \textit{
  JHEP} \textbf{ 05} (2012) 063,
  \href{http://dx.doi.org/10.1007/JHEP05(2012)063}{\doi{10.1007/JHEP05(2012)063}},
\href{http://www.arXiv.org/abs/1201.5069}{\texttt{arXiv:1201.5069}}.
%%CITATION = ARXIV:1201.5069;%%.

\bibitem{Adam:2015rba}
\hrefCMSnoop {}{{ALICE} Collaboration, ``{Inclusive, prompt and non-prompt
  \JPsi production at mid-rapidity in Pb-Pb collisions at $\sqrtsnn =
  2.76$\TeV}'',} \textit{ JHEP} \textbf{ 07} (2015) 051,
  \href{http://dx.doi.org/10.1007/JHEP07(2015)051}{\doi{10.1007/JHEP07(2015)051}},
\href{http://www.arXiv.org/abs/1504.07151}{\texttt{arXiv:1504.07151}}.
%%CITATION = ARXIV:1504.07151;%%.

\bibitem{Abelev:2012rv}
\hrefCMSnoop {}{{ALICE} Collaboration, ``{\JPsi suppression at forward rapidity
  in Pb-Pb collisions at $\sqrtsnn=2.76$\TeV}'',} \textit{ Phys. Rev. Lett.}
  \textbf{ 109} (2012) 072301,
  \href{http://dx.doi.org/10.1103/PhysRevLett.109.072301}{\doi{10.1103/PhysRevLett.109.072301}},
\href{http://www.arXiv.org/abs/1202.1383}{\texttt{arXiv:1202.1383}}.
%%CITATION = ARXIV:1202.1383;%%.

\bibitem{Adare:2011yf}
\hrefCMSnoop {}{{PHENIX} Collaboration, ``\JPsi suppression at forward rapidity
  in AuAu collisions at $\sqrtsnn = 200$\GeV'',} \textit{ Phys. Rev. C}
  \textbf{ 84} (2011) 054912,
  \href{http://dx.doi.org/10.1103/PhysRevC.84.054912}{\doi{10.1103/PhysRevC.84.054912}},
\href{http://www.arXiv.org/abs/1103.6269}{\texttt{arXiv:1103.6269}}.
%%CITATION = 1103.6269;%%.

\bibitem{Adam:2015isa}
\hrefCMSnoop {}{{ALICE} Collaboration, ``{Differential studies of inclusive
  \JPsi and \psiP production at forward rapidity in Pb-Pb collisions at $
  \sqrtsnn=2.76$\TeV}'',} \textit{ JHEP} \textbf{ 05} (2016) 179,
  \href{http://dx.doi.org/10.1007/JHEP05(2016)179}{\doi{10.1007/JHEP05(2016)179}},
\href{http://www.arXiv.org/abs/1506.08804}{\texttt{arXiv:1506.08804}}.
%%CITATION = ARXIV:1506.08804;%%.

\bibitem{Zhao:2011cv}
\hrefCMSnoop {}{X.~Zhao and R.~Rapp, ``{Medium modifications and production of
  charmonia at LHC}'',} \textit{ Nucl. Phys. A} \textbf{ 859} (2011) 114,
  \href{http://dx.doi.org/10.1016/j.nuclphysa.2011.05.001}{\doi{10.1016/j.nuclphysa.2011.05.001}},
  \href{http://www.arXiv.org/abs/1102.2194}{\texttt{arXiv:1102.2194}}.

\bibitem{Andronic:2011yq}
\hrefCMSnoop {}{A.~Andronic, P.~Braun-Munzinger, K.~Redlich, and J.~Stachel,
  ``{The thermal model on the verge of the ultimate test: particle production
  in Pb-Pb collisions at the LHC}'',} \textit{ J. Phys. G} \textbf{ 38} (2011)
  124081,
  \href{http://dx.doi.org/10.1088/0954-3899/38/12/124081}{\doi{10.1088/0954-3899/38/12/124081}},
\href{http://www.arXiv.org/abs/1106.6321}{\texttt{arXiv:1106.6321}}.
%%CITATION = ARXIV:1106.6321;%%.

\bibitem{Ferreiro:2012rq}
\hrefCMSnoop {}{E.~G. Ferreiro, ``{Charmonium dissociation and recombination at
  LHC: Revisiting comovers}'',} \textit{ Phys. Lett. B} \textbf{ 731} (2014)
  57,
  \href{http://dx.doi.org/10.1016/j.physletb.2014.02.011}{\doi{10.1016/j.physletb.2014.02.011}},
\href{http://www.arXiv.org/abs/1210.3209}{\texttt{arXiv:1210.3209}}.
%%CITATION = ARXIV:1210.3209;%%.

\bibitem{Ollitrault:1992bk}
\hrefCMSnoop {}{J.-Y. Ollitrault, ``{Anisotropy as a signature of transverse
  collective flow}'',} \textit{ Phys. Rev. D} \textbf{ 46} (1992) 229,
\href{http://dx.doi.org/10.1103/PhysRevD.46.229}{\doi{10.1103/PhysRevD.46.229}}.
%%CITATION = PHRVA,D46,229;%%.

\bibitem{Abelev:2014ipa}
\hrefCMSnoop {}{{ALICE} Collaboration, ``{Azimuthal anisotropy of D meson
  production in Pb-Pb collisions at $\sqrtsnn = 2.76$\TeV}'',} \textit{ Phys.
  Rev. C} \textbf{ 90} (2014) 034904,
  \href{http://dx.doi.org/10.1103/PhysRevC.90.034904}{\doi{10.1103/PhysRevC.90.034904}},
\href{http://www.arXiv.org/abs/1405.2001}{\texttt{arXiv:1405.2001}}.
%%CITATION = ARXIV:1405.2001;%%.

\bibitem{ALICE:2013xna}
\hrefCMSnoop {}{{ALICE} Collaboration, ``{\JPsi elliptic flow in \PbPb
  collisions at $\sqrtsnn = 2.76$\TeV}'',} \textit{ Phys. Rev. Lett.} \textbf{
  111} (2013) 162301,
  \href{http://dx.doi.org/10.1103/PhysRevLett.111.162301}{\doi{10.1103/PhysRevLett.111.162301}},
\href{http://www.arXiv.org/abs/1303.5880}{\texttt{arXiv:1303.5880}}.
%%CITATION = ARXIV:1303.5880;%%.

\bibitem{BraunMunzinger:2000px}
\hrefCMSnoop {}{P.~Braun-Munzinger and J.~Stachel, ``{(Non)thermal aspects of
  charmonium production and a new look at \JPsi suppression}'',} \textit{ Phys.
  Lett. B} \textbf{ 490} (2000) 196,
  \href{http://dx.doi.org/10.1016/S0370-2693(00)00991-6}{\doi{10.1016/S0370-2693(00)00991-6}},
\href{http://www.arXiv.org/abs/nucl-th/0007059}{\texttt{arXiv:nucl-th/0007059}}.
%%CITATION = NUCL-TH/0007059;%%.

\bibitem{Liu:2009nb}
\hrefCMSnoop {}{Y.~Liu, Z.~Qu, N.~Xu, and P.~Zhuang, ``{\JPsi transverse
  momentum distribution in high energy nuclear collisions at RHIC}'',} \textit{
  Phys. Lett. B} \textbf{ 678} (2009) 72,
  \href{http://dx.doi.org/10.1016/j.physletb.2009.06.006}{\doi{10.1016/j.physletb.2009.06.006}},
\href{http://www.arXiv.org/abs/0901.2757}{\texttt{arXiv:0901.2757}}.
%%CITATION = ARXIV:0901.2757;%%.

\bibitem{bib_CMS}
\hrefCMSnoop {}{{CMS} Collaboration, ``The {CMS} experiment at the {CERN}
  {LHC}'',} \textit{ JINST} \textbf{ 3} (2008) S08004,
  \href{http://dx.doi.org/10.1088/1748-0221/3/08/S08004}{\doi{10.1088/1748-0221/3/08/S08004}}.

\bibitem{Chatrchyan:2011sx}
\hrefCMSnoop {}{{CMS} Collaboration, ``{Observation and studies of jet
  quenching in \PbPb collisions at $\sqrtsnn = 2.76$\TeV}'',} \textit{ Phys.
  Rev. C} \textbf{ 84} (2011) 024906,
  \href{http://dx.doi.org/10.1103/PhysRevC.84.024906}{\doi{10.1103/PhysRevC.84.024906}},
  \href{http://www.arXiv.org/abs/1102.1957}{\texttt{arXiv:1102.1957}}.

\bibitem{CMS-PAS-LUM-13-002}
\href {http://cds.cern.ch/record/1643269}{{CMS} Collaboration, ``Luminosity
  calibration for the 2013 proton-lead and proton-proton data taking'',} CMS
  Physics Analysis Summary CMS-PAS-LUM-13-002, 2013.

\bibitem{Roland:2006kz}
\hrefCMSnoop {}{C.~Roland, ``{Track reconstruction in heavy ion collisions with
  the CMS silicon tracker}'',} in \textit{ {TIME} 20005 --- Proceedings of the
  1st Workshop on Tracking in High Multiplicity Environments}.
\newblock 2006.
\newblock [{Nucl. Instrum. Meth. A 566 (2006) 123}].
  \href{http://dx.doi.org/10.1016/j.nima.2006.05.023}{\doi{10.1016/j.nima.2006.05.023}}.

\bibitem{Chatrchyan:2014fea}
\hrefCMSnoop {}{{CMS} Collaboration, ``{Description and performance of track
  and primary-vertex reconstruction with the CMS tracker}'',} \textit{ JINST}
  \textbf{ 9} (2014) P10009,
  \href{http://dx.doi.org/10.1088/1748-0221/9/10/P10009}{\doi{10.1088/1748-0221/9/10/P10009}},
\href{http://www.arXiv.org/abs/1405.6569}{\texttt{arXiv:1405.6569}}.
%%CITATION = ARXIV:1405.6569;%%.

\bibitem{Miller:2007ri}
\hrefCMSnoop {}{M.~L. Miller, K.~Reygers, S.~J. Sanders, and P.~Steinberg,
  ``{Glauber modeling in high-energy nuclear collisions}'',} \textit{ Ann. Rev.
  Nucl. Part. Sci.} \textbf{ 57} (2007) 205,
  \href{http://dx.doi.org/10.1146/annurev.nucl.57.090506.123020}{\doi{10.1146/annurev.nucl.57.090506.123020}},
  \href{http://www.arXiv.org/abs/nucl-ex/0701025}{\texttt{arXiv:nucl-ex/0701025}}.

\bibitem{Sjostrand:2006za}
\hrefCMSnoop {}{T.~Sj{\"o}strand, S.~Mrenna, and P.~Z. Skands, ``{PYTHIA 6.4
  physics and manual}'',} \textit{ JHEP} \textbf{ 05} (2006) 026,
  \href{http://dx.doi.org/10.1088/1126-6708/2006/05/026}{\doi{10.1088/1126-6708/2006/05/026}},
  \href{http://www.arXiv.org/abs/hep-ph/0603175}{\texttt{arXiv:hep-ph/0603175}}.

\bibitem{Lange:2001uf}
\hrefCMSnoop {}{D.~J. Lange, ``{The EvtGen particle decay simulation
  package}'',} \textit{ Nucl. Instrum. Meth. A} \textbf{ 462} (2001) 152,
\href{http://dx.doi.org/10.1016/S0168-9002(01)00089-4}{\doi{10.1016/S0168-9002(01)00089-4}}.
%%CITATION = NUIMA,A462,152;%%.

\bibitem{Barberio:1993qi}
\hrefCMSnoop {}{E.~Barberio and Z.~W{\c{a}}s, ``{PHOTOS --- A universal Monte
  Carlo for QED radiative corrections: version 2.0}'',} \textit{ Comput. Phys.
  Commun.} \textbf{ 79} (1994) 291,
  \href{http://dx.doi.org/10.1016/0010-4655(94)90074-4}{\doi{10.1016/0010-4655(94)90074-4}}.

\bibitem{Abelev:2011md}
\hrefCMSnoop {}{{ALICE} Collaboration, ``{\JPsi polarization in \pp collisions
  at $\sqrts=7$\TeV}'',} \textit{ Phys. Rev. Lett.} \textbf{ 108} (2012)
  082001,
  \href{http://dx.doi.org/10.1103/PhysRevLett.108.082001}{\doi{10.1103/PhysRevLett.108.082001}},
\href{http://www.arXiv.org/abs/1111.1630}{\texttt{arXiv:1111.1630}}.
%%CITATION = ARXIV:1111.1630;%%.

\bibitem{Chatrchyan:2013cla}
\hrefCMSnoop {}{{CMS} Collaboration, ``{Measurement of the prompt \JPsi and
  \psiP polarizations in \pp collisions at $\sqrts = 7$\TeV}'',} \textit{ Phys.
  Lett. B} \textbf{ 727} (2013) 381,
  \href{http://dx.doi.org/10.1016/j.physletb.2013.10.055}{\doi{10.1016/j.physletb.2013.10.055}},
\href{http://www.arXiv.org/abs/1307.6070}{\texttt{arXiv:1307.6070}}.
%%CITATION = ARXIV:1307.6070;%%.

\bibitem{Aaij:2013nlm}
\hrefCMSnoop {}{{LHCb} Collaboration, ``{Measurement of \JPsi polarization in
  \pp collisions at $\sqrts=7$\TeV}'',} \textit{ Eur. Phys. J. C} \textbf{ 73}
  (2013) 2631,
  \href{http://dx.doi.org/10.1140/epjc/s10052-013-2631-3}{\doi{10.1140/epjc/s10052-013-2631-3}},
\href{http://www.arXiv.org/abs/1307.6379}{\texttt{arXiv:1307.6379}}.
%%CITATION = ARXIV:1307.6379;%%.

\bibitem{Chatrchyan:2011pe}
\hrefCMSnoop {}{{CMS} Collaboration, ``Indications of suppression of excited
  \PgU\ states in \PbPb collisions at $\sqrtsnn = 2.76$\TeV'',} \textit{ Phys.
  Rev. Lett.} \textbf{ 107} (2011) 052302,
  \href{http://dx.doi.org/10.1103/PhysRevLett.107.052302}{\doi{10.1103/PhysRevLett.107.052302}},
\href{http://www.arXiv.org/abs/1105.4894}{\texttt{arXiv:1105.4894}}.
%%CITATION = 1105.4894;%%.

\bibitem{Chatrchyan:2012lxa}
\hrefCMSnoop {}{{CMS} Collaboration, ``{Observation of sequential Upsilon
  suppression in PbPb collisions}'',} \textit{ Phys. Rev. Lett.} \textbf{ 109}
  (2012) 222301,
  \href{http://dx.doi.org/10.1103/PhysRevLett.109.222301}{\doi{10.1103/PhysRevLett.109.222301}},
\href{http://www.arXiv.org/abs/1208.2826}{\texttt{arXiv:1208.2826}}.
%%CITATION = ARXIV:1208.2826;%%.

\bibitem{Chatrchyan:2013nza}
\hrefCMSnoop {}{{CMS} Collaboration, ``{Event activity dependence of
  $\Upsilon\mathrm{(nS)}$ production in $\sqrtsnn=5.02$\TeV pPb and
  $\sqrts=2.76$\TeV pp collisions}'',} \textit{ JHEP} \textbf{ 04} (2014) 103,
  \href{http://dx.doi.org/10.1007/JHEP04(2014)103}{\doi{10.1007/JHEP04(2014)103}},
\href{http://www.arXiv.org/abs/1312.6300}{\texttt{arXiv:1312.6300}}.
%%CITATION = ARXIV:1312.6300;%%.

\bibitem{Khachatryan:2014bva}
\hrefCMSnoop {}{{CMS} Collaboration, ``{Measurement of prompt $\psiP \to \JPsi$
  yield ratios in \PbPb and \pp collisions at $\sqrtsnn=2.76$\TeV}'',} \textit{
  Phys. Rev. Lett.} \textbf{ 113} (2014) 262301,
  \href{http://dx.doi.org/10.1103/PhysRevLett.113.262301}{\doi{10.1103/PhysRevLett.113.262301}},
\href{http://www.arXiv.org/abs/1410.1804}{\texttt{arXiv:1410.1804}}.
%%CITATION = ARXIV:1410.1804;%%.

\bibitem{Lokhtin:2005px}
\hrefCMSnoop {}{I.~P. Lokhtin and A.~M. Snigirev, ``{A model of jet quenching
  in ultrarelativistic heavy ion collisions and high-\pt hadron spectra at
  RHIC}'',} \textit{ Eur. Phys. J. C} \textbf{ 45} (2006) 211,
  \href{http://dx.doi.org/10.1140/epjc/s2005-02426-3}{\doi{10.1140/epjc/s2005-02426-3}},
\href{http://www.arXiv.org/abs/hep-ph/0506189}{\texttt{arXiv:hep-ph/0506189}}.
%%CITATION = HEP-PH/0506189;%%.

\bibitem{Agostinelli:2002hh}
\hrefCMSnoop {}{{GEANT4} Collaboration, ``{GEANT4 --- A simulation toolkit}'',}
  \textit{ Nucl. Instrum. Meth. A} \textbf{ 506} (2003) 250,
\href{http://dx.doi.org/10.1016/S0168-9002(03)01368-8}{\doi{10.1016/S0168-9002(03)01368-8}}.
%%CITATION = NUIMA,A506,250;%%.

\bibitem{Chatrchyan:2012xi}
\hrefCMSnoop {}{{CMS} Collaboration, ``{Performance of CMS muon reconstruction
  in \pp collision events at $\sqrts=7$ TeV}'',} \textit{ JINST} \textbf{ 7}
  (2012) P10002,
  \href{http://dx.doi.org/10.1088/1748-0221/7/10/P10002}{\doi{10.1088/1748-0221/7/10/P10002}},
\href{http://www.arXiv.org/abs/1206.4071}{\texttt{arXiv:1206.4071}}.
%%CITATION = ARXIV:1206.4071;%%.

\bibitem{Chatrchyan:2011pb}
\hrefCMSnoop {}{{CMS} Collaboration, ``{Dependence on pseudorapidity and
  centrality of charged hadron production in \PbPb collisions at $\sqrtsnn =
  2.76$\TeV}'',} \textit{ JHEP} \textbf{ 08} (2011) 141,
  \href{http://dx.doi.org/10.1007/JHEP08(2011)141}{\doi{10.1007/JHEP08(2011)141}},
\href{http://www.arXiv.org/abs/1107.4800}{\texttt{arXiv:1107.4800}}.
%%CITATION = 1107.4800;%%.

\bibitem{Buskulic:1993vi}
\hrefCMSnoop {}{{ALEPH} Collaboration, ``{Measurement of the anti-B$^{0}$ and
  B$^{-}$ meson lifetimes}'',} \textit{ Phys. Lett. B} \textbf{ 307} (1993)
  194--208,
  \href{http://dx.doi.org/10.1016/0370-2693(93)90211-Y}{\doi{10.1016/0370-2693(93)90211-Y}}.
[Erratum: \DOI{10.1016/0370-2693(94)90054-X}].
%%CITATION = PHLTA,B307,194;%%.

\bibitem{Chatrchyan:2011kc}
\hrefCMSnoop {}{{CMS} Collaboration, ``{\JPsi and \psiP production in \pp
  collisions at $\sqrts=7$\TeV}'',} \textit{ JHEP} \textbf{ 02} (2012) 011,
  \href{http://dx.doi.org/10.1007/JHEP02(2012)011}{\doi{10.1007/JHEP02(2012)011}},
\href{http://www.arXiv.org/abs/1111.1557}{\texttt{arXiv:1111.1557}}.
%%CITATION = ARXIV:1111.1557;%%.

\bibitem{Oreglia:1980cs}
\href
  {http://www-public.slac.stanford.edu/sciDoc/docMeta.aspx?slacPubNumber=slac-r-236.html}{M.~Oreglia,
  ``{A Study of the Reactions $\psi^\prime \to \gamma \gamma \psi$}''}.
\newblock PhD thesis, SLAC,
1980.
\newblock
%%CITATION = SLAC-0236;%%.

\bibitem{Chatrchyan:2012ta}
\hrefCMSnoop {}{{CMS} Collaboration, ``{Measurement of the elliptic anisotropy
  of charged particles produced in PbPb collisions at nucleon-nucleon
  center-of-mass energy = 2.76 TeV}'',} \textit{ Phys. Rev. C} \textbf{ 87}
  (2013) 014902,
  \href{http://dx.doi.org/10.1103/PhysRevC.87.014902}{\doi{10.1103/PhysRevC.87.014902}},
\href{http://www.arXiv.org/abs/1204.1409}{\texttt{arXiv:1204.1409}}.
%%CITATION = ARXIV:1204.1409;%%.

\bibitem{Poskanzer:1998yz}
\hrefCMSnoop {}{A.~M. Poskanzer and S.~A. Voloshin, ``{Methods for analyzing
  anisotropic flow in relativistic nuclear collisions}'',} \textit{ Phys. Rev.
  C} \textbf{ 58} (1998) 1671,
  \href{http://dx.doi.org/10.1103/PhysRevC.58.1671}{\doi{10.1103/PhysRevC.58.1671}},
\href{http://www.arXiv.org/abs/nucl-ex/9805001}{\texttt{arXiv:nucl-ex/9805001}}.
%%CITATION = NUCL-EX/9805001;%%.

\bibitem{Chatrchyan:2012xq}
\hrefCMSnoop {}{{CMS} Collaboration, ``{Azimuthal anisotropy of charged
  particles at high transverse momenta in PbPb collisions at
  $\sqrtsnn=2.76$\TeV}'',} \textit{ Phys. Rev. Lett.} \textbf{ 109} (2012)
  022301,
  \href{http://dx.doi.org/10.1103/PhysRevLett.109.022301}{\doi{10.1103/PhysRevLett.109.022301}},
\href{http://www.arXiv.org/abs/1204.1850}{\texttt{arXiv:1204.1850}}.
%%CITATION = ARXIV:1204.1850;%%.

\bibitem{ALICE:2012ab}
\hrefCMSnoop {}{{ALICE} Collaboration, ``{Suppression of high transverse
  momentum D mesons in central Pb-Pb collisions at $\sqrtsnn=2.76$\TeV}'',}
  \textit{ JHEP} \textbf{ 09} (2012) 112,
  \href{http://dx.doi.org/10.1007/JHEP09(2012)112}{\doi{10.1007/JHEP09(2012)112}},
\href{http://www.arXiv.org/abs/1203.2160}{\texttt{arXiv:1203.2160}}.
%%CITATION = ARXIV:1203.2160;%%.

\bibitem{Adam:2015nna}
\hrefCMSnoop {}{{ALICE} Collaboration, ``{Centrality dependence of high-\pt D
  meson suppression in Pb-Pb collisions at $\sqrtsnn=2.76$\TeV}'',} \textit{
  JHEP} \textbf{ 11} (2015) 205,
  \href{http://dx.doi.org/10.1007/JHEP11(2015)205}{\doi{10.1007/JHEP11(2015)205}},
\href{http://www.arXiv.org/abs/1506.06604}{\texttt{arXiv:1506.06604}}.
%%CITATION = ARXIV:1506.06604;%%.

\bibitem{Arleo:2013zua}
\hrefCMSnoop {}{F.~Arleo, R.~Kolevatov, S.~Peign{\'e}, and M.~Rustamova,
  ``{Centrality and $p_\perp$ dependence of \JPsi suppression in proton-nucleus
  collisions from parton energy loss}'',} \textit{ JHEP} \textbf{ 05} (2013)
  155,
  \href{http://dx.doi.org/10.1007/JHEP05(2013)155}{\doi{10.1007/JHEP05(2013)155}},
\href{http://www.arXiv.org/abs/1304.0901}{\texttt{arXiv:1304.0901}}.
%%CITATION = ARXIV:1304.0901;%%.

\bibitem{Ferreiro:2014bia}
\hrefCMSnoop {}{E.~G. Ferreiro, ``{Excited charmonium suppression in
  proton-nucleus collisions as a consequence of comovers}'',} \textit{ Phys.
  Lett. B} \textbf{ 749} (2015) 98,
  \href{http://dx.doi.org/10.1016/j.physletb.2015.07.066}{\doi{10.1016/j.physletb.2015.07.066}},
\href{http://www.arXiv.org/abs/1411.0549}{\texttt{arXiv:1411.0549}}.
%%CITATION = ARXIV:1411.0549;%%.

\bibitem{Fujii:2013gxa}
\hrefCMSnoop {}{H.~Fujii and K.~Watanabe, ``{Heavy quark pair production in
  high energy pA collisions: Quarkonium}'',} \textit{ Nucl. Phys. A} \textbf{
  915} (2013) 1,
  \href{http://dx.doi.org/10.1016/j.nuclphysa.2013.06.011}{\doi{10.1016/j.nuclphysa.2013.06.011}},
\href{http://www.arXiv.org/abs/1304.2221}{\texttt{arXiv:1304.2221}}.
%%CITATION = ARXIV:1304.2221;%%.

\bibitem{Adam:2015jsa}
\hrefCMSnoop {}{{ALICE} Collaboration, ``{Centrality dependence of inclusive
  \JPsi production in p-Pb collisions at $ \sqrtsnn=5.02$\TeV}'',} \textit{
  JHEP} \textbf{ 11} (2015) 127,
  \href{http://dx.doi.org/10.1007/JHEP11(2015)127}{\doi{10.1007/JHEP11(2015)127}},
\href{http://www.arXiv.org/abs/1506.08808}{\texttt{arXiv:1506.08808}}.
%%CITATION = ARXIV:1506.08808;%%.

\bibitem{Adam:2015iga}
\hrefCMSnoop {}{{ALICE} Collaboration, ``{Rapidity and transverse-momentum
  dependence of the inclusive \JPsi nuclear modification factor in p-Pb
  collisions at $\sqrtsnn = 5.02$\TeV}'',} \textit{ JHEP} \textbf{ 06} (2015)
  055,
  \href{http://dx.doi.org/10.1007/JHEP06(2015)055}{\doi{10.1007/JHEP06(2015)055}},
\href{http://www.arXiv.org/abs/1503.07179}{\texttt{arXiv:1503.07179}}.
%%CITATION = ARXIV:1503.07179;%%.

\bibitem{Faccioli:2008ir}
\hrefCMSnoop {}{P.~Faccioli, C.~Lourenco, J.~Seixas, and H.~K. Woehri, ``{Study
  of \psiP and $\chi_c$ decays as feed-down sources of \JPsi
  hadro-production}'',} \textit{ JHEP} \textbf{ 10} (2008) 004,
  \href{http://dx.doi.org/10.1088/1126-6708/2008/10/004}{\doi{10.1088/1126-6708/2008/10/004}},
\href{http://www.arXiv.org/abs/0809.2153}{\texttt{arXiv:0809.2153}}.
%%CITATION = ARXIV:0809.2153;%%.

\bibitem{LHCb:2012af}
\hrefCMSnoop {}{{LHCb} Collaboration, ``{Measurement of the ratio of prompt
  $\chi_{c}$ to \JPsi production in pp collisions at $\sqrt{s}=7$ TeV}'',}
  \textit{ Phys. Lett. B} \textbf{ 718} (2012) 431,
  \href{http://dx.doi.org/10.1016/j.physletb.2012.10.068}{\doi{10.1016/j.physletb.2012.10.068}},
\href{http://www.arXiv.org/abs/1204.1462}{\texttt{arXiv:1204.1462}}.
%%CITATION = ARXIV:1204.1462;%%.

\bibitem{He:2014cla}
\hrefCMSnoop {}{M.~He, R.~J. Fries, and R.~Rapp, ``{Heavy flavor at the Large
  Hadron Collider in a strong coupling approach}'',} \textit{ Phys. Lett. B}
  \textbf{ 735} (2014) 445,
  \href{http://dx.doi.org/10.1016/j.physletb.2014.05.050}{\doi{10.1016/j.physletb.2014.05.050}},
\href{http://www.arXiv.org/abs/1401.3817}{\texttt{arXiv:1401.3817}}.
%%CITATION = ARXIV:1401.3817;%%.

\bibitem{Adam:2015sza}
\hrefCMSnoop {}{{ALICE} Collaboration, ``{Transverse momentum dependence of
  D-meson production in Pb-Pb collisions at $ \sqrtsnn=2.76$\TeV}'',} \textit{
  JHEP} \textbf{ 03} (2016) 081,
  \href{http://dx.doi.org/10.1007/JHEP03(2016)081}{\doi{10.1007/JHEP03(2016)081}},
\href{http://www.arXiv.org/abs/1509.06888}{\texttt{arXiv:1509.06888}}.
%%CITATION = ARXIV:1509.06888;%%.

\bibitem{Dokshitzer:1991fd}
\hrefCMSnoop {}{Y.~L. Dokshitzer, V.~A. Khoze, and S.~I. Troian, ``{On specific
  QCD properties of heavy quark fragmentation (`dead cone')}'',} \textit{ J.
  Phys. G} \textbf{ 17} (1991) 1602,
\href{http://dx.doi.org/10.1088/0954-3899/17/10/023}{\doi{10.1088/0954-3899/17/10/023}}.
%%CITATION = JPAGA,G17,1602;%%.

\bibitem{Armesto:2003jh}
\hrefCMSnoop {}{N.~Armesto, C.~A. Salgado, and U.~A. Wiedemann, ``{Medium
  induced gluon radiation off massive quarks fills the dead cone}'',} \textit{
  Phys. Rev. D} \textbf{ 69} (2004) 114003,
  \href{http://dx.doi.org/10.1103/PhysRevD.69.114003}{\doi{10.1103/PhysRevD.69.114003}},
\href{http://www.arXiv.org/abs/hep-ph/0312106}{\texttt{arXiv:hep-ph/0312106}}.
%%CITATION = HEP-PH/0312106;%%.

\bibitem{Djordjevic:2003zk}
\hrefCMSnoop {}{M.~Djordjevic and M.~Gyulassy, ``{Heavy quark radiative energy
  loss in QCD matter}'',} \textit{ Nucl. Phys. A} \textbf{ 733} (2004) 265,
  \href{http://dx.doi.org/10.1016/j.nuclphysa.2003.12.020}{\doi{10.1016/j.nuclphysa.2003.12.020}},
\href{http://www.arXiv.org/abs/nucl-th/0310076}{\texttt{arXiv:nucl-th/0310076}}.
%%CITATION = NUCL-TH/0310076;%%.

\bibitem{Chatrchyan:2013exa}
\hrefCMSnoop {}{{CMS} Collaboration, ``Evidence of $b$-Jet Quenching in {PbPb}
  Collisions at {$\sqrtsnn=2.76$\TeV}'',} \textit{ Phys. Rev. Lett.} \textbf{
  113} (2014) 132301,
  \href{http://dx.doi.org/10.1103/PhysRevLett.113.132301}{\doi{10.1103/PhysRevLett.113.132301}},
  \href{http://www.arXiv.org/abs/1312.4198}{\texttt{arXiv:1312.4198}}.
[Erratum: \DOI{10.1103/PhysRevLett.115.029903}].
%%CITATION = ARXIV:1312.4198;%%.

\bibitem{Djordjevic:2013pba}
\hrefCMSnoop {}{M.~Djordjevic, ``Heavy Flavor Puzzle at {LHC}: A Serendipitous
  Interplay of Jet Suppression and Fragmentation'',} \textit{ Phys. Rev. Lett.}
  \textbf{ 112} (2014) 042302,
  \href{http://dx.doi.org/10.1103/PhysRevLett.112.042302}{\doi{10.1103/PhysRevLett.112.042302}},
\href{http://www.arXiv.org/abs/1307.4702}{\texttt{arXiv:1307.4702}}.
%%CITATION = ARXIV:1307.4702;%%.

\end{thebibliography}\endgroup
